\documentclass[manuscript]{aastex631}
\usepackage{gensymb}
\usepackage{rotating}

\begin{document}

\title{Multiple overspill flood channels from young craters require surface melting and hundreds of meters of mid-latitude ice late in Mars history}

\author{Alexandra O. Warren}
\affiliation{Department of the Geophysical Sciences, University of Chicago\\
5734 S. Ellis Ave\\
Chicago, IL 60637, USA}

\author{Sharon A. Wilson}
\affiliation{Center for Earth and Planetary Studies, National Air and Space Museum\\
Smithsonian Institution, 6th at Independence SW \\
Washington, DC 20560, USA}

\author{Alan Howard}
\affiliation{Planetary Science Institute\\
1700 East Fort Lowell, \\
Tucson, AZ 85719, USA}


\author{Axel Noblet}
\affiliation{Department of Earth Sciences, University of Western Ontario \\ 1151 Richmond St. N. \\ London, ON N6A 5B7,
Canada}

\author{Edwin S. Kite}
\affiliation{Department of the Geophysical Sciences, University of Chicago\\
5734 S. Ellis Ave\\
Chicago, IL 60637, USA}



\section{Abstract}
Mars' tadpole craters are small, young craters whose crater rims are incised by one or more exit breaches but lack visible inlets. The tadpole forming climate records the poorly understood drying of Mars since the Early Hesperian. A third of tadpole craters have multiple breaches, therefore a process is needed that was able to generate crater rim incision in multiple locations. We use HiRISE data for four multiple breach tadpole craters to measure their crater fill, rims, and exit breaches. We compare these measurements and other data to our calculations of liquid water supply by rain, surface melting, groundwater discharge, and basal ice sheet melting to discriminate between four proposed formation hypotheses for tadpole breaches, favoring scenarios with ice-filled craters and supraglacial melting. We conclude that multiple breach tadpole craters record hundreds of meters of mid-latitude ice and climate conditions enabling intermittent melting in the Late Hesperian and Amazonian, suggesting that liquid water on Mars has been available in association with water ice for billions of years.

\section{Introduction} \label{sec:intro}

The surface of Mars records a climate catastrophe. A planet with enough liquid water to carve canyons and overspill lakes in the Noachian ($\sim$3.6~Ga) \citep{fassett2008valley} became a cold, arid desert where water is so unstable it seldom occurs as anything but ice or vapor \citep{murphy_review_2005}. This great drying of Mars \citep{kite_geologic_2019,kite2022high,scheller2021longterm} is recorded by water-worn features that become increasingly spatially patchy and imply increasingly arid conditions over time \citep{holo2021timing,grant2011late,kite2019persistence,wilson2021global}. This evolution has been driven by the progressive loss of the atmosphere to space or carbonate formation \citep{hu_tracing_2015-1}, changes in the available surface and atmospheric water inventory \citep{mahaffy2015imprint}, loss of non-CO$_{2}$ greenhouse forcing \citep{Wordsworth2021coupled,kite2022changing}, and changes in orbital parameters \citep{haberle2017atmosphere}. However, the precise causes, timing, duration, and spatial pattern of this drying out of Mars are still poorly constrained \citep{kite2022high}. In this paper, we examine the climate constraints on the cooling and drying of Mars provided by ``tadpole craters.''  

We define tadpole craters (also known as ``pollywog craters'') as small craters, extending up to 32~km across but typically ranging from 0.5-15~km in diameter, with at least one outlet breach leading away from their rims and no visible inlets. Outlet breaches indicate that at one time the rims of these craters were incised by liquid water \citep{wilson_cold-wet_2016,warren2021overspilling}. Unlike other preserved overspilling (i.e. hydrologically open) paleolakes on Mars \citep{bamber2022constraining,goudge2016insights,fassett2008valley,bamber2022paleolake,goudge2023assessing}, tadpole craters do not have inlet breaches and do not appear to form integrated chains of lakes. Tadpole craters occur in the mid-latitudes in both hemispheres \citep{wilson_cold-wet_2016,howard2022distribution}, with the best-studied group of tadpole craters in Arabia Terra \citep{wilson_cold-wet_2016,warren2021overspilling}. Tadpole craters exist close to the equator \citep{warren2021overspilling} but these appear to be unusual. 

\begin{figure}[t!]
\centering
\includegraphics[width=0.8\linewidth]{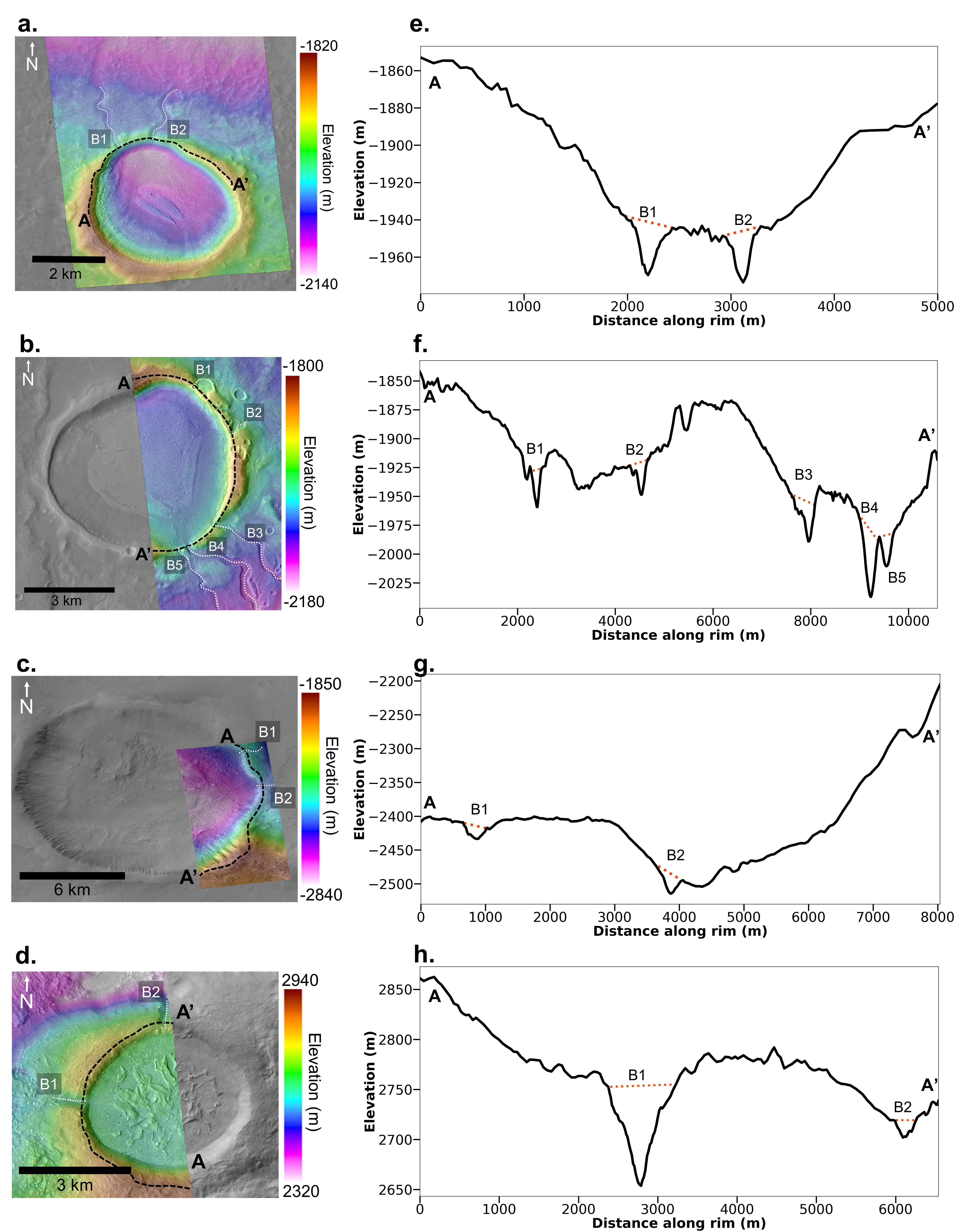}
\caption{a-d) Digital Elevation Models (DEMs) and e-h) topographic rim profiles of multiple breach tadpole craters. \textbf{a.} \textit{Arabia Terra~1}, \textbf{b.} \textit{Arabia Terra~2}, \textbf{c.} \textit{Hellas}, \textbf{d.} \textit{Noachis} (Tadpole crater nicknames, latitude/longitude, and image numbers given in Table~\ref{tab:mebtps}). B1-B5 mark breaches incised into the rim. Solid black lines show measured rim topography from A to A', dotted red lines show interpolated pre-incision topography between fresh-appearing (i.e. unincised) rim sections either side of each breach.}
\label{fig:profs}
\end{figure}

In this paper, we focus on tadpole craters with multiple breaches. These represent about one-third of the overall tadpole crater population in the latitude band 24-52$^{\circ}$~S. Multiple breach tadpole craters are difficult to understand because the most common cause of dam breaches on Earth is overtopping of the dam wall by water \citep{costa1988formation,neupane2019review}. The lowest point on the crater rim is the most likely location for overtopping to occur, however, we observe tadpole craters with tens of meters of vertical separation between the breach floors of their multiple outlets (Fig.~\ref{fig:profs}). Therefore, a mechanism must exist that can fill craters, generate liquid water, and route that water to multiple elevations around a crater rim. This constrains the climate during multiple breach tadpole crater overspill.

Although tadpole craters are too small to apply crater dating methods \citep{michael2010planetary,warner2015minimum}, few $\leq$4~km craters pre-date the Noachian-Hesperian boundary \citep{irwin2013distribution}, whereas many tadpole craters have diameters of $\leq$4~km. Additionally, tadpole crater rim breaches appear fresher than the more degraded valley networks \citep{wilson_cold-wet_2016}, and tadpole craters have topographically elevated, and sharp to rounded, crater rims \citep{wilson_cold-wet_2016,warren2021overspilling}. These lines of evidence suggest that tadpole craters post-date the most intense episodes of fluvial erosion on Mars, and therefore formed since the Late Hesperian \citep{wilson_cold-wet_2016}. Alongside other Hesperian and Amazonian fluvial and glacial features such as alluvial fans \citep{holo2021timing,grant2011late}, Fresh Shallow Valleys (FSVs, \citealt{wilson_cold-wet_2016,hobley2014fresh}), midlatitude eskers \citep{butcher2017recent}, and buried glaciers \citep{butcher2023eskers,holt2008radar,head2005tropical,fassett2010supraglacial}, tadpole craters record a step in the long-term evolution of Mars’ climate from the warmer, wetter Noachian to the planet's cold and arid present state \citep{kite2022high,mangold2012chronology,jakosky_mars_2001}.

To carve breaches, tadpole craters must first fill with either water or ice, and then generate enough liquid water at least once \citep{warren2021overspilling} to erode the observed rim breach(es). Therefore, the Martian climate must have been able to generate liquid water in the mid-latitudes during the Late Hesperian/Amazonian despite a colder, drier climate than that of the Late Noachian/Early Hesperian. A global aquifer protected by a cryosphere has been suggested as a mechanism for sustaining subsurface water under colder conditions (e.g. \citealt{clifford2010depth,hanna2005hydrological}). Additionally, climate models predict that high obliquity (e.g. 45$^{\circ}$) can enable accumulation of ice in the midlatitudes, and even further equatorward, as late as the Amazonian \citep{forget2006formation,madeleine2009amazonian,steele2017regolith,mischna2003orbital}.

\begin{figure}[t!]
\centering
\includegraphics[width=0.8\textwidth]{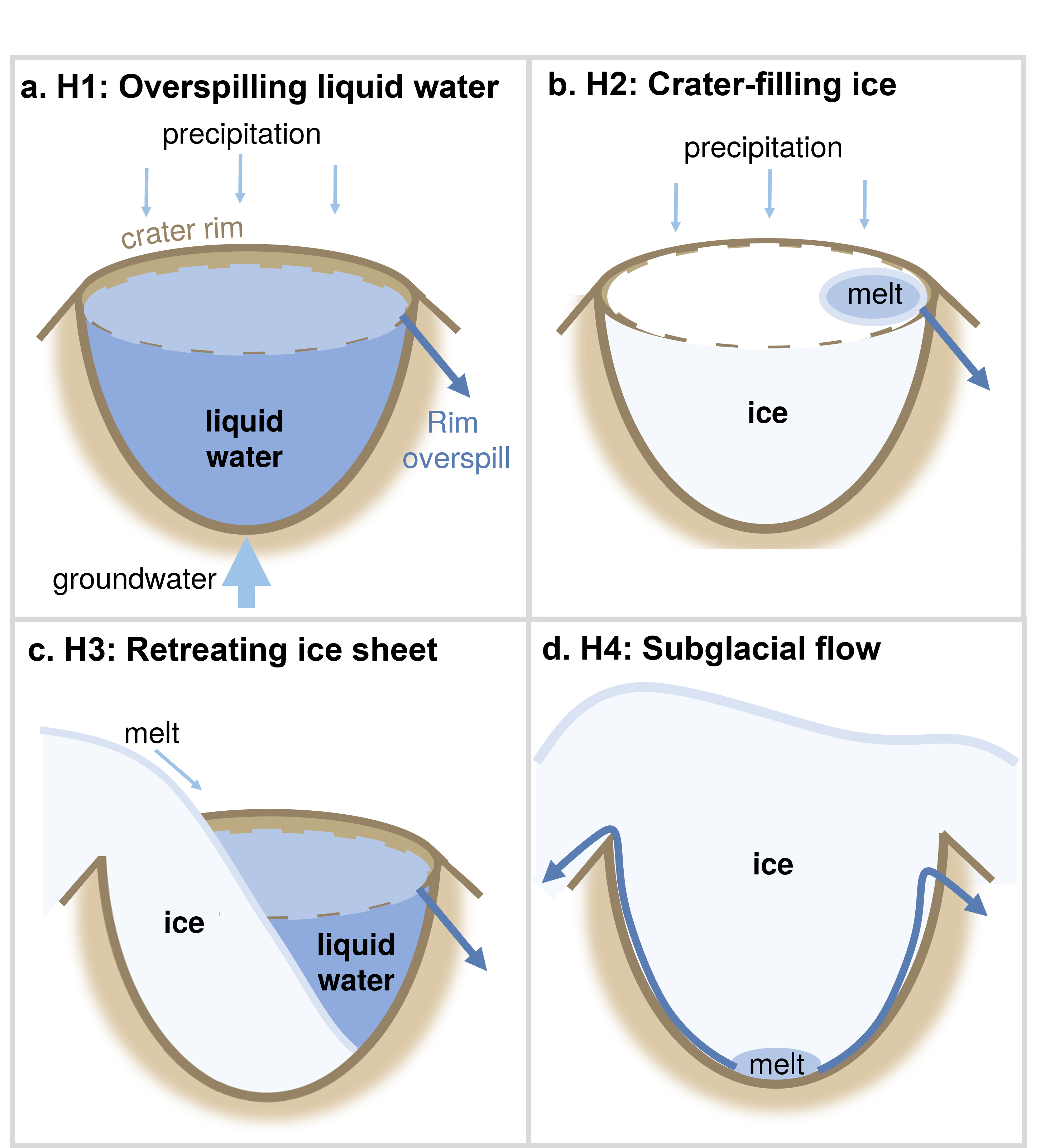}
\caption{Schematic illustration of our four overspilling hypotheses for tadpole craters. a. H1: Crater filled to rim with liquid water where breaches form when water level exceeds topographic lows on rim. b. H2: Ice-filled crater where breaches form when localized melt pond(s) overspill the crater rim. c. H3: Crater at the margins of a retreating regional ice sheet where breaches form when meltwater in crater exceeds topographic lows on rim, new topographic lows are exposed as the ice sheet retreats. d. H4: Crater buried beneath regional ice sheet where breaches form as subglacial channels. All breaches may form in a single overspilling event, or sequentially as rim topography is modified or exposed by changing ice extent and topography or erosion.} 
\label{fig:mech}
\end{figure}

To better understand the climate requirements of multiple breach tadpole craters, we consider four hypotheses (H1-H4) for tadpole crater overspill that can also explain how multiple breaches form (Fig.~\ref{fig:mech}):
\begin{itemize}
    \item \textbf{H1. Overflow of liquid water:} Craters fill to the brim with liquid water from precipitation, meltwater, or groundwater flow and overspill their rims. Either the lowest points on the rims were modified over time by erosion, creating new low points, or water inputs were sufficient to activate multiple breaches simultaneously (Sections \ref{sec:melt} \& \ref{sec:gwat}).
    \item \textbf{H2. Shallow melt ponds on crater-filling ice:} Craters fill to the brim with ice. Transient melting forms ice surface ponds of sufficient volume to overspill the crater rim and carve breaches (Section~\ref{sec:ovsp}).
    \item \textbf{H3: Retreating ice sheet:} A retreating ice sheet partially fills craters with ice, routing meltwater into the crater in sufficient volumes to overspill exposed sections of the rim. The ice sheet then retreats further, exposing lower areas of the rim and causing previous, higher breaches to be abandoned. (Section~\ref{sec:retreat})
    \item \textbf{H4. Subglacial flow:} A regional ice sheet covers the tadpole craters. Water from basal or surface melting pools at its base to drive subglacial streams out of craters (Section~\ref{sec:subg}).
    
\end{itemize}

In this paper, we aim to test these hypotheses and determine the climate constraints implied by tadpole craters with multiple breaches. To do this, we first analyze the size and elevation distribution of all tadpole craters, including those with both single and multiple breaches, in the latitude band 24-52$^{\circ}$~S (Section~\ref{sec:distribution}). Next, we focus on four multiple breach tadpole craters (Fig.~\ref{fig:profs}). We provide detailed descriptions of the rim, breach, and crater floor morphologies of these craters. We use observations of these four craters (Section~\ref{sec:morph}) to test our four proposed tadpole breach formation hypotheses (Fig.~\ref{fig:mech}). We use breach morphometry to calculate water flow rate under subaerial (Section~\ref{sec:sim}) and subglacial conditions (Section~\ref{sec:subg}), and the eroded volumes of each exit breach. We compare the results of these calculations to: 
\begin{enumerate}
    \item Predicted precipitation and melting rates (Section~\ref{sec:melt})
    \item Water supply rates from meltwater routed from an ice sheet larger than the crater (Section~\ref{sec:retreat})
    \item Discharge from a radially symmetrical, homogeneous, cryosphere-confined pressurized aquifer (Section~\ref{sec:gwat})
    \item The rim incising potential of meltwater ponds smaller than the crater area draining from the surface of a crater-filling ice body (Section~\ref{sec:ovsp})
    \item Melt volumes generated by supraglacial meltwater (Section~\ref{sec:supr})
    \item Melting rates and total melt thicknesses caused by a magma body intruded beneath a subglacial crater (Section~\ref{sec:int})
\end{enumerate}
Finally, we discuss potential terrestrial analogs for multiple breach tadpole craters (Section~\ref{sec:analogs}).

The main basis for rejecting a formation hypothesis is if our calculations show that it does not generate a water flow rate that can explain the multiple breach tadpole craters. In Section~7, we summarize the results of our hypothesis tests, and discuss the hypotheses that survive those tests (subsets of H2-H4). We find that only hundreds of meters of ice accumulation, and surface melting, can explain the data.

\section{Data and Methods}
\begin{sidewaystable}
\begin{tabular}{rcccccccll}
\hline
\textbf{Nickname}       & \textbf{Lat ($^{\circ}$N)} & \textbf{Lon ($^{\circ}$W)} & \textbf{$n_{b}$} & \textbf{$D_{c}$ (km)*} & \textbf{$d_{c}$ (m)} & \textbf{$d_{c,0}$ (m)} & \textbf{$V_{c,0}$ (km$^{3}$)} & \textbf{Crater ID*} & \textbf{HiRISE ID (left)} \\ \hline \hline
\textit{Arabia Terra~1} & 34.808       & 16.485       & 2                           & 6.4     & 224 &              630                  & 5.8                                            & 05\_002270         & ESP\_052510\_2150               \\
\textit{Arabia Terra~2} & 37.625       & 21.347       & 5                           & 3.6       & 195 &             390                  & 1.2                                            & 05-001364          & ESP\_053736\_2180               \\
\textit{Hellas}         & -45.881      & 45.818       & 2                           & 14.7      & 616 &             1290                 & 53.6                                           & 27\_000899         & ESP\_012014\_1335               \\
\textit{Noachis}        & -34.229      & 25.515       & 2                           & 4.8       & 252 &             500                  & 2.7                                            & 27\_002296         & ESP\_074112\_1455               \\ 
\hline
\end{tabular}
\caption{Summary of multiple breach tadpole craters investigated in this study. $n_{b}$ is number of exit breaches, $D_{c}$ is crater diameter, $d_{c}$ is measured crater depth (errors $<1$m), $d_{c,0}$ and $V_{c,0}$ are calculated original crater depth and volume respectively (Eqns. \ref{eq:d} \& \ref{eq:v}; \citealt{garvin2000north}). *Database of \cite{robbins2012global}.}
\label{tab:mebtps}
\end{sidewaystable}

\begin{figure}[t!]
\centering
\includegraphics[width=\linewidth]{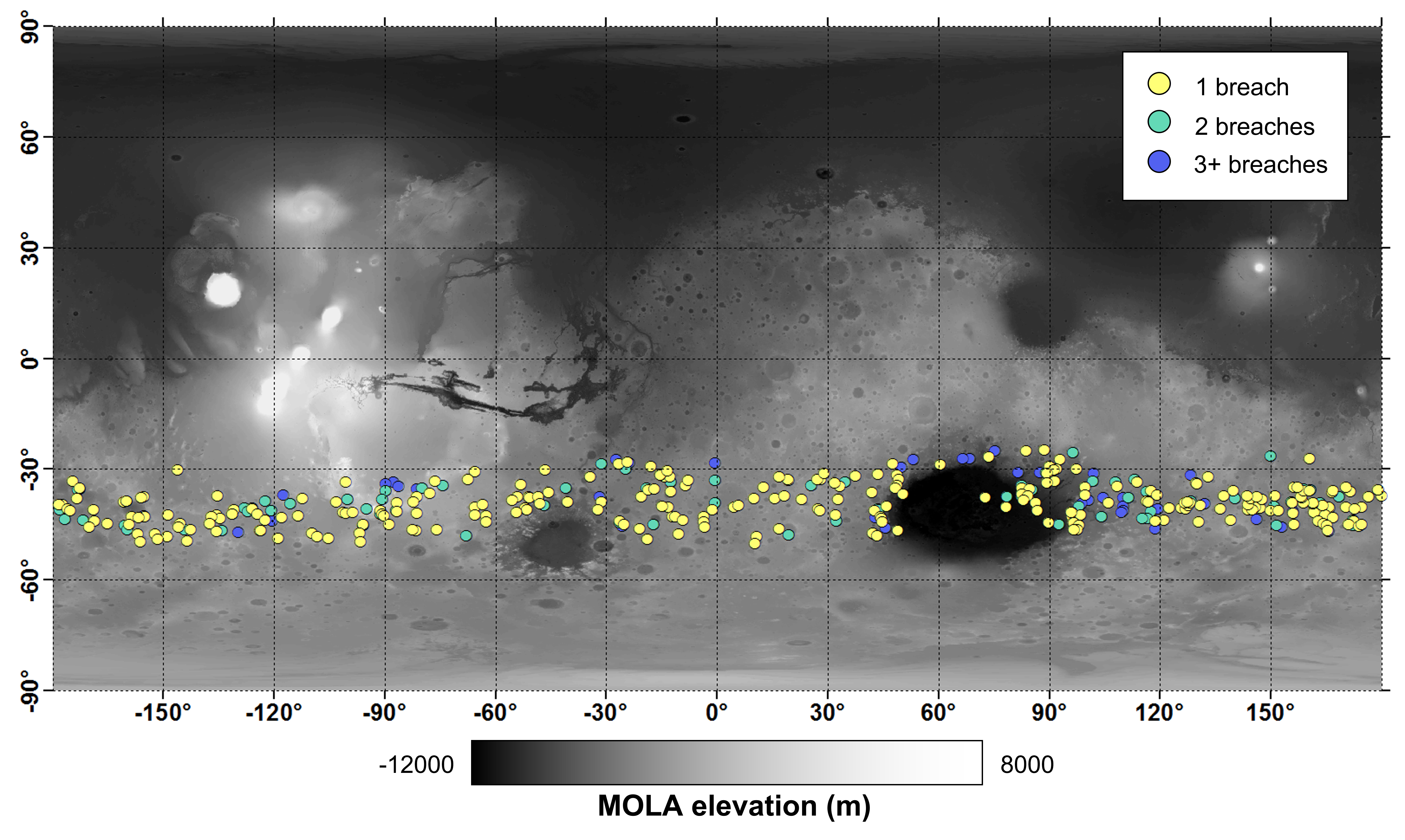}
\caption{Map of reliable tadpole craters in the latitude band 24-52$^{\circ}$~S overlaid on MOLA topography. Yellow circles indicate craters with a single exit breach, teal circles craters with 2 exit breaches, and blue circles craters with $\geq$3 exit breaches.}
\label{fig:map}
\end{figure} 

\subsection{Tadpole crater dataset}

Tadpole craters were observed in a latitude band between 24 and 52$^{\circ}$~S and assigned a reliability score between `a' (most reliable) and `c' (least reliable) by using a global CTX mosaic \textbf{(for details of tadpole crater database, see \citet{howard2022distribution} and  \citet{wilson2021distribution})} (Fig.~\ref{fig:map}). Every tadpole crater with $D_{c}\geq1$~km was cross-referenced with the Robbins crater database \citep{robbins2012global} to find its crater ID, circular fit diameter, latitude/longitude coordinates, and degradation state \citep{robbins2012global}. A second person independently reviewed all tadpole craters with reliability `a' to remove any less reliable observations and count the number of rim breaches using a global CTX mosaic \citep{dickson2018global} and global merged MOLA 128ppd and 64ppd mosaic \citep{smith2001mars}. Fig.~\ref{fig:map} shows a preferential occurrence of tadpole craters to the east of Hellas relative to the west of Hellas, which is consistent with preferential deposition of ice to the east of Hellas (due to a stationary, topographically-forced atmospheric wave) that is predicted by some global climate models at high obliquity \citep{forget2006}.

\subsection{Diameter Dependence of Tadpole Craters}
\label{sec:diam}

To determine whether there is a crater diameter preference for tadpole breach formation, we compare the tadpole crater dataset to all Robbins crater database craters in the same latitude band (24-52$^{\circ}$~S). Following \cite{head2010global}, we first performed a 2-sample Kolmogorov-Smirnov test using the \texttt{scipy.stats} package to determine whether the crater size-frequency distribution (SFD) of tadpole craters is consistent with being drawn from the same distribution as the SFD of all craters in the region. A p-value of $1.14\times10^{-4}$ suggests with $>$98\% confidence that the SFD of tadpole craters is distinct from that of all craters. We conclude that tadpole craters form by a process that does not uniformly affect craters of all sizes present in the studied latitude band. Rather, tadpole breaches preferentially form in smaller craters. 

\begin{figure}[t!]
\centering
\centering
\includegraphics[width=0.5\textwidth]{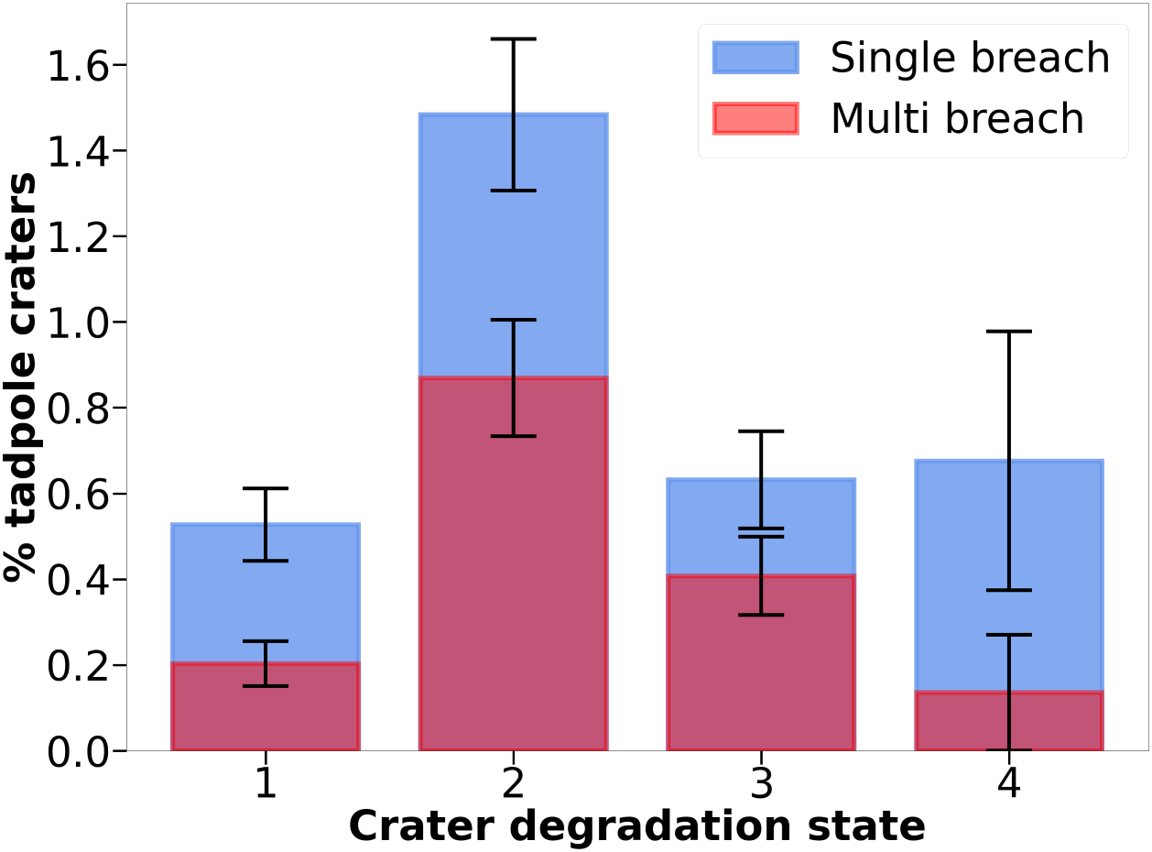}
\caption{Percentage of craters in each degradation state classification \citep{robbins2012global} that are tadpole craters, separated into tadpole craters with a single exit breach (blue), and those with multiple breaches (red). A degradation state of 4 corresponds to a pristine, fresh crater, and a degradation state of 1 corresponds to the most degraded craters which are highly infilled/modified by fluvial and aeolian processes with no preserved topographic rims or ejecta. The total tadpole crater fraction is the sum of the two bars, i.e. for craters that have degradation state 2 the percentage of craters that have multiple exit breaches is 0.9\%, and the percentage of craters that have single breaches is 1.5\%, for a total of 2.4\% of craters in  degradation state 2 that have exit breaches. Error bars show fractional error corresponding to $\sqrt{n}$ where $n$ is number of single breach (blue) or multiple breach (red) tadpole craters in each degradation state category.}  
\label{fig:degs}
\end{figure}

To further examine the difference between SFDs, we then normalized the Robbins crater dataset by degradation state \citep{robbins2012global} to match the degradation state distribution of the tadpole craters (Fig.~\ref{fig:degs}). Older crater populations have a greater proportion of larger craters \citep{hartmann2005martian} and older craters tend to be more degraded \citep{tanaka2014digital}. A degradation state of 4 corresponds to a pristine, fresh crater, and a degradation state of 1 corresponds to the most degraded craters which are highly infilled/modified by fluvial and aeolian processes with no preserved topographic rims or ejecta. Normalizing by degradation state allows us to subtract the effect of changing crater size-frequency distributions over time and determine whether the tadpole crater forming process(es) had a crater size preference, rather than just seeing size patterns associated with tadpole formation during a particular epoch in Mars history. We normalize for degradation state by calculating the proportion of tadpole craters in each degradation state category (1-4, i.e. most to least degraded), then dividing these proportions by the highest proportion value to find weights such that the most common degradation state has a weight of 1, and the other degradation states have weights $<1$. The largest proportion of pollywogs occurs for craters with a degradation state of 2 (Fig.~\ref{fig:degs}). We then assign the calculated degradation state weight to each individual crater. When subsequently binning the Robbins crater database (e.g. in crater diameter or elevation), we calculate the mean degradation state weight for each bin, and multiply the total number of craters in each bin by this mean weight. 

We divide the tadpole crater dataset and Robbins crater database into 11 equally-sized diameter bins for all craters with $D_{c}\geq$4~km. We choose this diameter cutoff as the database may be incomplete for smaller craters \citep{holo2021timing}. The number of bins is chosen using the Rice rule ($k = 2n_{a}^{1/3}$) which is suitable~for avoiding over-smoothing in non-normally distributed, continuous data \citep{terrell1985oversmoothed}. Here, $n_{a}=233$ is the total number of tadpole crater observations with $D_{c}\geq4$~km. We sum the number of craters in each diameter bin, with Robbins crater database craters multiplied by their degradation state-dependent weights. We calculate a $\sqrt{n}$ error for each tadpole crater bin to account for shot-noise error, where $n$ is the number of tadpole craters in that bin. We find the proportion of tadpole craters per bin and find a line of best fit to the proportion distribution per bin using the \texttt{scipy.optimize} package function \texttt{curve\_fit} (Fig.~\ref{fig:diam}). We then find the 95\% confidence interval for each line of best fit using bootstrapping, generating 5000 samples of size $n_{R,4}=$13,538 (the total number of craters in the Robbins database \citep{robbins2012global} within the latitude band 24-52$^{\circ}$~S with $D_{c}\geq4$~km).

\subsubsection{Elevation Dependence of Tadpole Craters}

To determine whether there is an elevation preference for tadpole breach formation, we investigate not only MOLA gridded topography (Fig.~\ref{fig:elev}a), but also smoothed topography on a variety of length scales to determine whether the occurrence of tadpole craters depends on local topography (e.g. a $<$100~km diameter hill) or regional topography (e.g. a several 100~km rise). We use the ArcGIS Geoprocessing Spatial Analyst Focal Statistics tool to generate smoothed MOLA topography with smoothing radii ($r_{s}$) of 50, 100, 200, 500, and 1000~km. This tool calculates the mean MOLA elevation of all cells within a circular radius $r_{s}$ for each cell in the mosaic. We then subtract the smoothed topography from the original MOLA mosaic to find the residual MOLA topography. We then extract the elevation and residual topography for the central point of each tadpole crater and Robbins crater database crater in the 24-52$^{\circ}$~S latitude band with $D_{c}\geq4$~km. We separate the craters into 11 evenly sized elevation bins and residual MOLA topography bins for each value of $r_{s}$. Again, the number of bins is chosen using the Rice rule: $k = 2n_{a}^{1/3}$\citep{terrell1985oversmoothed} where $n_{a}=233$ is the total number of tadpole crater observations with $D_{c}\geq4$~km. We restrict the residual MOLA topography bin range for each smoothing radius to exclude values beyond which there are 0 craters.

Following the same procedure as described in Section~\ref{sec:diam}, we fit a linear best-fit line to the degradation state normalized tadpole crater percentage in each binned elevation and residual topography dataset, and find the 95\% confidence intervals using bootstrapping. In our fitting procedure, the tadpole crater percentage in each bin is weighted proportional to the inverse of the $\sqrt{n}$ error such that the bin with the smallest $\sqrt{n}$ error has a weight of 1, and the bin with the largest error has a weight of $\mathrm{min}\left(\sqrt{n}\right)/\mathrm{max}\left(\sqrt{n}\right)$. This procedure reduces the impact of bins with very few craters on our fits. We repeat this process for the single exit breach and multiple breach tadpole crater datasets.

\section{Results and Interpretations}

\subsection{Tadpole crater distribution}
\label{sec:distribution}
\subsubsection{Diameter Dependence of Tadpole Craters}

\begin{figure}[t!]
\centering
\includegraphics[width=\linewidth]{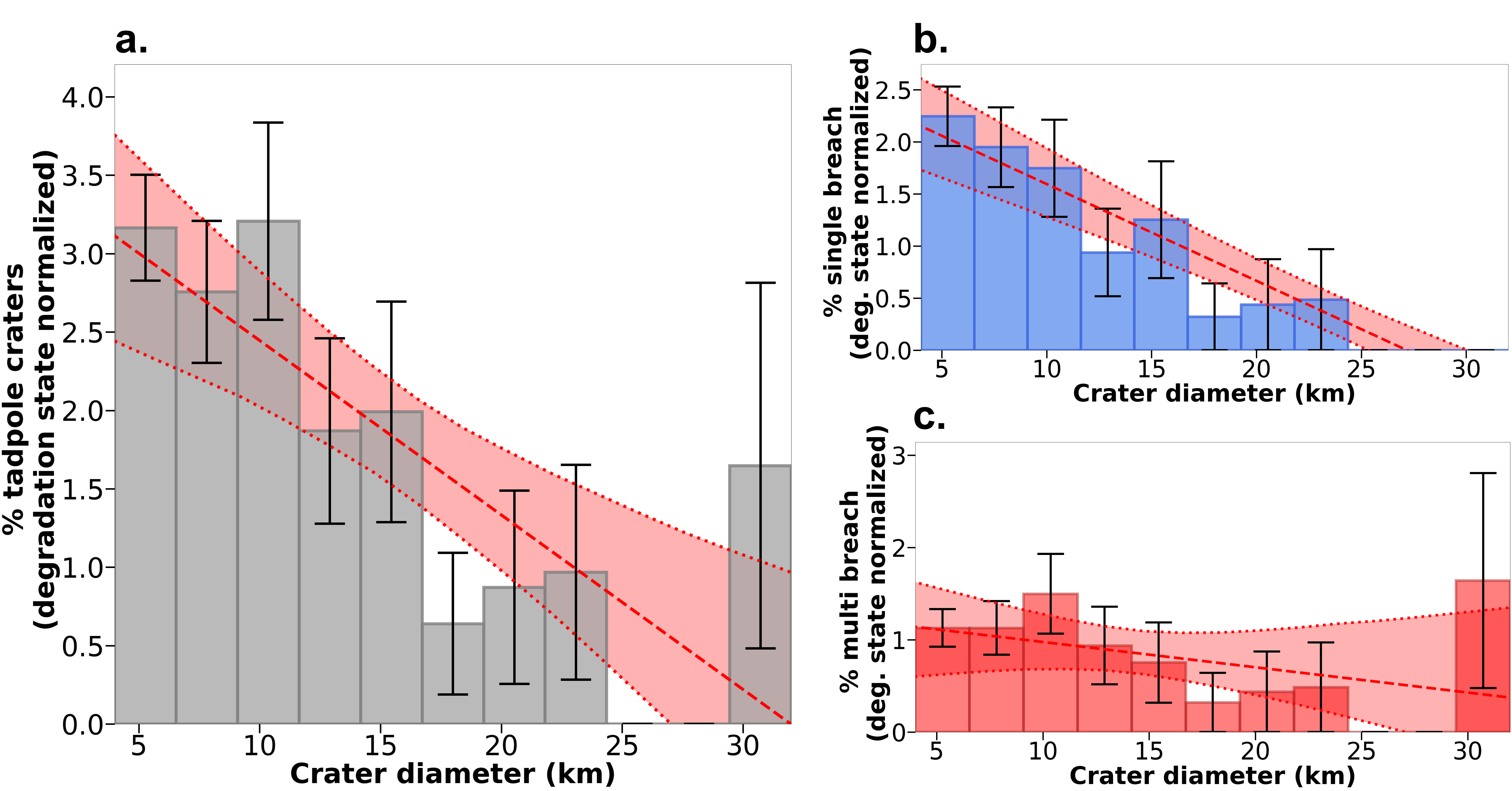}
\caption{Tadpole craters show a statistically significant preference for smaller craters. \textbf{a.} Histogram of the proportion of all craters that are tadpole craters in the latitude band 24-52$^{\circ}$~S  as a function of diameter, normalized for degradation state. \textbf{b.} Histogram of the proportion of all craters that are tadpole craters with a single exit breach by diameter. \textbf{c.} Histogram of the proportion of all craters that are tadpole craters with multiple breaches by diameter. For all panels, black bars show $\sqrt{n}$ error bars where $n$ is the total number of tadpole craters in each bin ($n_{a}=233$ is the total number of tadpole craters with $D_{c}\geq4$~km for the full tadpole crater population, $n_{s}=145$ for single exit breach tadpole craters, $n_{m}=77$ for multiple breach tadpole craters). Dashed red lines show a linear best fit. Dotted red lines bounding red shaded areas indicate 99.8\% confidence intervals for linear best fit found using 5000 bootstrapped samples of size $n_{R}=$13,538 (the total number of craters in latitude band 24-52$^{\circ}$~S with $D_{c}\geq4$~km in the Robbins database \citep{robbins2012global}) with replacement.}
\label{fig:diam}
\end{figure}

Tadpole breaches preferentially form in craters with smaller diameters to better than 99.98\% confidence (Fig.~\ref{fig:diam}a, total number of tadpole craters with $D_{c}\geq$4~km $n_{a,4}=233$). Of 5000 bootstrap fits of percentage tadpole craters as a function of binned diameter, 0 had a slope $\geq0$. To test that this result is not just a product of binning, we also used the Python package \textsc{skikit-learn} to fit a linear least squares regression to tadpole occurrence against diameter (tadpole craters assigned a score of 1, non-tadpole craters a score of -1) for each of the 5000 bootstrap-generated crater populations with all craters unbinned. 0 linear regressions had a slope $\geq0$. The small crater preference is dominated by the size preference of craters with a single exit breach (Fig.~\ref{fig:diam}b, total single exit breach craters $n_{s,4}=145$). Tadpole craters with multiple breaches show no clear crater diameter trend (Fig.~\ref{fig:diam}, total multiple breach craters $n_{m,4}=77$). However, the distribution of multiple breach craters is affected by small-number statistics, and a larger sample may reveal a clear trend not visible among the 77 samples used in this analysis.

We interpret the preferential occurrence of tadpole breaches in smaller craters as an indication that the overspilling mechanism is limited by the column thickness of water (as either liquid or ice) available during the tadpole breach-forming period. The relative paucity of large tadpole craters, when corrected for crater degradation state, suggests that insufficient water (as liquid or ice) was available to fill deep topographic depressions. For example, a regionally compartmentalized groundwater system \citep{harrison2009regionally} may have limited the water volume available to flow into craters from discharge of a cryosphere-pressurized aquifer. Alternatively, deposition of a widespread but vertically thin ice sheet in the midlatitudes may have filled smaller craters to the rim with snow and ice, but left larger craters underfilled, so that meltwater ponds on the surface of the ice did not overtop crater rims. The largest tadpole crater in our dataset ($D_{c}=$~31.5~km) suggests the ice cover must have been $\lesssim2500$~m thick (Eqn. \ref{eq:d}). However, local thickening of ice due to favorable conditions for precipitation or direct condensation is possible. 90\% of tadpole craters in the latitude band 24-52$^{\circ}$~S have $D_{c}<12$~km, which corresponds to an ice thickness $<1000$~m (Eqn. \ref{eq:d}). The low average height of mid-to-high latitude Amazonian pedestal craters (40.4~m in the S hemisphere), a proxy for ice-rich deposit thickness \citep{kadish2010pedestal}, has been used to suggest that the total ice inventory available to fill craters and mantle high latitudes may have been limited by the combined water inventory of the polar layered deposits, shallow ice deposits, and regolith \citep{fastook2013amazonian}

\subsubsection{Elevation dependence of tadpole craters} 

\begin{figure}[t!]
\centering
\includegraphics[width=0.6\linewidth]{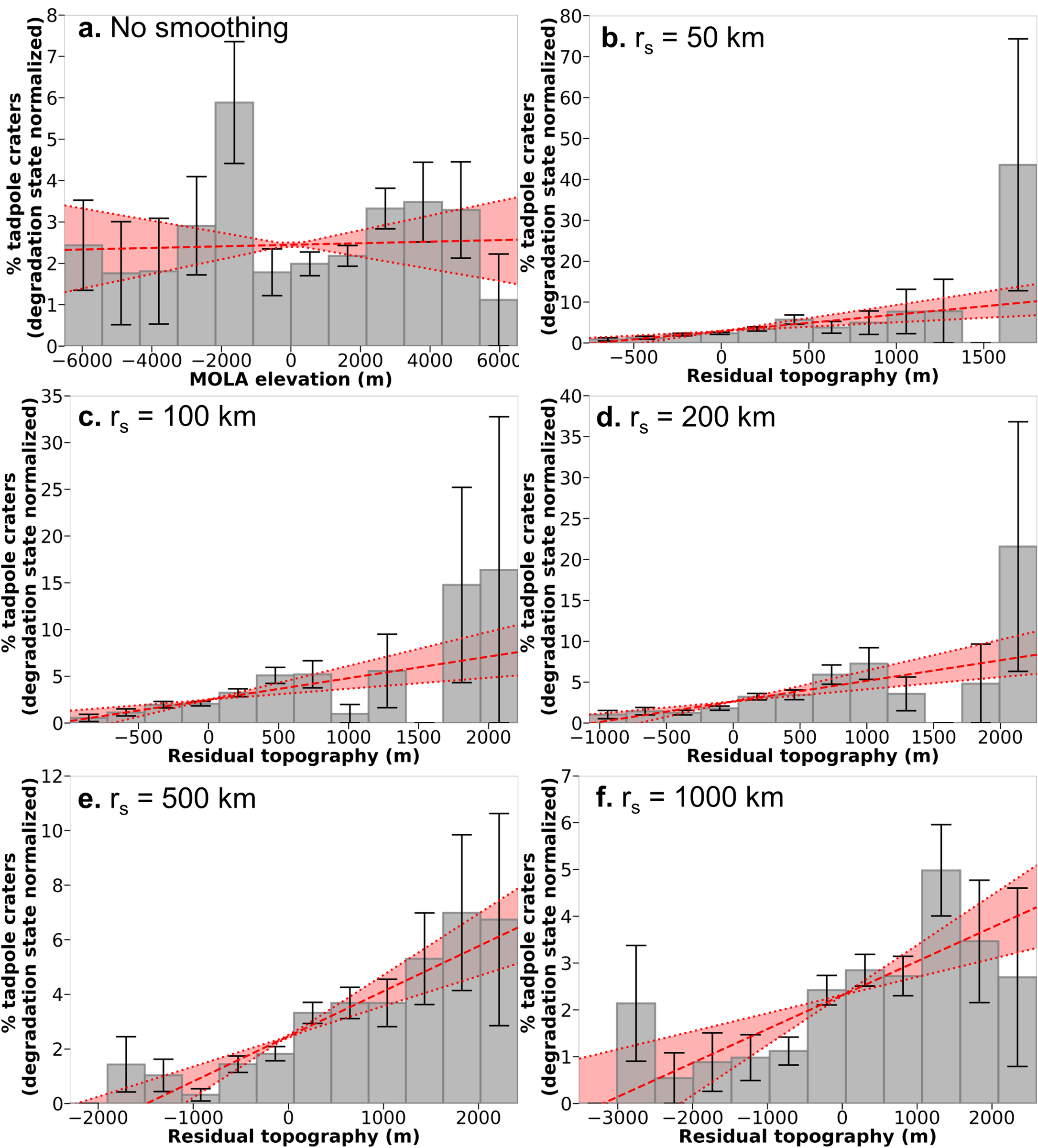}
\caption{Tadpole craters do not show a clear elevation trend, but preferentially occur on relative highs across all length scales. \textbf{a.} Histogram of degradation-state-normalized (see Section~\ref{sec:diam}) percentage of total crater population that is tadpole craters with MOLA elevation. b-f. Histogram of degradation-state-normalized percentage tadpole crater percentage with residual MOLA topography. We generate smoothed MOLA topography with smoothing radii ($r_{s}$) of 50, 100, 200, 500, and 1000~km. We then subtract the smoothed topography from the original MOLA mosaic to find the residual MOLA topography. Black bars show $\sqrt{n}$ error bars where $n$ is the total number of tadpole craters in each bin ($n_{a}=233$ is the total number of tadpole craters with $D_{c}\geq4$~km for the full tadpole crater population). Dashed red lines show a linear best fit to inverse $\sqrt{n}$ weighted bin totals. Dotted red lines bounding red shaded areas indicate 95\% confidence intervals for linear best fit found using 5000 bootstrapped samples of size $n_{R}=$13,538 (the total number of craters in latitude band 24-52$^{\circ}$~S with $D_{c}\geq4$~km in the Robbins database \citep{robbins2012global}) with replacement). Elevation data corresponds to crater floor elevation as in the Robbins database \citep{robbins2012global}.} 
\label{fig:elev}
\end{figure}

We find no elevation trend (Fig.~\ref{fig:elev}). However, there is a peak in the percentage of tadpole craters at elevations between -2670 and -890~m which corresponds to the elevations of an apparent cluster of tadpole craters on the slopes of the Hellas Basin (Fig.~\ref{fig:map}). In the residual topography data, there is a consistent trend towards a greater tadpole crater fraction at higher elevations (Fig.~\ref{fig:elev}b-f) in the full tadpole crater dataset, as well as the single exit breach and multiple breach subsets of the tadpole crater population (all 99.98\% confidence; 0 of 5000 bootstrap fits for each smoothing radius have slopes $<0$). This trend is strongest for smaller smoothing radii, with a maximum best fit slope of 4.1\%~km$^{-1}$ for a 50~km smoothing radius. We interpret this as a greater likelihood of tadpole breach formation for craters located on ``local'' highs within tens of kilometers of the crater, in addition to a slight preference for tadpole breach formation on regional highs, as indicated by the trend remaining present at a 1000~km smoothing radius (Fig.~\ref{fig:elev}f). This interpretation disfavors a groundwater hypothesis for tadpole crater water source, as higher surface elevations act to reduce the pressure head driving groundwater discharge \citep{clifford_evolution_2001} into a crater. A higher local elevation preference may also favor a precipitation or atmospheric deposition control on tadpole breach formation, for example, if local highs are cold enough to favor snow/ice accumulation as occurs at high elevations in climate models of a cold and icy early Mars under a thicker CO$_{2}$ atmosphere \citep{wordsworth_global_2013} and models of Amazonian ice accumulation under high obliquity \citep{forget2006formation}. Another possibility is that a preference for local highs indicates an orographic precipitation control on tadpole breach formation, where the local topography drives moisture-bearing parcels of atmosphere upwards, generating precipitation, e.g. of water ice, on relative highs \citep{scanlon2013orographic}.

The tadpole crater dataset used in this study is restricted to a narrow latitude band (24-52$^{\circ}$~S). However, tadpole craters also occur outside this band \citep{warren2021overspilling,wilson_cold-wet_2016}, particularly in the northern midlatitudes, therefore a more complete global tadpole crater dataset might reveal elevation trends not distinguishable using the regional data presented here. 

\subsection{The morphology of tadpole crater rims, fill, and breaches}
\label{sec:morph}

\begin{figure}[t!]
\centering
\includegraphics[width=0.7\linewidth]{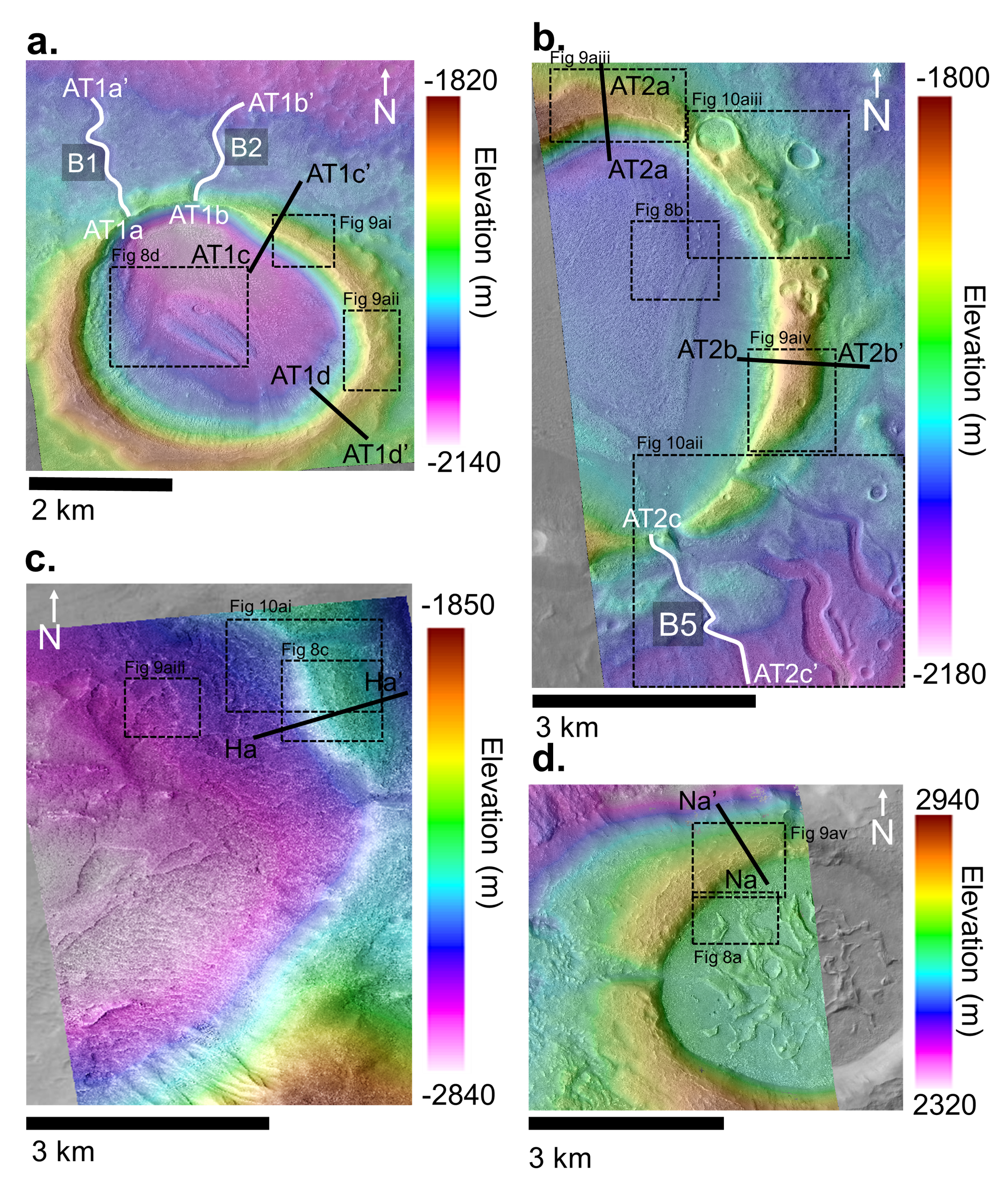}
\caption{Digital Elevation Models (DEMs) of multiple breach tadpole craters showing measurement areas for other figures. \textbf{a.} \textit{Arabia Terra~1} (left: ESP\_052510\_2150), \textbf{b.} \textit{Arabia Terra~2} (left: ESP\_053736\_2180), \textbf{c.} \textit{Hellas} (, left: ESP\_012014\_1335), \textbf{d.} \textit{Noachis} (, left: ESP\_074112\_1455) (Table~\ref{tab:mebtps}). White solid lines indicate breach long profiles (Fig.~\ref{fig:breaches}) with breach labels corresponding to those in Fig.~\ref{fig:profs}, black solid lines indicate rim topographic profiles (Fig.~\ref{fig:rims}), and black dashed boxes indicate regions shown in other figures.} 
\label{fig:key}
\end{figure}

We analysed four multiple breach tadpole craters using HiRISE digital elevation models (DEMs, Figs~\ref{fig:key}-\ref{fig:breaches})  to characterize the morphology of the crater fill, rims, and breaches. These are the only multiple breach tadpole craters with available HiRISE stereopairs. Two fall within the latitude band 24-52$^{\circ}$~S analyzed in Section~\ref{sec:diam}, and two are in Arabia Terra and have been previously studied \citep{warren2021overspilling}. Throughout this paper, each multiple breach tadpole crater is referred to by a nickname corresponding to the general region of Mars in which it occurs, listed in Table~\ref{tab:mebtps}.

In this subsection, we present evidence that:
\begin{enumerate}
    \item All multiple breach tadpole craters are partially filled with debris-mantled, ice-rich deposits
    \item Crater rims have been modified by ice
    \item Several breaches are filled with debris-mantled, ice-rich deposits, and therefore their depths may be underestimated
    \item The two breaches from \textit{Arabia Terra 1} have sections of valley floor that go upslope, which, if primary, would suggest a subglacial origin
\end{enumerate}
This modern ice-fill and ice-related rim modification may post-date tadpole breach incision, forming instead during subsequent ice ages on Mars \citep{butcher2017recent,Head2003,fassett2014extended}.

\subsubsection{Nature of Tadpole Crater Floor Fill}
\label{sec:fill}

Using empirical equations for the depth and volume of craters on Mars based on MOLA observations \citep{garvin2000north}, we can calculate the expected depths ($d_{c}$, km) and volumes ($V_{c}$, km$^{3}$) for each tadpole crater (Table~\ref{tab:mebtps}):

\begin{equation}
\label{eq:d}
    d_{c} = 0.131D_{c}^{0.85}
\end{equation}
\begin{equation}
\label{eq:v}
    V_{c} = 0.04 D_{c}^{2.68}
\end{equation}
where $D_{c}$ is the crater diameter in km. These equations span the simple-to-complex transition \citep{garvin2000north}, and are therefore appropriate for all four multiple breach tadpole craters in this study (Table~\ref{tab:mebtps}).

Comparing the measured maximum crater depths from the DEMs with their predicted original depths \citep{garvin2000north} (Table~\ref{tab:mebtps}), all four craters have been infilled with hundreds of meters of material (\textit{Arabia Terra~2}: 190~m (2 s.f.); \textit{Hellas} (left: ESP\_012014\_1335): 670~m (2 s.f.)). Infilling is caused by processes including mass wasting, fluvial modification, and aeolian deposition, as sediment is moved from the crater walls or beyond to the crater floor over time \citep{mangold2012chronology,craddock1997crater}. Many mid-latitude craters also show evidence for debris-covered ice fill (e.g. \citealt{levy2009concentric,mangold2003geomorphic,levy2014sequestered,butcher2023eskers}), associated with other debris-covered glacial landforms poleward of around 30$^{\circ}$ that represent a water inventory of $>1.25\times 10^{5}$~km$^{3}$ \citep{levy2014sequestered}. We also find evidence for debris-covered ice fill in these four multiple breach tadpole craters.

\begin{figure}[t!]
\centering
\includegraphics[width=0.8\textwidth]{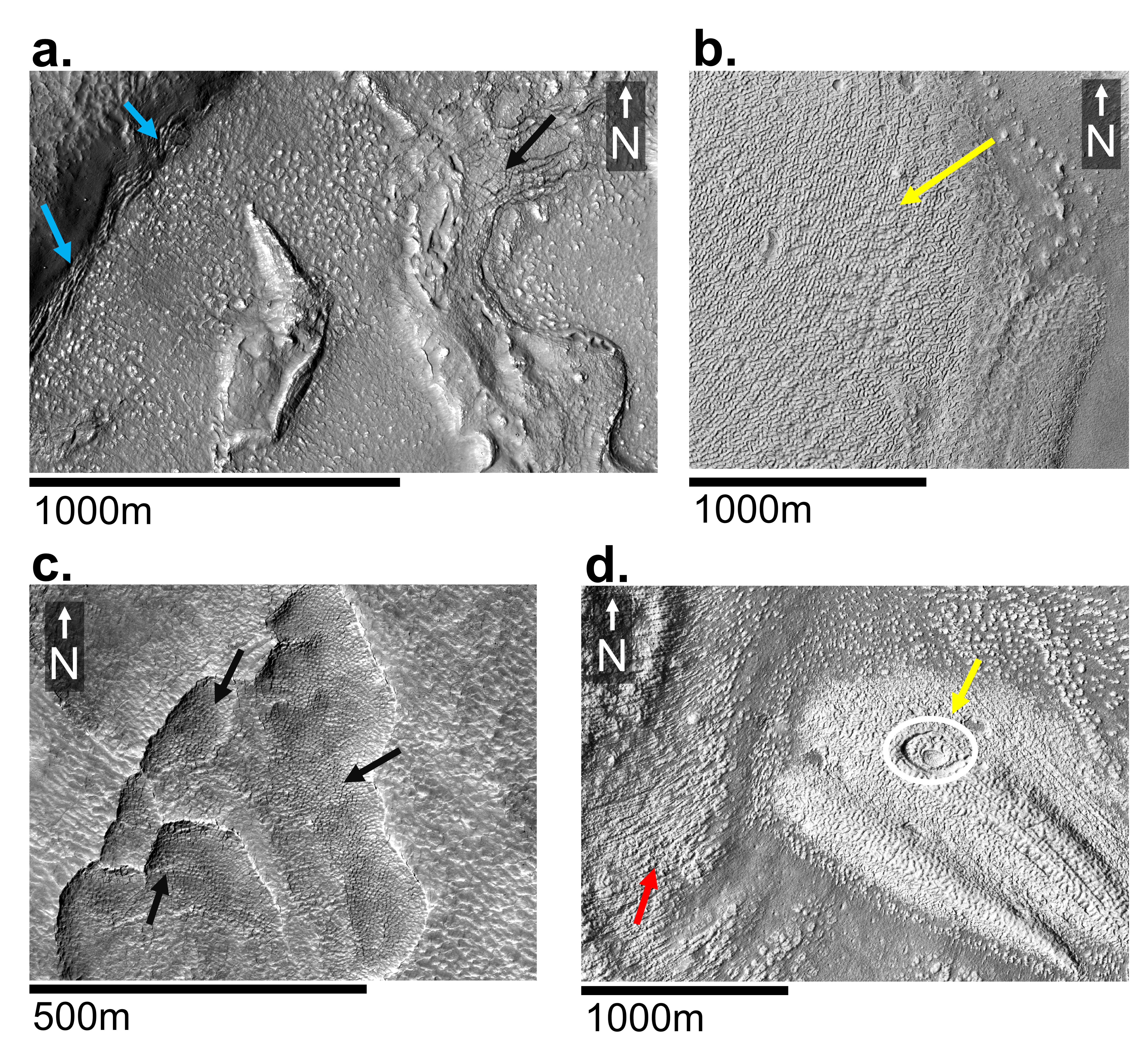}
\caption{HiRISE imagery of crater floor fill, interpreted as crater-filling ice. Blue arrows indicate wrinkling at the boundary between the steep topography and texturally altered bedrock-like \citep{conway2018glacial} appearance of the preserved crater rim and the flatter crater fill. Black arrows indicate decameter-scale polygonally-dissected crater fill resembling polygons observed in latitude-dependent mantle occurring alongside concentric crater fill and brain terrain \citep{levy2009concentric}, and in transitions between different brain terrain textures in ice and LDM-filled craters \citep{mangold2003geomorphic}. White ellipses indicate candidate ring mold craters \citep{kress2008ring}. Yellow arrows indicate areas of closed-cell brain terrain \citep{levy2009concentric}, interpreted as indicators of crater-filling ice or ice-rich debris beneath LDM. Red arrows indicate arcuate, parallel ridges of light-toned, knobby material, interpreted as arcuate ridges \citep{conway2018glacial}, likely associated with ice fill. \textbf{a.} \textit{Hellas} (left: ESP\_012014\_1335): decameter-scale bright mounds surrounded by mid-toned relatively smooth material throughout the image resemble buttes of partially-dissected brain terrain \citep{mangold2003geomorphic} and knobby textures observed on lobate debris aprons (LDAs) \citep{PIERCE200346}, \textbf{b.} \textit{Arabia Terra~2} (left: ESP\_053736\_2180), \textbf{c.} \textit{Noachis} (left: ESP\_074112\_1455), \textbf{d.} \textit{Arabia Terra~1} (left: ESP\_052510\_2150): ring mold crater indicating impact into ice-rich substrate in center left of image \citep{kress2008ring}.}
\label{fig:fill}
\end{figure}

In HiRISE imagery, large areas of the crater floor fill of \textit{Arabia Terra~1} and \textit{Arabia Terra~2} show the distinctive interconnected knobs characteristic of ``brain terrain'' \citep{mangold2003geomorphic,levy2009concentric,butcher2023eskers,levy2014sequestered,levy2010concentric} (Fig.~\ref{fig:fill}b \& d). This texture is typically associated with lobate debris aprons (LDAs), lineated valley fill (LVF), and concentric crater fill (CCF), all of which show viscous flow features interpreted as the result of buried ice 10-100~Myr old \citep{mangold2003geomorphic}. \cite{mangold2003geomorphic} and \cite{levy2009concentric} propose differential sublimation of debris-mantled ice as the mechanism driving brain terrain formation. Periglacial freeze-thaw processes have also been proposed as a mechanism for forming brain terrain \citep{mellon2008periglacial,hibbard2021surface}, however, this interpretation is difficult to reconcile with the cold, arid conditions expected on recent Mars \citep{LEVY2016enhanced}. Additionally, \textit{Arabia Terra~1} has a candidate ``ring mold'' crater at its center (Fig.~\ref{fig:rims}). Ring mold craters are characteristic of impacts into ice rich substrate \citep{kress2008ring}. We interpret the crater fill of \textit{Arabia Terra~1} and \textit{Arabia Terra~2} as between 30\% (pore-filling) and 90\% ice mixed with debris \citep{levy2014sequestered}.

The crater floor fills observed in \textit{Hellas} and \textit{Noachis} lack clearly identifiable brain terrain (Fig.~\ref{fig:fill}c \& a respectively). The floor fill in \textit{Hellas} is instead characterized by decameter-scale polygons visible in shallow depressions that act as windows into crater floor spanning mottled terrain (Fig.~\ref{fig:fill}c). The sharp edges of these depressions and their scale are similar to sublimation pits observed surrounding mid- to high-latitude pedestal craters, indicating the presence of snow/ice in the substrate \citep{kadish2010pedestal}. Their interior polygons resemble homogeneous patterned ground observed in the high latitudes \citep{MANGOLD2005high}. Formation of decameter-scale polygons likely requires the presence of water ice. Polygonal patterned ground can form through desiccation and thermal variations in and/or sublimation from an ice-bearing layer, and may also suggest freeze-thaw processes \citep{MANGOLD2005high,balme2013morphological}. Similar polygonal textures are also observed in latitude-dependent mantle in areas associated with concentric crater fill and brain terrain \citep{levy2009concentric}, and in transitions between different brain terrain textures in ice-filled craters \citep{mangold2003geomorphic}. We interpret the \textit{Hellas} multiple breach tadpole crater floor textures as ice-rich fill.

In \textit{Noachis}, the crater floor fill has a distinctive large-scale morphology, with irregularly shaped, discontinuous curvilinear ridges (Fig.~\ref{fig:key}). Up close, some surfaces of these topographically elevated regions have decameter-scale polygons resembling those in \textit{Hellas} (Fig.~\ref{fig:fill}a), and their margins do not have a sharp contact with adjacent crater fill. Areas without polygons closely resemble material surrounding the brain terrain in \textit{Arabia Terra~1} and \textit{Arabia Terra~2} (e.g. Fig.~\ref{fig:fill}d, top left). Additionally, at the boundary between the steep crater rim and the flatter crater floor, the infilling material exhibits wrinkles, which we interpret as indicative of viscous material, in this case, most likely ice. Therefore, our interpretation is that the crater fill in \textit{Noachis} also includes ice.

Although we interpret all four multiple breach tadpole craters as filled with a mixture of ice and debris, this ice was not necessarily present at the time of crater overspill. The debris-covered glacial features observed in the Martian mid-latitudes may be as young as tens of millions of years \citep{mangold2003geomorphic}, but there is evidence for multiple ice ages on Mars \citep{butcher2017recent,Head2003,fassett2014extended}. If the breaches were incised on the earlier end of the tadpole crater age range in the late Hesperian or early Amazonian, then there may have been billions of years available for the contemporary crater-filling ice-rich deposits to accumulate (and potentially for any earlier crater-filling ice/ice-rich debris to be lost if the tadpole craters were ice-filled or subglacial, e.g. Fig.~\ref{fig:mech}b \& d). Therefore, the modern fill of multiple breach tadpole craters does not necessarily favor a particular mechanism for crater overspill.

\subsubsection{Nature of Tadpole Crater Rims}
\label{sec:rims}
\begin{figure}[t!]
\centering
\includegraphics[width=\linewidth]{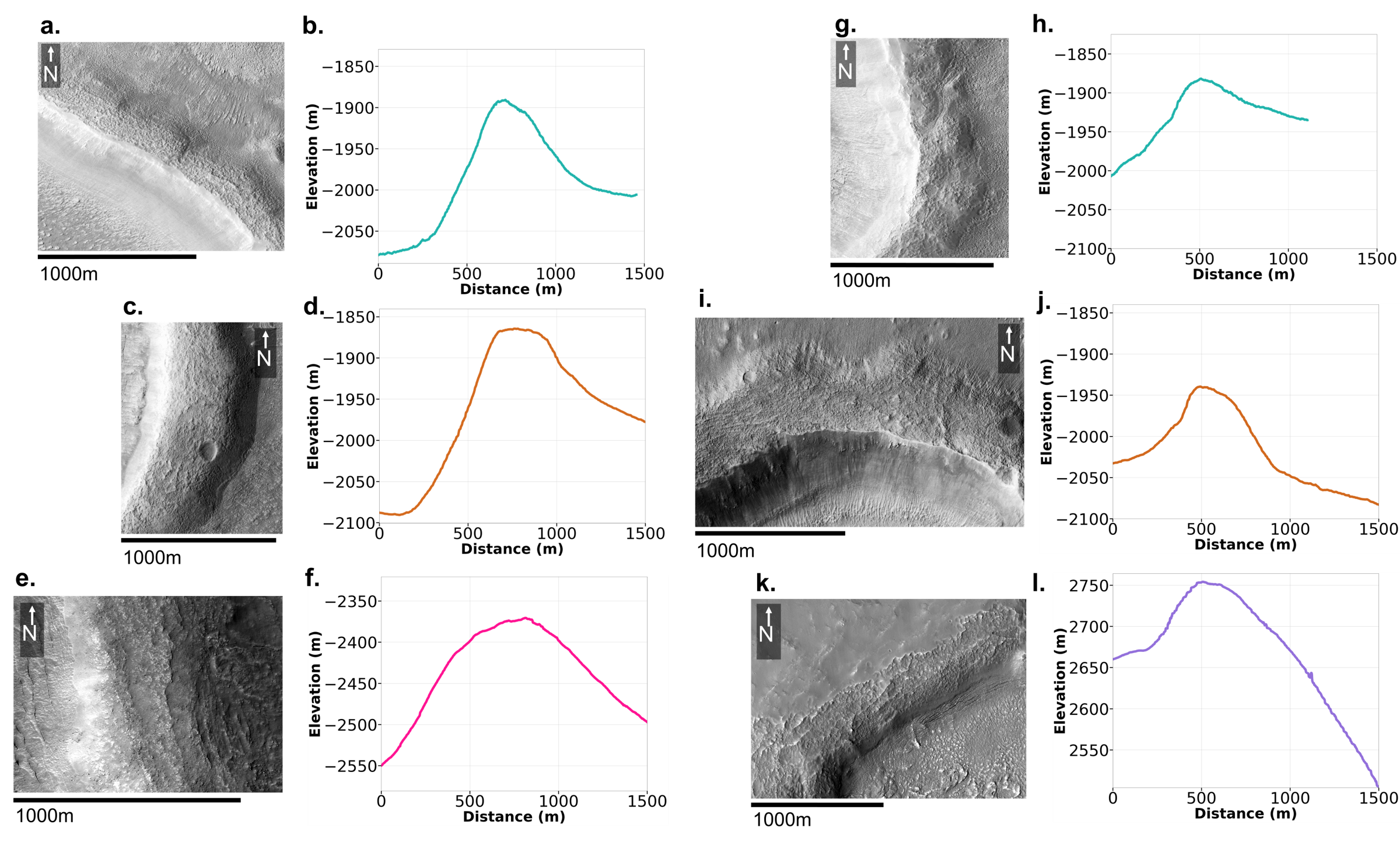}
\caption{\textbf{a.} HiRISE imagery of unincised portions of multiple breach crater rims showing mottled rim textures resembling texturally altered bedrock \citep{conway2018glacial} along the most topographically elevated portions of the rim in a. \& g. \textit{Arabia Terra~1} (left: ESP\_053736\_2180), c. \& i. \textit{Arabia Terra~2} (left: ESP\_053736\_2180), e. \textit{Hellas} (left: ESP\_012014\_1335), and k. \textit{Noachis} (left: ESP\_074112\_1455) (Table~\ref{tab:mebtps}). \textbf{b.} Topographic profiles of crater rims (see Fig.~\ref{fig:key} for profile transects) in b. \& h. \textit{Arabia Terra~1}, d. \& j. \textit{Arabia Terra~2}, f. \textit{Hellas}, and l. \textit{Noachis}, showing topographically rounded rims (e.g. h, d, l) and sharper rim sections (e.g. b, j) more closely resembling the steep, narrow shape of fresh crater rims (e.g. \citealt{robbins2012new}). In all profiles, the crater interior is to the left. }
\label{fig:rims}
\end{figure}

Tadpole craters have topographically elevated and sometimes fresh-appearing rims (Fig.~\ref{fig:rims}). In our four DEMs, the highest portions of the crater rims have  a ``mottled'' texture comprised of small meter to decameter scale, irregularly shaped pits with sharp edges (Fig.~\ref{fig:rims}a). Occasionally, these small pits and their bounding ridges appear to form linear features persisting for a few hundred meters (e.g. Fig.~\ref{fig:rims}e, i,~\&~k). The tops of the observed tadpole crater rims also tend to be rounded (Fig.~\ref{fig:rims}b, f,~\&~l), although the inner edge of the rim where there is a break in slope towards the crater interior may remain sharp and appear fresh (Fig.~\ref{fig:rims}i~\&~j). The combined mottled appearance and rounded rim profiles strongly resembles the distinctive ``texturally altered bedrock'' documented by \cite{conway2018glacial} in mid-high latitude craters and interpreted as an indicator of glacial processes, as texturally altered bedrock visually resembles terrestrial surfaces that have experienced ice-shattering and frost-segregation processes \citep{conway2018glacial}. On Earth, ice-shattering and frost-segregation typically require thin films of liquid water during the growth of ice lenses that break down rock \citep{matsuoka2008frost}. However, accumulation of ice can proceed without liquid water in dry soils at the base of Antarctic ice \citep{lacelle2013excess}, and dry ice lens growth may be able to occur under Martian conditions \citep{sizemore2015initiation}. We interpret the observed multiple breach tadpole crater rim textures as a signature of modification by ice. However, these textures need not be caused by the breach incising climate, as the craters also show evidence for contemporary ice fill (Section~\ref{sec:fill}), therefore it is plausible that the rim texture is formed during a later episode of ice deposition.

Few craters of a similar size to typical tadpole craters (0.5-15~km) are preserved on Noachian terrains \citep{craddock2002case}; craters with $D_{c}<4$~km are absent \citep{irwin2013distribution}. Therefore it is difficult to make a direct textural comparison between tadpole crater rims and the rims of similar sized craters modified under enhanced fluvial activity. However, tadpole crater rims are not flattened as much as typical Noachian craters, definitionally lack fluvial channels superposing their rims, and do not show evidence for widespread small-scale gullies on the interior crater rim slope as observed for many degraded Noachian craters \citep{weiss2015crater}. Therefore, although our multiple breach tadpole crater rim observations indicate modification relative to fresh craters, our observations are consistent with the conclusion of \cite{wilson_cold-wet_2016} that these tadpole craters are Late Hesperian/Amazonian.

\subsubsection{Rim Breaches}
\label{sec:breaches}

\begin{figure}[t!]
\centering
\includegraphics[width=\linewidth]{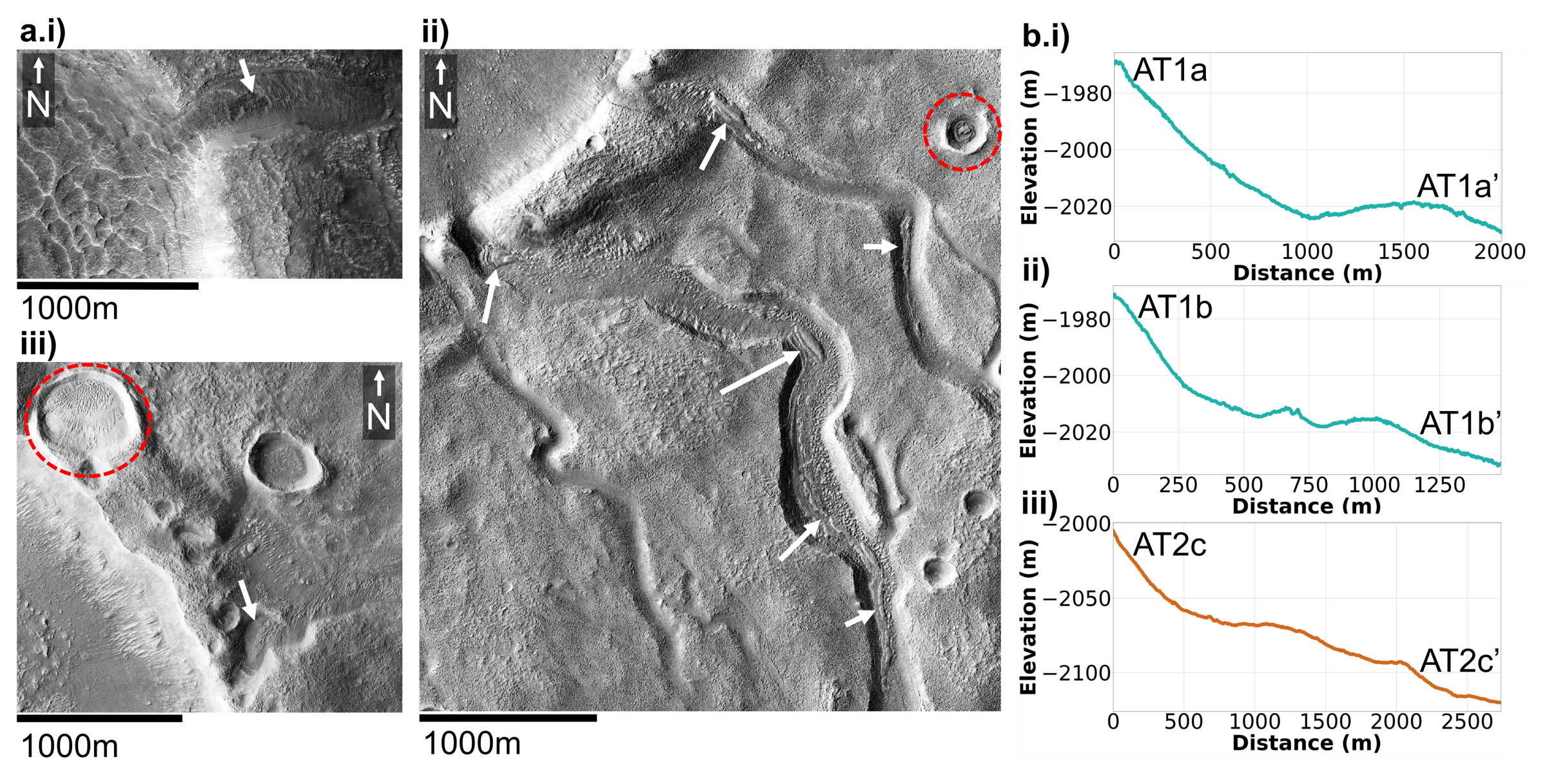}
\caption{\textbf{a-c.} HiRISE imagery of crater rim breaches infilled with discontinuous material resembling crater-filling ice-rich deposits (white arrows, Section~\ref{sec:fill}) in a) \textit{Hellas} (left: ESP\_012014\_1335), b \& c) \textit{Arabia Terra~2} (left: ESP\_053736\_2180). Red dashed circles highlight small craters filled with material morphologically consistent with ice-rich deposits. \textbf{d-f.} Downstream topographic profiles of ice-rich deposit-free exit breaches (traces of profiles shown in Fig.~\ref{fig:key}) for d \& e) \textit{Arabia Terra~1}, and f) \textit{Arabia Terra~2}.}  
\label{fig:breaches}
\end{figure}

We measured the dimensions of the exit breaches incising the rims of the four multiple breach tadpole craters using the measured rim topographic profiles (Fig.~\ref{fig:profs}). The breaches have v-shaped cross sections (Fig.~\ref{fig:profs}e-h) and are typically tens of meters deep and hundreds of meters wide where they intersect the rim (Table~\ref{tab:breach}). We measure the length of the breaches by taking 3 downstream profiles of each breach until the point where the breach valley is either no longer discernible from the surrounding terrain, or until another valley intersects the breach valley (e.g. Fig.~\ref{fig:breaches}b., lower left). 

Although some breach valleys have smooth floors and do not appear to be infilled, about half of the breaches of the four multiple breach tadpole craters have material morphologically resembling and/or continuous with the crater floor ice-rich deposits identified in Section~\ref{sec:fill} (Fig.~\ref{fig:breaches}a-d). As a result, we likely underestimate the depths of these breaches, and their downstream topographic profiles do not reflect the incised valley floor resulting from crater rim breaching. For breaches without ice-rich infill, we measure downstream topographic profiles (Fig.~\ref{fig:breaches}e-h). For \textit{Arabia Terra~1}, the valley floor elevation increases downstream for some segments of both exit breaches (\ref{fig:breaches}d \& e). In terrestrial environments, such valley floor ``undulations'' with upslope valley reaches are diagnostic of subglacial drainage \citep{galofre2018subglacial}. Undulations have also been used as supporting evidence for subglacial drainage on Mars \citep{butcher2017recent,galofre2018subglacial,GrauGalofre2020valley}. The breach valley walls and floors are smooth and featureless, and do not show evidence for overlying crater ejecta or mass wasting from the valley walls. However, if tadpole crater overspill was associated with widespread mid-latitude ice cover, there may have been subsequent loss of ice from the material at the base of and beneath the subglacial channel, leading to later modification of the downstream profile. In the absence of visible evidence for sublimation features or disruption of the breach valleys, we favor the interpretation that the undulations in the \textit{Arabia Terra~1} breaches are a primary feature, but do not consider them to be diagnostic of a subglacial origin by themselves given the possibility of post-incision changes to topography that may not be detectable~at HiRISE resolution. The downstream profiles of the remaining ice-free breaches (Fig.~\ref{fig:breaches}) do not show evidence for upslope reaches, which is consistent with either subglacial or subaerial erosion \citep{galofre2018subglacial}.

\begin{table}[t!]

\begin{tabular}{lcccccc}
\hline
\textbf{Tadpole} & \textbf{Breach} & \textbf{$d_{b}$} & \textbf{$w_{b}$} & \textbf{$L$ (m)} & \textbf{$V_{b}$} & \textbf{$V_{w,min}$}
 \\
\textbf{Crater} &   &   & \textbf{(m)} & \textbf{$L$ (m)} & \textbf{(m$^{3}$} & \textbf{(m$^{3}$)}
 \\ \hline \hline
\textit{Arabia Terra~1}          & B1              & 29$\pm0.5$       & 442$\pm0.3$          & 2140$\pm4$       & 4.07$\times 10^{6}$  & $1.36\times10^{7}$     \\
                                 & B2              & 28$\pm0.3$       & 338$\pm0.2$          & 1550$\pm1$       & $3.21\times10^{7}$   & $1.07\times10^{7}$      \\ \hline
\textit{Arabia Terra~2}          & B1              & 32$\pm0.4$       & 199$\pm0.5$          & 400$\pm4$        & $6.05\times10^{5}$     & $2.02\times10^{6}$     \\
                                 & B2              & 28$\pm0.5$       & 362$\pm0.1$          & 810$\pm3$        & $3.64\times10^{6}$   & $1.21\times10^{7}$   \\
                                 & B3              & 34$\pm0.1$       & 466$\pm0.1$          & 3140$\pm2$       & $7.20\times10^{6}$   & $2.40\times10^{7}$     \\
                                 & B4              & 57$\pm0.2$       & 390$\pm1$            & 1210$\pm1$       & $6.32\times10^{6}$   & $2.11\times10^{7}$     \\
                                 & B5              & 27$\pm0.1$       & 282$\pm0.5$          & 580$\pm4$        & $1.32\times10^{6}$    & $4.41\times10^{7}$    \\ \hline
\textit{Hellas}                  & B1              & 19$\pm0.1$       & 351$\pm0.1$          & 1070$\pm2$       & $9.35\times10^{6}$  & $3.12\times10^{7}$     \\
                                 & B2              & 30$\pm0.4$       & 426$\pm0.3$          & 780$\pm3$        & $2.64\times10^{6}$ & $8.80\times10^{6}$    \\ \hline
\textit{Noachis}                 & B1              & 100$\pm0.1$      & 837$\pm0.1$          & 1020$\pm2$       & $2.66\times10^{7}$  & $8.87\times10^{7}$    \\
                                 & B2              & 17$\pm0.4$       & 277$\pm0.3$          & 980$\pm1$        & $5.93\times10^{6}$    & $1.98\times10^{7}$      \\ \hline
\end{tabular}
\caption{Breach measurements for each tadpole crater in this study (Table~\ref{tab:mebtps}, see Fig.~\ref{fig:profs} for breach labels). $d_{b}$ is maximum breach depth at the crater rim measured from profiles in Fig.~\ref{fig:profs}, $w_{b}$ is breach width on the crater rim measured from profiles in Fig.~\ref{fig:profs}. $L$ is the traced length of each breach until it either intersects another valley or becomes invisible. $V_{b}$ is the breach volume calculated from the mean of at least 3 breach area profiles found in ArcGIS multiplied by $L$. All error bars come from the range of triplicate measurements. $V_{w,min}$ is the minimum water volume required to erode $V_{b}$ assuming a maximum sediment:water ratio of 0.3. }
\label{tab:breach}
\end{table}

\section{Modeling Results and Interpretations: Analysis of Hypotheses for Tadpole Crater Overspill}

\subsection{Subaerial Overspill (H1-H3)}
\label{sec:sim}
In this section, we consider subaerial overspill hypotheses for tadpole crater breach formation (H1-H3, Fig.~\ref{fig:mech}), where flow of water over the crater rim is driven by the water level reaching or exceeding the height of the rim and the rim topography.

For a crater filled with water to form multiple exit breaches over the rim, the most straightforward explanation is that the crater rim was modified between overspilling episodes, creating a new lowest point on the rim. The preservation of tadpole craters with diameters $<$4~km precludes modification by the high ($10^{2}$-$10^{4}$~nm~yr$^{-1}$) erosion rates during a wet climate predicted for Noachian Mars \citep{craddock2002case,salese2016sedimentary}, which erased craters of these sizes from Noachian-aged terrains \citep{irwin2013distribution}. Using studies of small crater degradation from orbital and rover measurements, slower, aeolian erosion rates during the Hesperian and Amazonian have been estimated to be much lower, of order 0.01-10~nm~yr$^{-1}$ \citep{golombek2000erosion,salese2016sedimentary,golombek2014small}. At an erosion rate of 10~nm~yr$^{-1}$, it would take 4.5~Gyr for aeolian erosion to erode the rim of \textit{Hellas} from the lowest point of B1 to the highest point on the pre-incision topography of B2 (Fig.~\ref{fig:profs}cii), which is unrealistic because tadpole craters must post-date the Noachian. Additionally, for \textit{Arabia Terra 1}, \textit{Arabia Terra 2}, and \textit{Noachis}, the lowest point in the highest incised breach is lower than the highest point on the pre-incision topography in another breach (e.g. \textit{Noachis} B1 and B2, Fig.~\ref{fig:profs}dii). This precludes the highest breach from incising first followed by a period of rim erosion because if the rim did not erode beneath the lowest point of an earlier breach, water should overspill from the first breach instead of forming a new breach above the highest point on the rim. Therefore, we exclude sequential overspill with periods of rim modification in between each breach incision as a multiple breach-forming mechanism.

\begin{figure}[t!]
\centering
\includegraphics[width=0.5\textwidth]{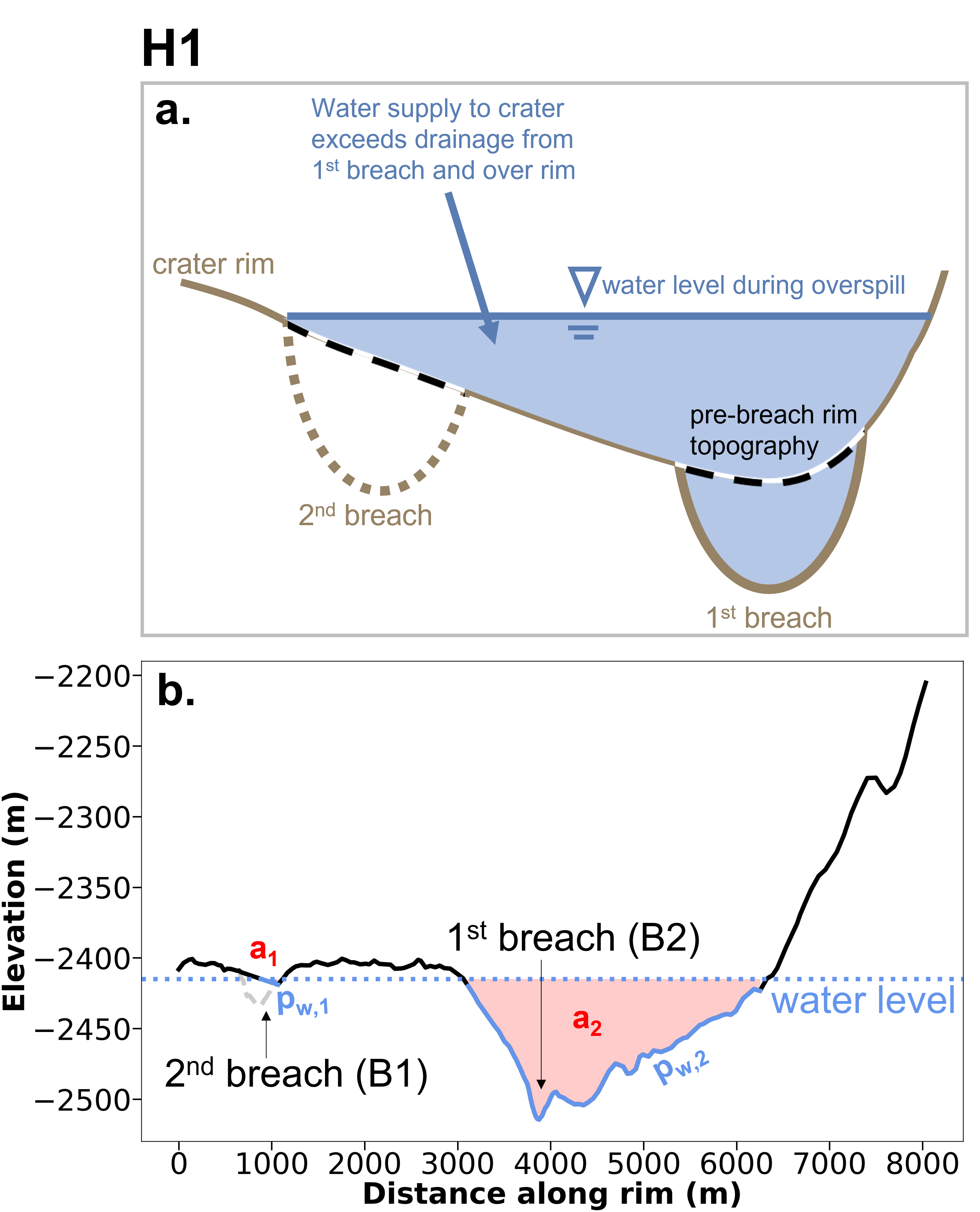}
\caption{\textbf{a.} Schematic illustration of simultaneous erosion of multiple rim breaches in a water-filled crater. The crater is supplied by a continuous water source such as precipitation or groundwater. Once the water level reaches the lowest point on the rim, an outlet breach begins eroding, but the flux of water out of that breach can’t keep up with the input flux. The water level continues to rise until new, higher breaches form to accommodate the overspilling water. In this scenario, the water source must be able to supply water at a rate greater than the combined flow rate out of the crater via the first breach and over the submerged crater rim. \textbf{b.} Topographic rim profile of \textit{Hellas} tadpole crater (Table~\ref{tab:mebtps}, Fig.~\ref{fig:profs}) with pre-incision topography for Breach 1 (B1). Dotted blue line shows water level. Solid blue line on topographic profile shows wetted perimeters for portions of flow out of the crater ($p_{w,1}$ and $p_{w,2}$). Red hashed regions show cross-sectional areas ($a_{1}$ and $a_{2}$) of portions of flow out of the crater. }
\label{fig:sim_ov}
\end{figure}

For multiple breaches to form simultaneously from a single crater filled with water, the water level must reach at least the lowest point of the pre-incision topography at the highest breach. In this scenario, water flows into the crater, filling it up. Once the water level reaches the lowest point on the rim, an outlet breach starts eroding. However, the flux of water into the crater exceeds the flux out of the first breach, and the water level continues to rise until new, higher elevation breaches form to accommodate the overspilling water (Fig.~\ref{fig:sim_ov}). In this scenario, water flows over the rim between breaches, and the total flux out of the crater through the eroding breaches and over the rim must be matched by the flux into the crater at least for a short time. Therefore, we can find the necessary water input flux to enable this multiple breach-forming mechanism to occur. 

Using the 4 available HiRISE DEMs of multiple breach tadpole craters (Fig.~\ref{fig:profs}), we measured topographic rim profiles by placing points on the highest elevation, well-preserved areas of the intact crater rim, excluding incised portions. Following \cite{Dussaillant2010}, we use this first topographic profile to calculate an approximate, simplified pre-incision rim topography using linear interpolation between fresh crater rim points either side of each breach. We then measured a second topographic rim profile for each crater including the points from the first profile, in addition to points placed on well-preserved areas of the incised breaches along a straight line between the edges of each breach. We combine these two topographic profiles to make a composite topographic profile where the lowest elevation breach has formed, but all higher breaches still have their pre-incision topography. We then select a ``water level'' ($z_{w}$) such that the lowest part of the pre-incision topography of the highest elevation breach would be submerged, and use this $z_{w}$ to calculate the wetted perimeter ($p_{w,i}$) and flow area ($a_{w,i}$) for each $i$th continuous Section~of the crater rim below the water level to calculate hydraulic radius ($R_{H,i}$):
\begin{equation}
    R_{H,i} = \frac{a_{w,i}}{p_{w,i}}
\end{equation}
this allows us to use Manning's equation to find the average flow velocity ($\bar{u}$) over each submerged rim section:
\begin{equation}
    \bar{u}_{i} = \frac{R_{H,i}^{2/3}S^{1/2}}{n}
\end{equation}
where $S$ is the slope for which we use values of 0.1 and 0.2 based on average measured rim slope values in \cite{warren2021overspilling}, and $n$ is Manning's roughness coefficient, also known as Manning's $n$. We use a standard gravity scaling for Manning's $n$ \citep{brownlie1983flow}:
\begin{equation}
    n = \frac{k^{1/6}}{8 g^{1/2}}
\end{equation}
where $g$ is surface gravity and $k$ is the channel bed roughness length scale which we set equal to grainsize $d_{g}$ from 1~mm to 1~m (Table~\ref{tab:param}) in each model run \citep{warren2021overspilling}.

We then calculate the total flux out of the crater:
\begin{equation}
    Q_{out} = R_{H,1}\bar{u}_{1} + ... + R_{H,i}\bar{u}_{i}
\label{eq:qout}
\end{equation}
where 1 to $i$ indicate the number of continuously submerged rim sections. 

We can then compare $Q_{out}$ for each crater (Table~\ref{tab:qin}) to potential water sources for a water-filled crater (Sections \ref{sec:melt} \& \ref{sec:gwat}).

\begin{table}[t!]
\begin{tabular}{rcccc}
\hline
 & & & \textbf{One-shot} & \textbf{Multi-shot} \\ \textbf{Tadpole} & \textbf{Q$_{out}$}         & \textbf{$A_{m}$ to reach}         & \textbf{Groundwater:} & \textbf{Melt pond overspill:}  \\
\multicolumn{1}{l}{}   \textbf{Crater} & \textbf{(m$^{3}$s$^{-1}$)} & \textbf{ Q$_{out}$ (km$^{2}$)} & \textbf{Q$_{in,max}$ (m$^{3}$s$^{-1}$)} & \textbf{$t_{melt,min}$ (days)} \\ \hline \hline
\textit{Arabia Terra~1} & 2.00$\times10^{4}$         & 157,000                                & 5.71$\times10^{5}$   & 77                             \\
\textit{Arabia Terra~2} & 8.82$\times10^{5}$         & 6,960,000                              & 4.63$\times10^{5}$    & 58                             \\
\textit{Hellas}         & 1.63$\times10^{6}$         & 12,900,000                             & 3.47$\times10^{5}$    & 4                              \\
\textit{Noachis}        & 1.86$\times10^{5}$         & 1,470,000                              & 5.13$\times10^{5}$    & 95                             \\ \hline
\end{tabular}
\caption{Modelling results for subaerial crater overspill hypotheses. $Q_{out}$ is the calculated flow out of the crater if all breaches are active simultaneously (Eqn. \ref{eq:qout}, Fig.~\ref{fig:sim_ov}, Section~\ref{sec:sim}). $A_{m}$ is the area over which melting would need to occur at $4$~m~yr$^{-1}$~m$^{-2}$ to supply $Q_{out}$. $Q_{in,max}$ is the maximum groundwater discharge rate found in our models, corresponding to groundwater flux from a pressurized aquifer with initial head $h_{a}=5000$~m, permeability $k_{h}=10^{-9}$~m$^{2}$ (Table~\ref{tab:param}), when the total volume of water discharged from the aquifer is 0.1 crater volumes (Section~\ref{sec:gwat}). $t_{melt,min}$ is the minimum time melting needs to occur at a rate of $4$~m~yr$^{-1}$~m$^{-2}$ to supply enough liquid water to melt ponds to repeatedly overspill and erode the observed breach depths for all tadpole crater rim breaches (Section~\ref{sec:ovsp}).}
\label{tab:qin}
\end{table}

\subsubsection{Precipitation \& Meltwater}
\label{sec:melt}

As tadpole craters lack inlets and have intact rims, liquid water cannot be routed over land to tadpole craters from a catchment area larger than the crater itself. To find the precipitation/melt rate within each multiple breach tadpole crater required to generate $Q_{out}$ required to activate all exit breaches simultaneously, we can simply divide $Q_{out}$ (Table~\ref{tab:qin}) by crater area. This gives values ranging from 0.6~mm~s$^{-1}$ for \textit{Arabia Terra~1} to 8.5~mm~s$^{-1}$ for \textit{Hellas}. Even the lowest value of 0.6~mm~s$^{-1}$ is equivalent to almost 52,000~mm~day$^{-1}$, which is 35 times the maximum daily rainfall rate recorded on Earth \citep{chien2011extreme} -- an unrealistically high precipitation or snowmelt rate for Earth or Mars. Such high precipitation rates would also be expected to dissect the surrounding landscape, which is not observed. We can therefore exclude precipitation within the crater and snowmelt leading to simultaneous overspill at multiple breach locations as a formation hypothesis for multiple breach tadpole craters. 

\textbf{\subsubsection{Retreating Ice Sheet: Simultaneous Overspill}}
\label{sec:retreat}

\begin{figure}[t!]
\centering
\includegraphics[width=0.9\textwidth]{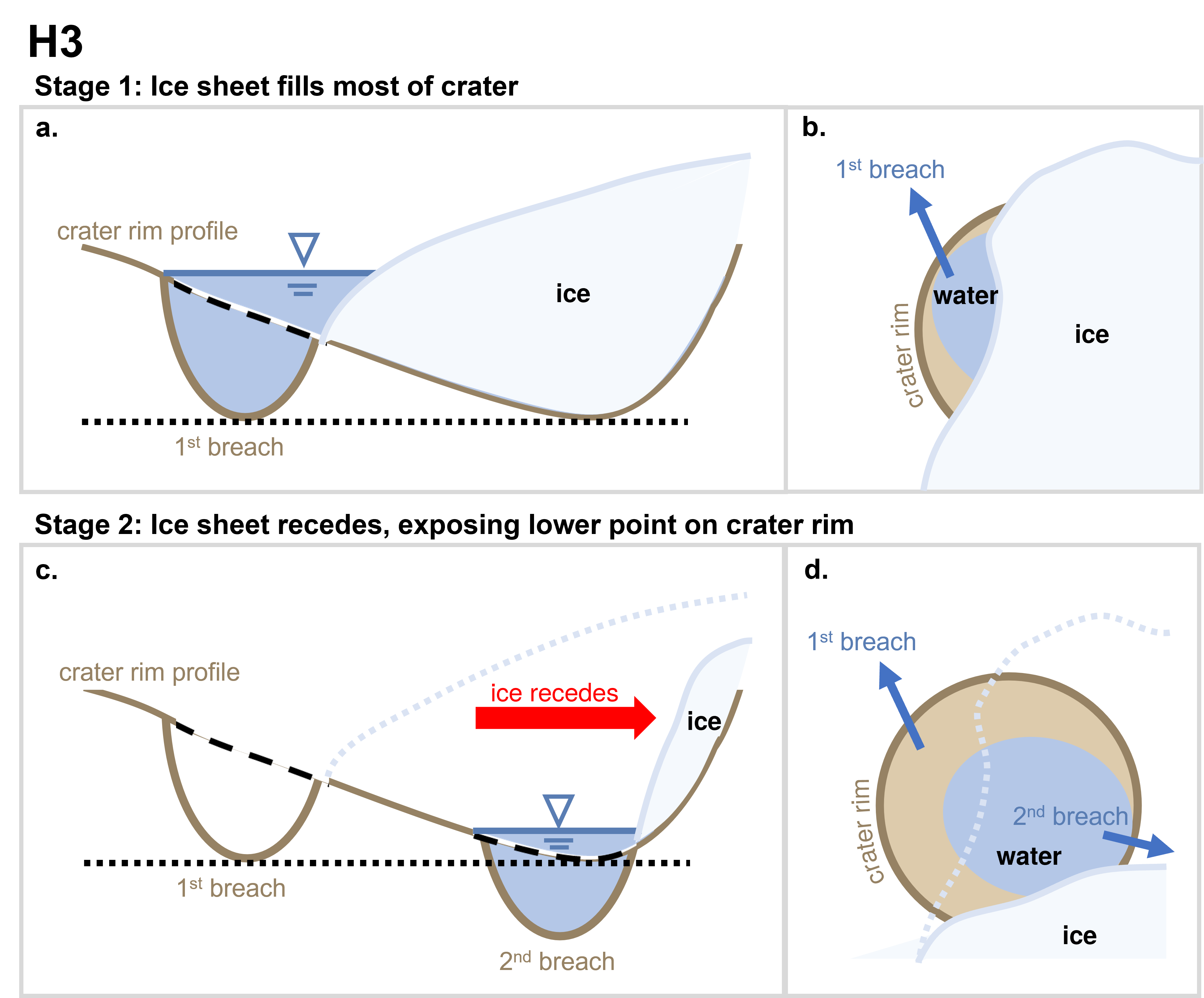}
\caption{Schematic illustration of an ice sheet retreating in two stages that progressively expose lower rim topography. First stage where ice sheet fills most of crater in a) profile view and b)  in birds eye view. Second stage where ice sheet recedes and exposes lower points on the crater rim in c) profile view and d) birds-eye view.}
\label{fig:retreat}
\end{figure}

A potential mechanism for expanding the catchment area of multiple breach tadpole craters without forming inlets is the intersection~of a retreating ice sheet margin with the crater interior (Fig.~\ref{fig:retreat}a). In this scenario, the high meltwater supply rates required to match $Q_{out}$ could be achieved by expanding the melting area rather than appealing to higher melting rates, for example by routing meltwater from a large ice sheet area into a crater. We calculate the required melting area for each multiple breach tadpole crater by dividing $Q_{out}$ by the melting rate on Mars, which is energy limited to around 4~m~yr$^{-1}$~m$^{-2}$ (\citealt{kite2013seasonal}; 100~W~m$^{-2}$ corresponds to $\sim1$~mm~hr$^{-1}$~m$^{-2}$ of melt).

The results of this calculation yield melting areas 1,700 to 67,000 times larger than the multiple breach tadpole crater areas (Table~\ref{tab:qin}), comparable to areas the size of Illinois and Russia for \textit{Arabia Terra~1} and \textit{Hellas} respectively. For \textit{Hellas}, this would be equivalent to routing meltwater from an ice sheet covering the entire Hellas basin into a single 14.6~km diameter crater. This would require unrealistic concave-up ice sheet topography funneling water to a single point, whereas ice sheets are typically convex-up, leading to melt runoff on all sides. Therefore we exclude the \textit{simultaneous} formation of multiple breaches by meltwater routed from a retreating ice sheet (H1, Fig~\ref{fig:mech}).

In principle, supraglacial meltwater could gradually accumulate in ponds and lakes on the ice surface and then experience a sudden release during ice sheet retreat. In practice however, \textbf{based on the physics of ice hydrofracture and observations of Greenland and of Antarctic ice shelves (e.g., \citealt{das2008,rice2015}}, it is likely that such supraglacial holding ponds would drain to the base of the ice sheet. Once at the base of the ice sheet, water  could not flow into the tadpole craters at the large fluxes required without forming inlet breaches, which are not observed.

\textbf{\subsubsection{Retreating Ice Sheet: Sequential Overspill}}

However, a rapidly retreating ice sheet is consistent with the \textit{sequential} formation of multiple breaches in a tadpole crater. As the ice sheet retreats, it provides meltwater to the crater for overspilling and progressively exposes new areas of the rim. As topographically lower portions of rim are exposed, water overspills there, and earlier, higher elevation breaches are abandoned (Fig.~\ref{fig:retreat}).
\textbf{This hypothesis is discussed in much more detail, including a terrestrial analog, in Section 6.}

\subsubsection{Groundwater}
\label{sec:gwat}

One mechanism proposed for generating extremely high discharge rates and water volumes to carve outflow channels is groundwater discharge from cryosphere-confined aquifers \citep{carr1979formation,harrison2004tharsis,manga2004martian,andrews2007hydrological}. Although tadpole crater outflow breaches are orders of magnitude smaller than these features, groundwater released from a pressurized aquifer might generate the high water fluxes required to sustain flow over the submerged rim of a tadpole crater, enabling multiple breaches to incise simultaneously.

\begin{figure}[t!]
\centering
\includegraphics[width=0.6\textwidth]{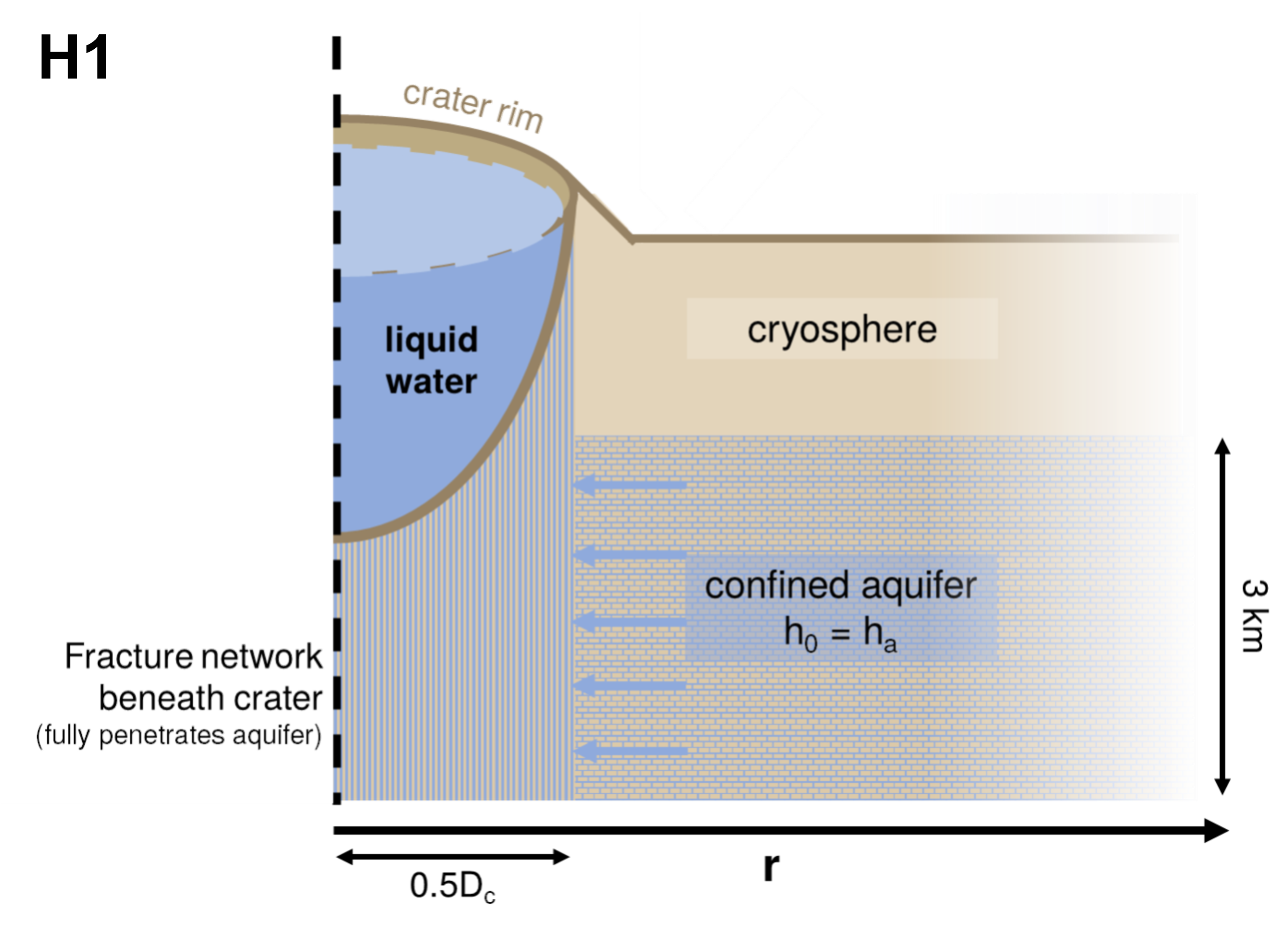}
\caption{Schematic illustration of our groundwater model. A homogeneous, radially symmetrical confined aquifer 3~km thick pressurized by an overlying cryosphere is fully penetrated by a network of fractures beneath the tadpole crater, extending to the crater rim at $r=0.5D_{c}$ where $D_{c}$ is the crater diameter. Dashed black line indicates axis of radial symmetry. We assume that the flux of water to the crater is limited by flow through the confined aquifer to the edge of the fracture network (blue arrows). The aquifer starts with an initial head ($h_{0}$) of $h_{a}$, which is a free parameter in our groundwater model (Table~\ref{tab:param}). }
\label{fig:gwat_mod}
\end{figure}

To test this hypothesis, we find the groundwater flux into a tadpole crater. We consider a case where there is a network of fractures beneath the crater, for example generated by impact fracturing \citep{mackinnon1989impacted,hanna2005hydrological}, that fully penetrates a homogeneous aquifer (Fig.~\ref{fig:gwat_mod}). When water from the aquifer reaches the edge of the fracture network at a distance of 0.5$D_{c}$ from the center of the crater, we assume it can begin filling the crater. 

The radial flow of groundwater through a confined, homogeneous aquifer in 1-D can be found using \citep{domenico1998physical}:

\begin{equation}
    \frac{\partial h}{\partial t} = \frac{k_{h}\rho_{w}g}{\mu S_{s}}\frac{\partial}{\partial r} \left( r\frac{\partial h}{\partial r}\right)
\end{equation}
where $h$ is hydraulic head in the aquifer, $t$ is time, $k_{h}$ is radial permeability, $\rho_{w}$ is the density of water, $\mu$ is the viscosity of water, $S_{s}$ is the specific storage of the aquifer, and $r$ is radial distance from a central point (i.e. the center of the crater) (Table~\ref{tab:param}). We solve this equation using an explicit finite differences method for an aquifer of fixed thickness (3~km) to find the maximum groundwater flux into each multi-exit-breach tadpole crater using the parameters listed in Table~\ref{tab:param}. To maximize the groundwater discharge in our aquifer model, which is conservative in terms of rejecting the groundwater hypothesis, we neglect the turbulent upper limit on flow rate out of the aquifer as water moves through fractures to the tadpole crater \citep{manga2004martian}. We investigate the groundwater flow rates for 10 evenly log-spaced values of both $k_{h}$ and the initial pressure head in the aquifer ($h_{a}$) to capture a range of possible pressurized aquifer conditions.

\begin{figure}[t!]
\centering
\includegraphics[width=0.7\textwidth]{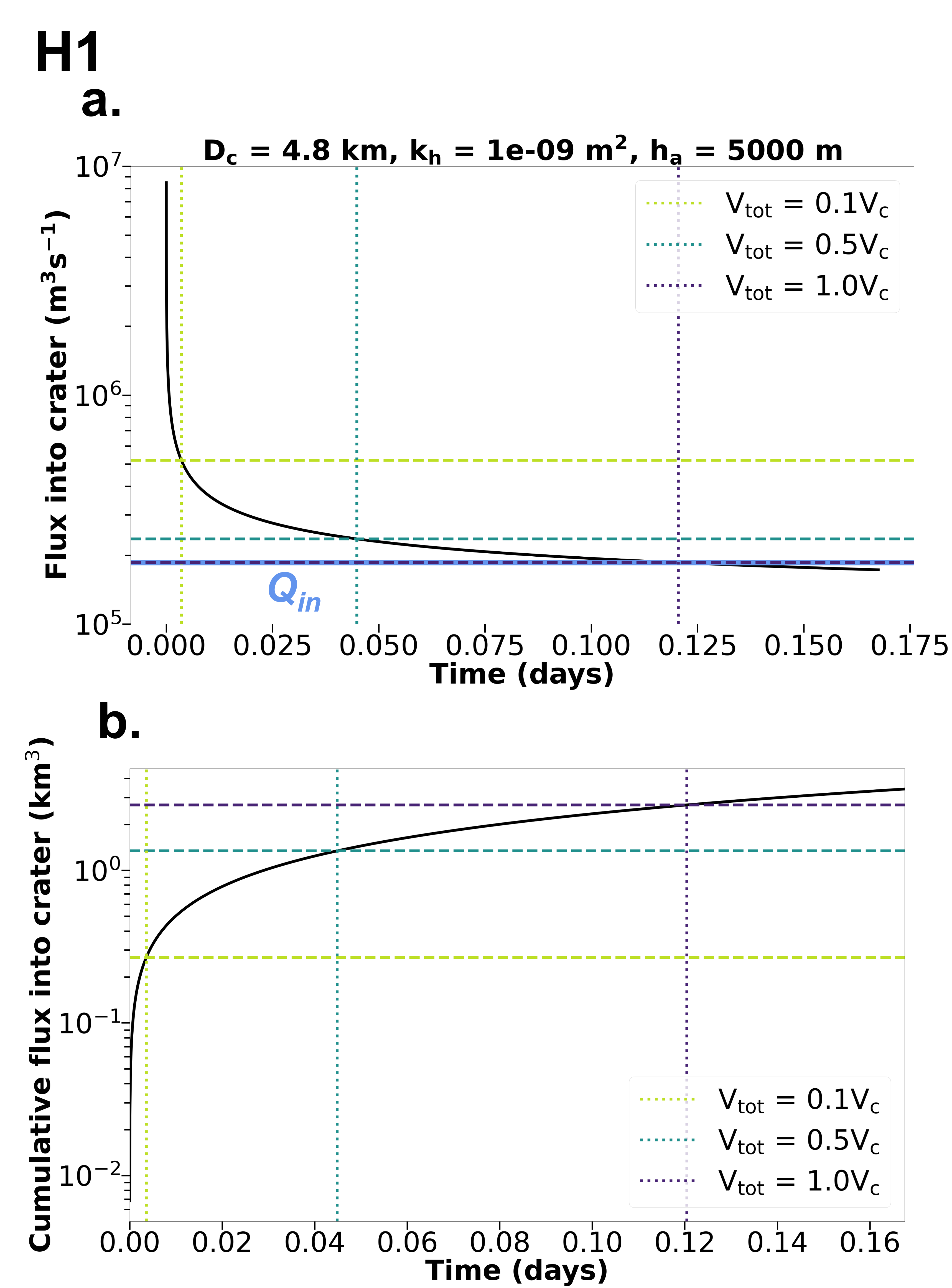}
\caption{Example water fluxes into the crater from our groundwater model shows that confined aquifers with high permeability ($k_{h}$) and high initial pressure head ($h_{a}$) can exceed the calculated flow out of the crater required to activate all channels simultaneously ($Q_{in}$; Table~\ref{tab:qin}, solid blue line) for \textit{Noachis} (Table~\ref{tab:mebtps}). This corresponds to Hypothesis 1, where liquid water overspills the crater rim. \textbf{a.} Time evolution of groundwater flux into tadpole crater and \textbf{b.} cumulative groundwater flux into crater. Dotted vertical lines indicate the times when the groundwater discharge has filled 0.1, 0.5, and 1 times the crater volume ($V_{c}$). Dashed lines indicate the instantaneous and cumulative flux at these times.}  
\label{fig:gwat_ex}
\end{figure}

As expected, groundwater flow is maximized for the highest values of $k_{h}$ ($10^{-9}$~m$^{2}$, corresponding to the average horizontal permeability of a basaltic aquifer composed of stacked lava flows at the surface; \citealt{hanna2005hydrological, manga2004martian}) and $h_{a}$ (5000~m). However, even these conditions are not able to generate sufficient groundwater flow to match $Q_{out}$ for tadpole craters \textit{Arabia Terra~2} and \textit{Hellas} for a 3~km thick aquifer (Table~\ref{tab:qin}). In the case of \textit{Noachis}, sufficient groundwater flow can only be sustained until the volume of water discharged from the aquifer has reached 0.9 crater volumes (Fig.~\ref{fig:gwat_ex}) for the highest values of $k_{h}$ ($10^{-9}$~m$^{2}$) and $h_{a}$ (5000~m). 

As our models are 1D, $Q_{in}$ scales linearly with aquifer thickness. For \textit{Arabia Terra~2} an aquifer thickness of approximately 6~km would be required, and a 14~km thick aquifer would be required to match $Q_{out}$ for \textit{Hellas}. On Earth, crustal permeabilities decrease with depth by several orders of magnitude within a few kilometers \citep{huenges1997physical}, and a factor 10$^{4}$ decrease in permeability is expected for the martian subsurface as increasing lithostatic stresses lead to the closure of pore space and open fractures \citep{hanna2005hydrological}. Additionally, aquifer heads of $h_{a}=5000$~m are equivalent to a pressure of 18.5~MPa, which are not achieved in models of aquifer pressurization by a downward-freezing aquifer following a rapid change in surface temperature from 270 to 220~K \citep{hanna2005hydrological}. Therefore, we conclude that discharge from a confined aquifer driving simultaneous overspill of multiple breaches is inconsistent with \textit{Arabia Terra~2} and \textit{Hellas}, and unlikely for \textit{Noachis}.

Our calculations cannot exclude groundwater-driven simultaneous overspill of both breaches at \textit{Arabia Terra~1}. However, the unrealistic combination of high permeabilities, aquifer pressures, and thick aquifers required to reconcile a groundwater source for $Q_{out}$ for \textit{Arabia Terra~2} and \textit{Hellas}, the preferential occurrence of tadpole craters at high elevations relative to surrounding topography (Fig.~\ref{fig:elev}, Section~\ref{sec:distribution}), and the absence of evidence for rim incision or fluvial modification between breaches (Section~\ref{sec:rims}) disfavors this mechanism for the overall tadpole crater population.

An additional consideration for a groundwater-driven crater overspilling mechanism is that a fracture network feeding the crater is required. Although we interpret the four multiple breach tadpole craters in this study to be ice filled (Section~\ref{sec:fill}), thus burying any potential crater floor fractures, an impact-generated fracture system would likely extend up the crater wall, through the crater rim, and potentially into surrounding terrain \citep{POLANSKEY1990impact,senthil2005structural,senthil2008impact}. No fractures are visible on the scale of our HiRISE images. Therefore, any fractures (if present), must have apertures smaller than approximately 1~m. While the absence of these features in four tadpole craters does not exclude groundwater as a mechanism for all tadpole craters, noting the presence/absence of groundwater-related features and textures is important for future observations of tadpole craters. 

\begin{table}[t!]
\begin{tabular}{rcll}
\hline
\multicolumn{1}{l}{\textbf{Symbol}} & \multicolumn{1}{l}{\textbf{Value(s)}} & \textbf{Units}                 & \textbf{Description}                           \\ \hline \hline
\multicolumn{4}{c}{\textit{General parameters}}                                                                                                               \\ \hline
$g$                                 & 3.71                                  & m~s$^{-2}$                & Mars gravity                                   \\
$\rho_{w}$                          & 1000                                  & kg~m$^{-3}$               & Density of water                               \\
$\rho_{i}$                          & 918                                   & kg~m$^{-3}$               & Density of ice                                 \\
$\rho_{r}$                          & 2500                                  & kg~m$^{-3}$               & Density of crust                               \\ \hline
\multicolumn{4}{c}{\textit{Groundwater model}}                                                                                                                \\ \hline
$k_{h}$                             & $10^{-15}$ - $10^{-9}$                & m$^{2}$                        & Permeability *,$^{+}$         \\
$h_{a}$                             & 100 - 5000                            & m                              & Initial head in pressurized aquifer            \\
$S_{s}$                             & $10^{-6}$                             & m$^{-2}$                       & Specific storage$^{+}$     \\
$z_{a}$                             & 3000                                  & m                              & Aquifer thickness                              \\
$\mu$                               & $1\times 10^{-6}$                      & m$^{2}$~s$^{-1}$          & Viscosity of water                             \\ \hline
\multicolumn{4}{c}{\textit{Intrusion model}}                                                                                                                  \\ \hline
$T_{int}$                           & 1300                                  & K                              & Initial temperature of intrusion               \\
$T_{surf}$                          & 230                                   & K                              & Surface temperature                            \\
$z_{i}$                             & 100 - 2000                            & m                              & Ice thickness                                  \\
$q_{geo}$                           & 25                                    & mW~m$^{-2}$               & Geothermal heat flux \\
$c_{p,w}$                           & 4180                                  & J~kg$^{-1}$~K$^{-1}$ & Specific heat capacity of water                \\
$c_{p,i}$                           & 2100                                  & J~kg$^{-1}$~K$^{-1}$ & Specific heat capacity of ice                  \\
$c_{p,r}$                           & 1100                                  & J~kg$^{-1}$~K$^{-1}$ & Specific heat capacity of crust                \\
$L_{i}$                             & 330                                   & kJ kg$^{-1}$                            & Latent heat of melting ice                     \\ \hline
\multicolumn{4}{c}{\textit{Melt pond overspill model}}                                                                                                                           \\ \hline
$h_{0}$                             & 0.1 - 1                               & m                              & Initial flow depth in breach \citep{warren2021overspilling}                   \\
$d_{g}$                             & 0.001 - 1                             & m                              & Grainsize in breach \citep{warren2021overspilling}    
                \\
$f_{a}$                             & 0.001 - 1                             & -                              & Fraction of crater area covered by melt pond
                \\
$S$                             & 0.1, 0.2                             & -                              & Slope \\ \hline

\end{tabular}
\caption{Parameter values used for models of radial groundwater flow, ice melting above an intrusion, and overspill from a melt pond on the surface of an ice-filled crater. *\cite{andrews2007hydrological}, $^{+}$\cite{manga2004martian} }
\label{tab:param}
\end{table}

\begin{figure}[t!]
\centering
\includegraphics[width=\linewidth]{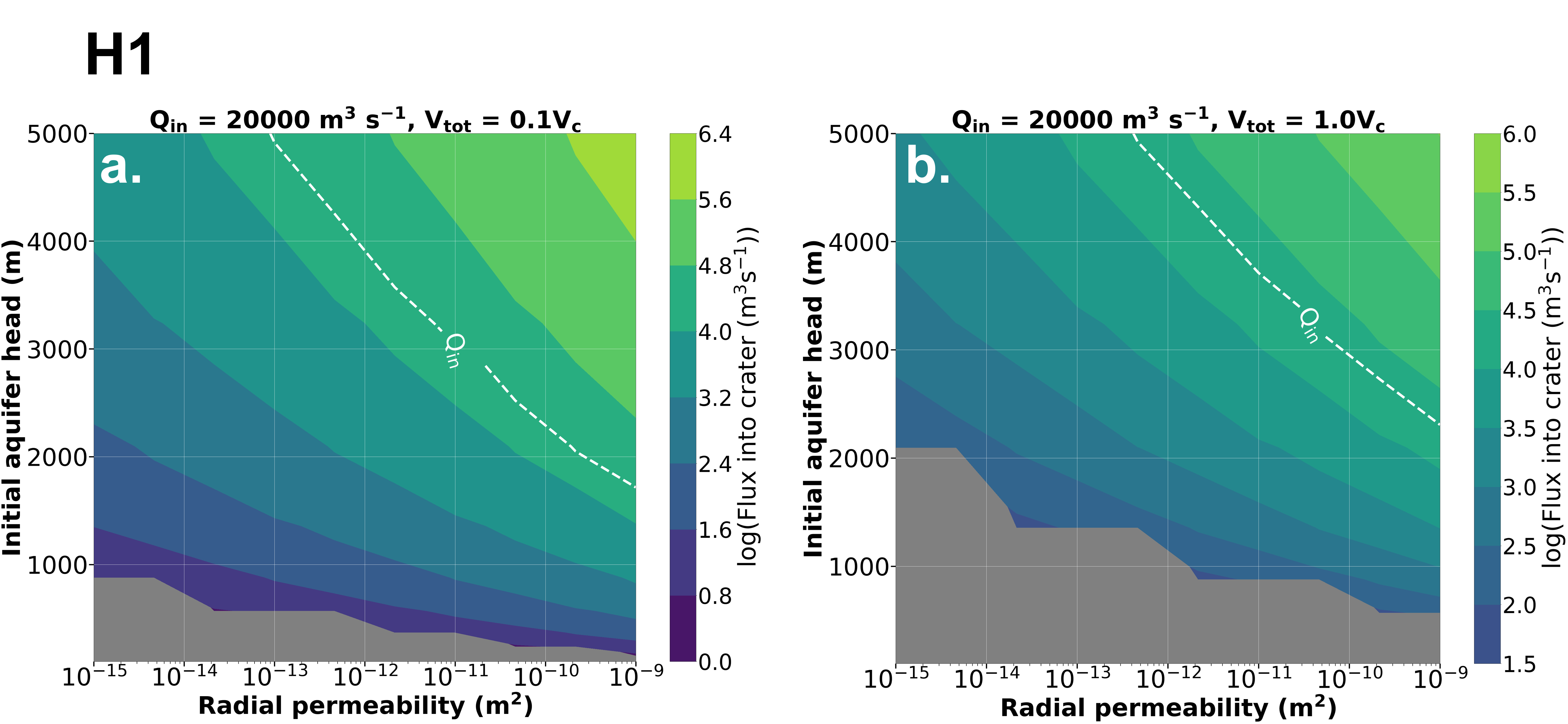}
\caption{Groundwater flow through a confined aquifer with initial head $>$2000~m and permeability $>10^{-12}$~m$^{2}$ can sustain the water supply rate necessary to overspill both breaches simultaneously for tadpole crater \textit{Arabia Terra~2} (Table~\ref{tab:qin}) from when the groundwater has supplied \textbf{a.} 0.1 crater volumes of water to \textbf{b.} 1 crater volume of water (i.e. the crater is filled to the brim with groundwater). This corresponds to Hypothesis 1, where liquid water overspills the crater rim. However, higher initial aquifer head and permeabilities are required to sustain fluxes greater than $Q_{in}$ until the crater is completely filled with water. Grey shaded areas indicate negligible groundwater flow into the crater on the timescale of one model timestep. Modeling results for other multiple breach tadpole craters are in Appendix~\ref{sec:gwat_full}.}
\label{fig:gwat_cont}
\end{figure}

\subsubsection{Ice-filled crater}
\label{sec:ovsp}

The 4 tadpole craters in this study all show features on the crater floor consistent with ice-fill, e.g. brain terrain and concentric ridges, and ice-filled craters are abundant poleward of around 30~degrees latitude. This ice fill may be a remnant of previous ice coverage, and the same ice (or an older generation of ice) may have infilled the crater at the time of breaching. If the crater was once ice-filled, subaerial drainage of supraglacial meltwater ponds could have drained over the crater rim \citep{warren2021overspilling}. In this scenario, breach location would be dependent on ice surface topography and melt routing, not crater rim topography. 

For each tadpole crater, we can calculate the number of meltwater overspill events needed to erode each crater rim breach to its observed depth as a function of meltwater pond diameter, and thus approximate the number of melt-years needed on Mars to generate multi-exit breach tadpole craters, constraining the Martian climate.

To do this, we implement the coupled model of lake draining and breach erosion described in \cite{warren2021overspilling} using our measured crater and breach properties in Tables \ref{tab:mebtps} \& \ref{tab:breach} respectively. This model assumes no flow into the crater during erosion, no downstream changes in channel/valley geometry, flow depth, or flow velocity, and perfect downstream advection and removal of all eroded sediment. In order to find the minimum number of melt ponding events required, we assume that pond depth does not limit erosion, and instead that the pond depth for each erosion event is simply equal to the total eroded depth for that event plus the initial water depth in the channel (Table~\ref{tab:param}). We set pond area for each breach as a fractional area ($f_{p}$) of the entire tadpole crater ($A_{c}$) and calculate the number of melting days ($t_{m}$) represented by each breaching event (Table~\ref{tab:qin}) using:

\begin{equation}
    t_{m} = \frac{f_{p}A_{c}z_{e}}{r_{m}A_{c}}
\end{equation}
where $r_{m}$ is melt rate (4~m~yr$^{-1}$~m$^{-2}$; \citealt{kite2013seasonal}). We assume no initial damming of the channel, which is conservative because it reduces the total erosion required to achieve the observed breach depths.

\begin{figure}[t!]
\centering
\includegraphics[width=\linewidth]{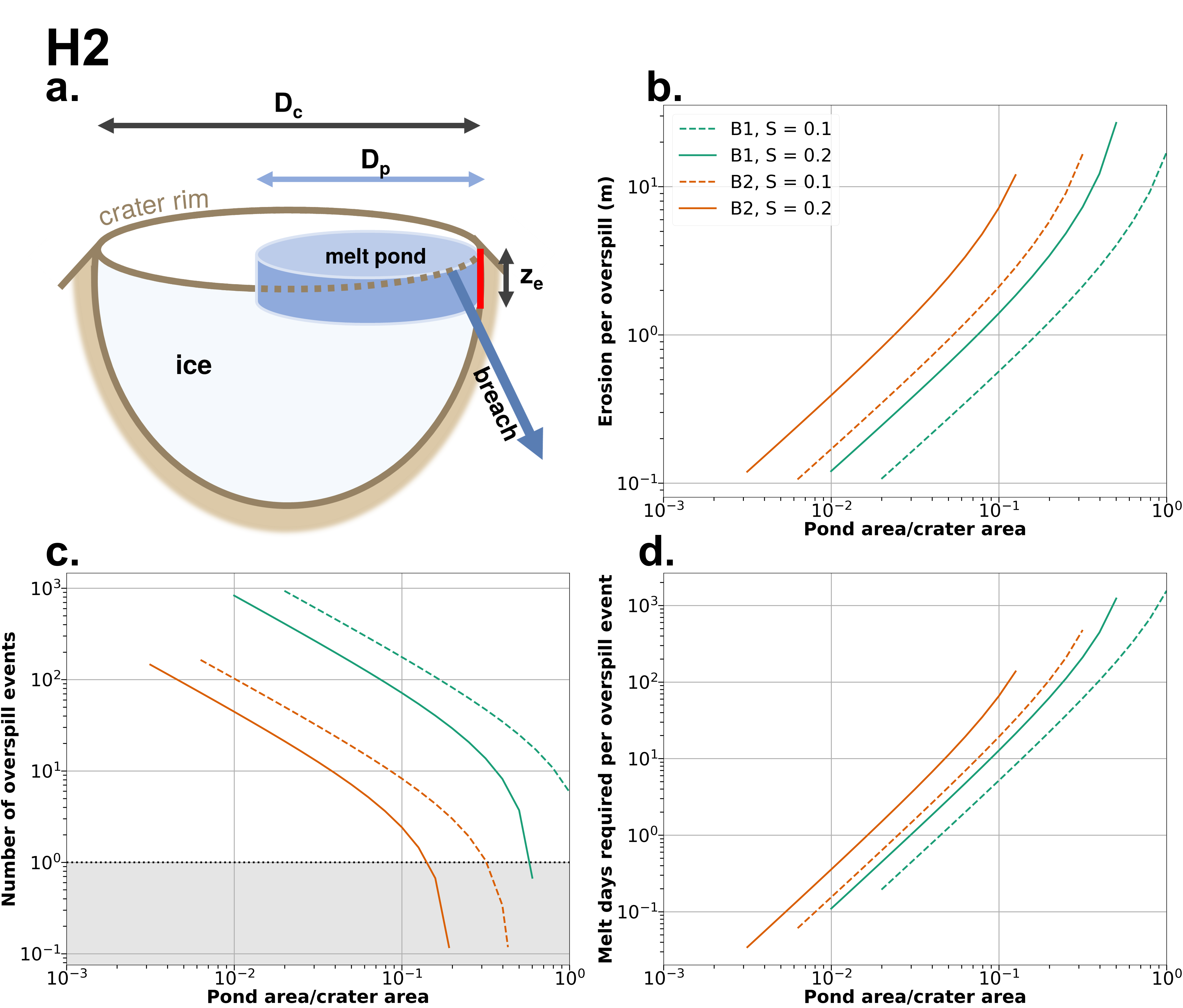}
\caption{Meltwater pond overspill modelling results for \textit{Noachis} B1 (teal) and B2 (orange) (Fig.~\ref{fig:profs}) for measured breach parameters in Table~\ref{tab:breach} as a function of fractional melt pond area. \textbf{a.} Schematic illustration of an ice-filled crater with a shallow melt pond smaller than the area of the crater with diameter $D_{p}$ overspilling the rim via a breach. For each overspill event, the initial water depth in the melt pond must be at least the depth eroded in the overspill model ($z_{e}$, \citealt{warren2021overspilling}).  \textbf{b.} Downward erosion of breach base per individual overspilling event. Smaller melt pond areas generate less erosion when they overspill the rim. \textbf{c.} The number of overspill events required to carve each observed breach. The number of breaches required decreases as melt pond area increases. \textbf{d.} Melt days required to generate the water volumes required per overspill event in \textbf{c.} at a melt rate of 4~m~yr$^{-1}$~m$^{2}$ \citep{kite2013seasonal} across the entire crater area. Modeling results for other multiple breach tadpole craters are in Appendix~\ref{sec:ovsp_full}.}
\label{fig:nevents}
\end{figure}

To find the minimum total magnitude of the climate excursion required to generate all exit breaches for all four multiple breach tadpole craters, we sum the minimum total number of melt days to carve each breach ($t_{melt,min}$, Table~\ref{tab:qin}). This represents the minimum number of days that the temperature must exceed 273~K to form the observed exit breaches. As this mechanism allows for each breach to form independently of any others, and each breach may be eroded by up to 10$^{3}$ small overspill events, these days do not necessarily need to be continuous.  Rather, melting conditions need only persist continuously for as long as is needed to generate a sufficient volume of meltwater, which can be as little as hours (Fig.~\ref{fig:nevents}d). 

Using the Mars Climate Database v6.1 \citep{forget1999improved,millour2018exploring}, the modern surface temperature at \textit{Noachis} exceeds 273~K for approximately 6~hrs a day during southern hemisphere summer, with peak temperatures of 310~K. Although peak temperatures vary, all four multiple breach tadpole craters in this study experience temperatures $>273$~K for multiple hours during the warmest parts of the year. However, the high albedo and thermal inertia of ice and evaporitic cooling would prevent these temperatures alone from generating liquid water on modern Mars \citep{ingersoll1970mars}, even if there were surface-exposed ice at these locations. Therefore, overspill of multiple breaches at this tadpole crater would require at least enough additional warming to overcome evaporitic cooling, in addition to stabilization of surface ice and ice accumulation in the southern mid-latitudes (which can occur under higher obliquity conditions in climate models, e.g. \citealt{forget2006formation,FASTOOK201460}), and potentially increased atmospheric pressure depending on the dust content and crystalline structure of crater-filling ice/snowpack \citep{CLOW1987generation}. However, if diurnal temperatures $>$273~K are required (for example to prevent re-freezing of a larger melt pond overnight), 50~K of warming is required at \textit{Arabia Terra~2} in the northern hemisphere, and 25~K of warming at \textit{Hellas} in the southern hemisphere. 50~K of warming is comparable to the warming invoked to explain the intense fluvial erosion and valley networks of the Noachian, and is difficult to achieve without either a much thicker ($>$1~bar) CO$_{2}$ atmosphere and additional greenhouse gases such as H$_{2}$ \citep{ramirez2014,wordsworth2017,kite2014,ramirez_volcanic_2017,batalha_testing_2015}, H$_{2}$O-ice cloud warming \citep{kite2021warm}, or another as yet unexplored warming mechanism. One potential warming mechanism that may reduce the required warming to generate melt is a local increase in air temperature caused by the low albedo of exposed crater rim rock. This mechanism aids the formation of lakes on the surfaces of Antarctic glaciers even when summer temperatures remain below freezing \citep{Stokes2019widespread,Kingslake2017,Arthur2022large}, and supraglacial melt is often found in close association with exposed bedrock ridges and other areas of low albedo such as blue ice \citep{Stokes2019widespread,bintanja_reijmer_2001}. Overall, we favor repeated overspill of small melt ponds that require volumes of water than can be generated in $<1$ day of melting, which is consistent with our overspill erosion results for all four multiple breach tadpole craters and can be achieved in a colder, more arid climate as expected for the Hesperian onwards on Mars.

\subsection{Subglacial Overspill (H4)}
\label{sec:subg}

In this subsection, we consider  our Hypothesis 4 (Fig.\ref{fig:mech}d). In this hypothesis, the flow of liquid water over the crater rims was driven by overpressure at the base of a regional ice sheet rather than by topography, as in the subaerial hypotheses (H1-3, Fig.\ref{fig:mech}a-c) discussed in the preceding subsection. We consider the driving pressures required to force water up the inner slope of the crater rim, as well as two potential water sources for subglacial streams. 

\subsubsection{Water Volume}

An expression for the mean velocity of flow in a tunnel within or at the base of a glacier is \citep{nye1976water}:

\begin{equation}
\label{eq:man_sub}
    \bar{u} = \frac{R_{H}^{2/3}}{n} \left[ \frac{1}{\rho_{w} g} \left( \rho_{w}g\sin{\alpha} - \frac{\partial p}{\partial x} \right) \right]^{1/2} 
\end{equation}
where $R_{H}$ is hydraulic radius, $n$ is Manning's roughness coefficient, $\alpha$ is the bedrock channel slope and $\frac{\partial p}{\partial x}$ is the downstream pressure gradient. 

The maximum driving pressure gradient for water exiting a subglacial channel is the difference in ice overburden pressure at the base of the tadpole crater and atmospheric pressure over the runout length of the measured outlet breaches, divided by the length of the outlet breach $L$ (Table~\ref{tab:breach}):

\begin{equation}
\label{eq:sub_pgrad}
    \frac{\partial p}{\partial x} \approx \frac{\rho_{i}g (d_{c} + h_{i})}{(L+0.5D_{c})}
\end{equation}
where $d_{c}$ is the depth of the tadpole crater (Table~\ref{tab:mebtps}), and $h_{i}$ is the ice thickness above the crater. This approach assumes that ice sheet thickness drops to zero by the end of the carved breach, maximizing the driving pressure gradient while also ensuring the full measured breach length remains subglacial. 

Using our measured topographic profiles for each multiple breach tadpole crater, we can calculate the hydraulic radius of each individual breach using the same procedure as described for the multiple simultaneously active floods but adding the width of the submerged area to the wetted perimeter ($p_{w}$) to account for the top surface of the flow being in contact with ice. We assume that the ice has the same roughness as the rest of the channel perimeter \citep{nye1976water}. Then, we can use Equation~\ref{eq:man_sub} to calculate discharge out of each individual breach.

Terrestrial glacial lake outburst floods (GLOFs) can transport large volumes of coarse sediment and may dominate bedrock erosion in mountainous regions such as the Himalayas (e.g. \citealt{cook2018glacial}). If we assume that tadpole crater breaches formed subglacially during a glacial outburst flood-like event, we can use this discharge to estimate the volume of water $V_{L}$ needed to erode each breach using the empirical relationship between peak tunnel channel discharge $Q_{p}$ during terrestrial j{\"o}kulhlaups \citep{walder1996outburst}:

\begin{equation}
\label{eq:jokul}
    Q_{p} = 46 V_{L}^{0.66}
\end{equation}

A caveat of using this equation is that it does not include a scaling for Mars gravity, and is based on terrestrial observations only. We find that for subglacial flow driven by slope only, the empirical terrestrial j{\"o}kulhlaup relationship generates water volumes ($V_{sub}$, Table~\ref{tab:subg_man_calc}) exceeding the initial fresh crater volume ($V_{c}$, Table~\ref{tab:mebtps}) for all breaches in \textit{Arabia Terra~2}, and one breach in \textit{Noachis}, representing more than half of all breaches. However, the water volume cannot exceed the crater volume. Therefore, if the exit breaches did form via j{\"o}kulhlaup-like GLOFs, we conclude that the terrestrial scaling between tunnel channel discharge and total water volume does not hold. Alternatively, the breaches may represent consistent subglacial drainage maintained over a longer period of time, rather than a catastrophic GLOF-like event. 

An alternative approach to calculate the minimum required subglacial water volume is to use the total eroded breach volumes (Table~\ref{tab:breach}) to calculate the volume of water required to move that volume of sediment, given a range of sediment to 
 water ratios. For the full range of dilute (i.e. clear water) to hyperconcentrated flows, sediment:water ratios typically vary from 0.0005 to 0.3 \citep{MANGOLD2012origin}. This factor of 600 difference yields a wide range of water volumes for each multiple breach tadpole crater breach, however, we only consider the minimum volume to constrain the minimum melting event necessary to generate sufficient water to erode each breach under subglacial conditions ($V_{w,min}$, Table~\ref{tab:breach}).

We recommend future quantitative modeling of subglacial erosion for a variety of ice sheet geometeries to better constrain the total volume of water needed.

\begin{table}[t!]
\begin{tabular}{lccc|cc|ccc}
                                 &                 &                      &                      & \multicolumn{2}{c|}{\textit{Slope only $\left(\frac{\partial p}{\partial x} = 0 \right)$}} & \multicolumn{3}{c}{\textit{Crater ice only ($h_{i} = 0$)}}                                                                              \\ \hline
\textbf{Tadpole} &  & \textbf{$p_{w,ice}$} & \textbf{$R_{h,ice}$} & \textbf{$Q_{out,ice}$}       & \textbf{$V_{sub}$}                      & \textbf{$\frac{\partial p}{\partial x}$} & \textbf{$Q_{out,ice}$}     & \textbf{$V_{sub}$}                                       \\
                                 \textbf{Crater} &                 & \textbf{(m)}         & \textbf{(m)}         & \textbf{(m$^{3}$s$^{-1}$)}   & \textbf{(km$^{3}$)}   & \textbf{(Pa m$^{-1}$)}                   & \textbf{(m$^{3}$s$^{-1}$)} & \textbf{(km$^{3}$)}  \\ \hline \hline
\textit{Arabia}          & B1              & 686                  & 6.9                  & 1.0$\times 10^{4}$                    & 4                                       & 1.0$\times 10^{3}$                                & 2.1$\times 10^{4}$                  & 10                                                                        \\
\textit{Terra 1}                        & B2              & 635                  & 6.4                  & 8.5$\times 10^{3}$                    & 3                                       & 1.4$\times 10^{3}$                                & 1.9$\times 10^{4}$                  & 9                                                                        \\ \hline
\textit{Arabia}          & B1              & 446                  & 7.8                  & 8.4$\times 10^{3}$                    & 3                                       & 3.3$\times 10^{3}$                                & 2.7$\times 10^{4}$                  & 16                                                                      \\
\textit{Terra~2}                        & B2              & 706                  & 6.4                  & 9.5$\times 10^{3}$                    & 3                                       & 1.6$\times 10^{3}$                                & 2.3$\times 10^{4}$                  & 12                                                                        \\
\textit{}                        & B3              & 922                  & 8                    & 1.8$\times 10^{4}$                    & 8                                       & 4.2$\times 10^{2}$                                & 2.7$\times 10^{4}$                  & 16                                                                        \\
\textit{}                        & B4              & 676                  & 16.6                 & 4.4$\times 10^{4}$                    & 33                                      & 1.1$\times 10^{3}$                                & 9.1$\times 10^{4}$                  & 99                                                                        \\
\textit{}                        & B5              & 605                  & 7.4                  & 1.0$\times 10^{4}$                    & 4                                       & 2.3$\times 10^{3}$                                & 2.9$\times 10^{4}$                  & 17                                                                         \\ \hline
\textit{Hellas}                  & B1              & 642                  & 6.6                  & 9.0$\times 10^{3}$                    & 3                                       & 4.1$\times 10^{3}$                                & 3.3$\times 10^{4}$                  & 21                                                                        \\
\textit{}                        & B2              & 489                  & 7.1                  & 7.8$\times 10^{3}$                    & 2                                       & 5.6$\times 10^{3}$                                & 3.3$\times 10^{4}$                  & 21                                                                         \\ \hline
\textit{Noachis}                 & B1              & 1813                 & 21                   & 1.8$\times 10^{5}$                    & 270                                     & 1.7$\times 10^{3}$                                & 4.3$\times 10^{5}$                  & 1000                                                                     \\
\textit{}                        & B2              & 624                  & 4.5                  & 4.7$\times 10^{3}$                    & 1                                       & 1.7$\times 10^{3}$                                & 1.2$\times 10^{4}$                  & 4                                                                         \\ \hline
\end{tabular}
\caption{Calculations for subglacial breach scenarios for all tadpole craters (Tables \ref{tab:param} \& \ref{tab:breach}, Fig. \ref{fig:key}), assuming ice surface level with the top of the modern observable breach incision. We consider 2 cases: 1. Flow in each subglacial channel is driven only by a 0.1 slope. 2. Flow in the subglacial channel is driven by pressure gradient between crater bottom and surrounding terrain, roughly equivalent to the pressure exerted by a column of ice of thickness $d_{c}$. $p_{w,ice}$ is wetted perimeter including ice surface ($p_{w}$ + $w_{v}$, Table~\ref{tab:breach}), $R_{H,ice}$ is the hydraulic radius calculated using $p_{w,ice}$. $Q_{out,ice}$ is peak flow rate (Eqns. \ref{eq:man_sub} \& \ref{eq:qout}), $V_{sub}$ is an empirical approximation of total subglacial water volume assuming j{\"o}kulhlaup-like drainage of crater (Eqn. \ref{eq:jokul}). $\frac{\partial p}{\partial x}$ is subglacial downstream pressure gradient (Eqn. \ref{eq:sub_pgrad}). All calculated water volumes and fluxes are given to 2 s.f.} 
\label{tab:subg_man_calc}
\end{table}

\subsubsection{Ice Sheet Slope}
Another factor in the subglacial routing of meltwater out of the tadpole craters is that the subglacial water pressure must be sufficient to overcome the steep topographic slope of the crater wall. Assuming that the basal meltwater pressure is static, which is appropriate for small discharges \citep{bjornsson2003subglacial} (i.e. persistent subglacial flow rather than a catastrophic j{\"o}kulhlaup-like event), the fluid potential gradient ($\frac{\partial p}{\partial x}$) at a given point beneath an ice sheet is given by the sum of the gravitational potential gradient and the overburden pressure gradient\citep{bjornsson2003subglacial}:

\begin{equation}
     \frac{\partial p}{\partial x} = (\rho_{w} - \rho_{i})g \frac{\partial z_{b}}{\partial x} + \rho_{i}g  \frac{\partial (z_{s} - z_{b})}{\partial x}
\end{equation}
where $\rho_{i}$ and $\rho_{w}$ are the densities of ice and water respectively, $z_{b}$ is the elevation at the base of the ice sheet relative to some datum (on Earth, sea level), and $z_{s}$ is the elevation of the ice sheet surface. At a single point, $\nabla z_{b}$ and $\nabla (z_{s} - z_{b})$ are equivalent to the slopes at the base of the ice sheet ($S_{b}$) and the surface of the ice sheet ($S_{s}$) respectively, and the driving pressure gradient at that point can be approximated by:

\begin{equation}
    \frac{\Delta p}{\Delta x} = (\rho_{w} - \rho_{i}) g S_{b} + \rho_{i} g S_{s}
\end{equation}

On the interior wall of the crater leading to the channel, $S_{b}$ is negative. Therefore for the flow of water to proceed out of the crater, it must be exceeded by $\rho_{i} g S_{s}$ such that $\Delta \phi$ is positive. This leads to an expression for the minimum ice sheet surface slope required:

\begin{equation}
    S_{s} > \frac{(\rho_{w} - \rho_{i})}{\rho_{i}}S_{b}
\end{equation}

The surface slope $S_{s}$ must therefore be at least 9\% of the basal slope in order to drive meltwater up and out of the crater. Therefore, for a basal slope of 0.6 ($\approx$30 degrees, or close to the angle of repose), the ice surface slope would need to be greater than $0.054$, or around 3 degrees. The slopes on the exterior edges of crater-filling ice mounds \citep{conway2012climate}, and at the termini of lobate debris aprons \citep{mangold2003geomorphic,FASTOOK2014formation} can be steeper than this. Therefore, we cannot exclude subglacial incision of exit breaches based on this simple calculation of the required ice surface slope.

\subsubsection{Water Sources for Subglacial Flow: Supraglacial Melting}
\label{sec:supr}
A lower limit on the duration of a climate event required to generate sufficient meltwater to form a multiple breach tadpole crater can be obtained from the volume of water required to move the total sediment volume of the breach channels (Table~\ref{tab:breach}). These volumes are equivalent to a water depth of between 0.06~m and 2.76~m per unit crater area, which is equivalent to between 6 and 252 days per breach of meltwater generation at the crater surface at a rate of 4~m~yr$^{-1}$~m$^{-2}$ \citep{kite2013seasonal}, with a mean of 60 melting days to erode each breach if meltwater is sourced solely from the crater area.  These values are comparable to the number of total melting days required in the surface melt pond hypothesis (Section~\ref{sec:ovsp}), and also need not be continuous as there is no requirement for the breaches to erode in a single subglacial event. 

Other potential heat sources for melting include re-entry of ejecta following an impact - i.e., radiative heating of the surface by the atmosphere, where the atmosphere has been heated by re-entry of ejecta \citep{goldin2009self,toon1997environmental}, or the climate effects of very large impacts (impactors $\geq100$~km diameter) \citep{turbet2020environmental,segura2012impact}. For heating by ejecta re-entry, around 70~minutes of heating at 5~kW~m$^{-2}$ would generate the equivalent of 1 year's worth of melt, however, models of the thermal pulse following the Chicxulub Impact on Earth only achieve heating rates $\geq5$~kW~m$^{-2}$ for a few minutes \citep{goldin2009self}. Therefore a single impact would not be able to generate the large melt ponds required to erode observed breaches in a single event (Fig.~\ref{fig:nevents}d), and tens to hundreds of Chicxulub-equivalent impacts (impactors tens of km in diameter) would be needed to explain breaches eroded by smaller ponds. For very large impacts, recent climate models suggest that short-lived high-intensity precipitation and high temperatures can last several years \citep{turbet2020environmental}. However, the intense, global precipitation caused by an impact on this scale would likely lead to intense fluvial weathering and crater degradation \citep{palumbo2018impact}, which is inconsistent with the small size and preserved rims of tadpole craters (Section~\ref{sec:morph}). Regardless of the melting heat source, however, if subglacial flow is responsible for the multiple exit breach craters then an ice sheet (not just crater-filling ice pods) would be required.

\subsubsection{Water Sources for Subglacial Flow: Magmatic Intrusions}
\label{sec:int}

\begin{figure}[t!]
\centering
\includegraphics[width=0.55\textwidth]{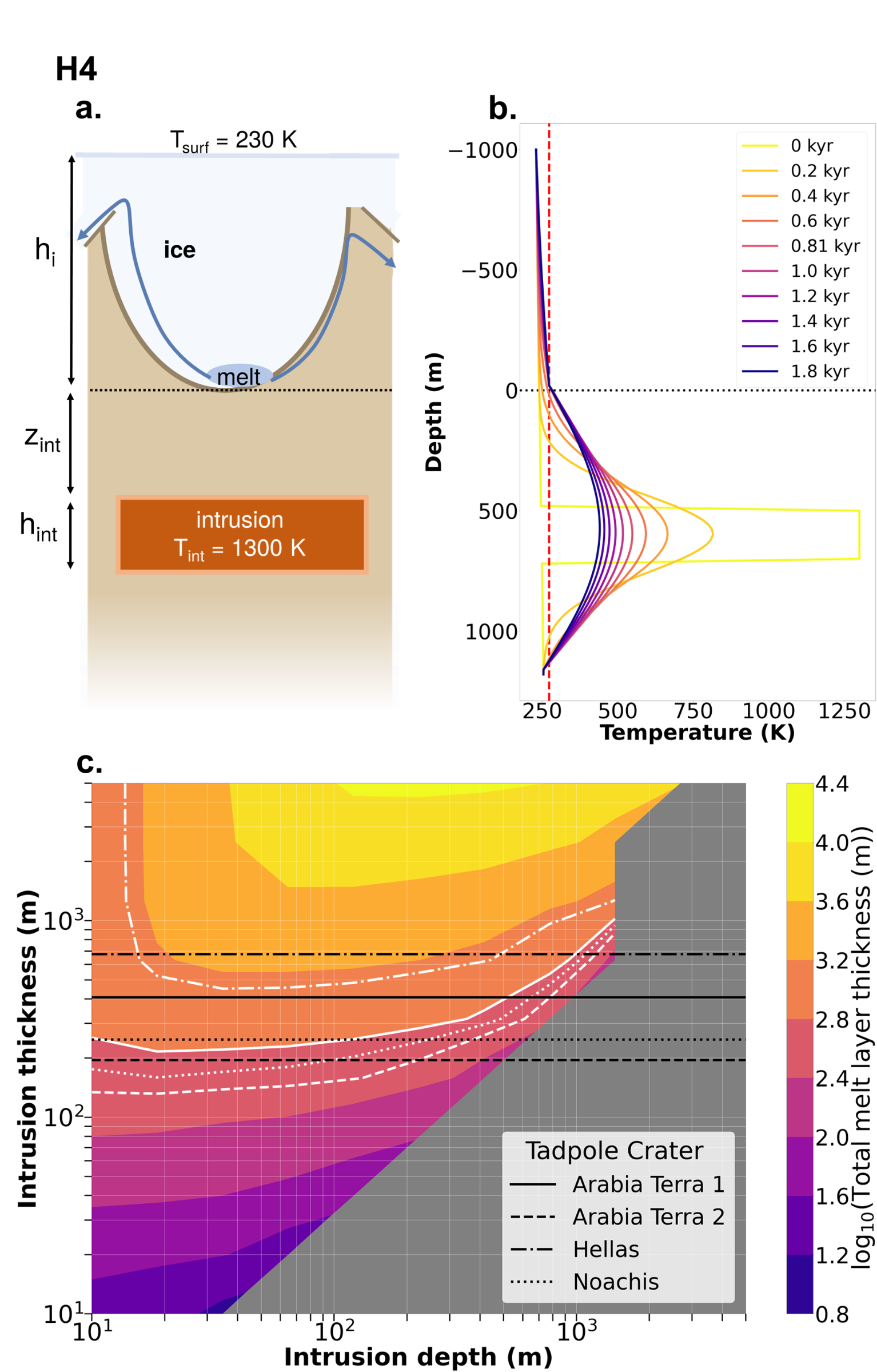}
\caption{\textbf{a.} Schematic illustration of 1D intrusion model. $h_{i}$ is ice thickness, $z_{int}$ is intrusion depth below the crater floor, $h_{int}$ is the vertical thickness of the intrusion (assumed to have same diameter as the crater). Dashed line indicates base of ice at $z=0$. \textbf{b.} Time evolution of vertical temperature profile (1D profile through central point of crater) for a 200~m thick intrusion with an initial temperature $T_{int}$~=~1300~K emplaced 500~m below the crater floor for a crater overlain by 1~km of ice. Vertical red dashed line indicates indicates $T=273$~K. Dashed line indicates base of ice at $z=0$. \textbf{c.} total thickness of water melted per unit area at the base of a crater for intrusions of different thicknesses at a range of depths beneath the crater floor (Table~\ref{tab:param}). Black lines indicate maximum intrusion depth calculated from difference between predicted and measured crater floor depths (Table~\ref{tab:mebtps}). White lines indicate predicted initial crater depth, i.e. where each crater would be filled to the brim with melt. Key shows line styles corresponding to different multiple breach tadpole craters.}
\label{fig:int_mod}
\end{figure}

Some recent glacial features on Mars, such as eskers, require wet-based glaciation to occur, such that there is meltwater available at the base of the ice \citep{GALLAGHER2015eskers,butcher2023eskers}. For the expected climate regime and geothermal heat flux during the Amazonian, cold-based formation is expected for viscous flow features such as lineated valley fill and lobate debris aprons \citep{WOODLEY2022multiple,GALLAGHER2015eskers}. Previous studies have proposed magmatic intrusions as a source of enhanced geothermal heat flux to drive melting \citep{arnold2019modeled}. We use a forward Euler finite differences method in 1D to solve for the cooling of an intrusion of thickness $H_{int}$ at depth $d_{int}$ in the crust beneath a surface ice layer of thickness $H_{ice}$ (Fig.~\ref{fig:int_mod}) to calculate the conditions under which meltwater can be generated at the base of a regional ice sheet through magmatic heat flux through the base of a tadpole crater. 

The initial temperature profile with depth is set by the geothermal gradient through an ice layer with thickness $h_{ice}$, surface temperature $T_{surf}$, and the instantaneous emplacement of an intrusion with thickness $h_{int}$ at temperature $T_{int}$ at a depth of $z_{int}$ (Table~\ref{tab:param}, Fig.~\ref{fig:int_mod}). 

Temperature then evolves diffusively in the crust and ice layer:
\begin{equation}
    \frac{\partial T}{\partial z} = \frac{1}{\rho_{z}c_{p,z}}\frac{\partial}{\partial z} \left(k_{z} \frac{\partial T}{\partial z} \right)
\end{equation}
where $\rho_{z}$, $c_{p,z}$, and $k_{z}$ are the density, specific heat capacity, and thermal conductivity of the material at depth $z$ (ice or rock). 

Once the temperature at the base of the ice reaches the melting point of water, we fix the temperature at the ice-rock interface to 273~K (analogous to if a water layer were forming). We then calculate the heat flux across this boundary ($q_{i-r}$) at each timestep, and assume that all heat provided by the cooling intrusion at depth minus the conductive losses to the overlying ice layer is used to melt additional ice. We calculate the melting rate ($\dot{z}_{m}$) at each timestep: 

\begin{equation}
    \dot{z}_{m} = \frac{q_{i-r}}{L_{i}\rho_{w}}
\end{equation}

We then calculate the maximum column thickness of meltwater produced by finding the cumulative sum of melt produced at each timestep (Fig.~\ref{fig:int_mod}c). This approach implicitly assumes that the meltwater is removed instantaneously from the base of the ice, and artificially holds ice thickness constant. Therefore, once the base of the ice reaches 273~K, the conductive heat losses are fixed by the ice thickness and surface temperature, which set the thermal gradient in the ice. In reality, however, the progressive thinning of the ice layer would steepen the thermal gradient in the ice and therefore reduce the heat flux across the rock-ice interface, leading to less melting.

To constrain the likelihood of magmatic intrusions as a melt-generating mechanism, we assume that all deformation related to intrusion emplacement is accommodated through fracturing and decoupling of the rock above the intrusion from the surrounding crust, leading to crater floor uplift \citep{walwer2021magma}. In this scenario, the intrusion diameter is limited to $D_{c}$ so that uplift due to intrusion inflation only occurs within the crater, and the total intrusion thickness is the difference between the predicted fresh crater depth and its current observed depth (Table~\ref{tab:mebtps}, \citealt{walwer2021magma}). To maximize the intrusion-driven melting, we neglect the cooling at the edges of the intrusion. Therefore our 1-D calculations best apply to the heating at the center of each crater, furthest from any additional conductive cooling at the intrusion's horizontal boundaries. This is conservative in the context of examining the feasibility of intrusion-driven basal melting because it maximizes the melt volume and melting duration for each intrusion.

\begin{figure}[t!]
\centering
\includegraphics[width=0.5\textwidth]{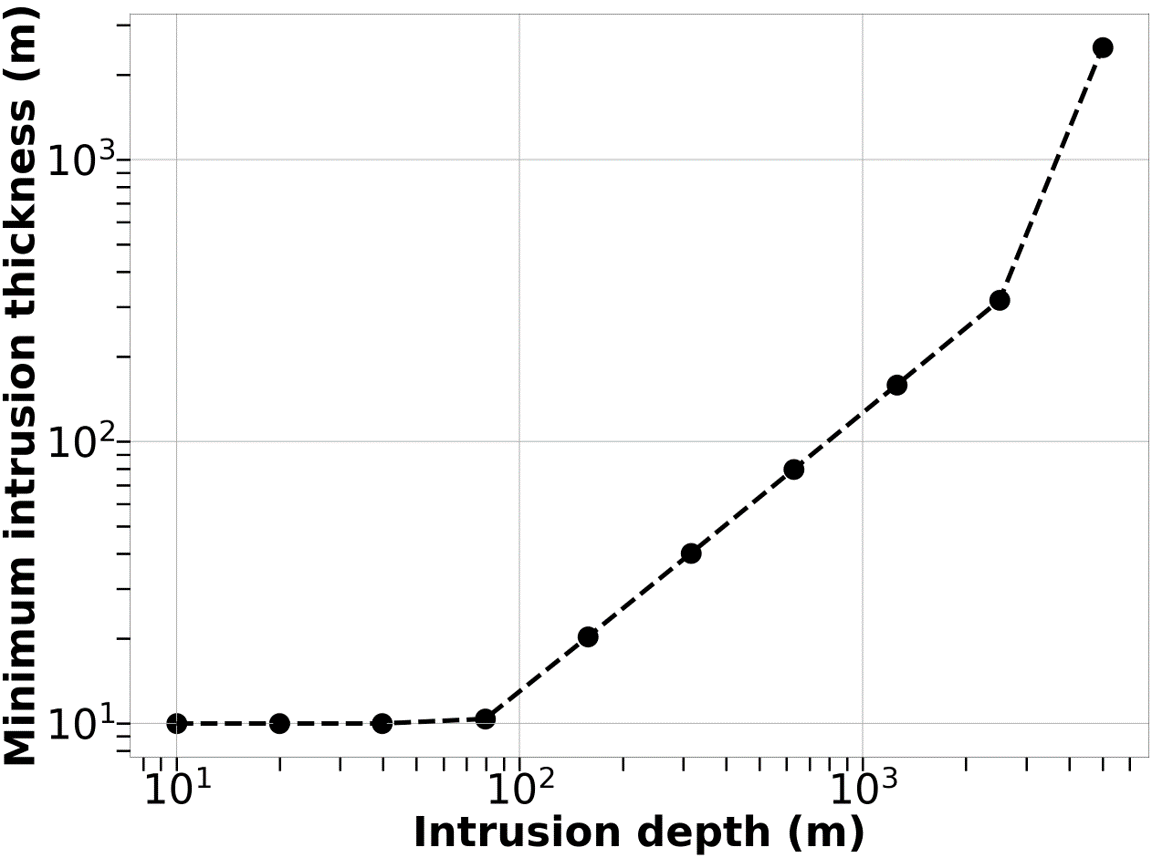}
\caption{Minimum required intrusion thickness as a function of intrusion depth to trigger basal melting of an ice sheet filling a multiple breach tadpole crater. Minimum thickness is found by interpolating our melting model results to find the intrusion thickness at which 0 melting occurs, i.e. the intrusion is too deep for heating around the margins to raise the surface temperature above 273~K.}
\label{fig:cutoff}
\end{figure}

We compare the total melt thickness to the depth of each multiple breach tadpole crater, and the crater area-averaged melt thickness required to erode the observed outlet breaches assuming a sediment:water ratio of 0.3 (Table~\ref{tab:breach}). We find that intrusions beneath craters can readily provide sufficient heat to melt enough water to erode all the breaches for every multiple breach crater, and even to fill each crater to the brim with melt water (Fig.~\ref{fig:int_mod}). Impact-induced decompression melting triggered by reduced pressure beneath craters when kilometers of material is removed has been proposed as a mechanism for generating volcanic activity on Martian and lunar crater floors \citep{edwards2014formation,elkinstanton2005giant,Schultz1976floor}, and there is evidence for impact-associated igneous activity in the Sudbury impact structure on Earth (e.g. \citealt{prevec2000evolution}). Infilled Martian craters with high thermal inertia flat floors consistent with volcanic fill by lavas or pyroclastic flows have an mean diameter $>40$~km \citep{edwards2014formation}. In a survey of 2800 craters $>1$~km in diameter with high thermal inertia floor fill, the smallest crater with features consistent with volcanic activity had a diameter of 8~km \citep{edwards2014formation}. Floor fractured craters, which can form through intrusions beneath crater floors \citep{walwer2021magma,Schultz1976floor,montigny2022origin}, similarly have a mean diameter of 43~km, and the size-frequency distribution of floor fractured craters in Xanthe Terra suggests that larger craters preferentially have fractured floors \citep{sato2010formation}. Therefore, if intrusions drive basal melting in tadpole craters, the preferential occurrence of tadpole breaches in small craters (Section~\ref{sec:diam}, Fig.~\ref{fig:diam}) may be inconsistent with intrusions beneath the craters themselves. However, larger, regional-scale intrusions could also trigger basal melting and may not lead to crater floor lava flows or fracture formation, therefore we cannot eliminate intrusions as a source for subglacial melt. Using our 1-D melting model, we can approximate the necessary intrusion depth to intrusion thickness relationship required to induce basal melting beneath a 1000~m thick ice sheet (Fig.~\ref{fig:cutoff}). Our results show that laccoliths emplaced beneath the four multiple breach tadpole craters can provide a sufficient boost to geothermal heat flux to drive basal melting at the base of a 1000~m thick ice sheet for shallow intrusions. If large intrusions are the melt source for tadpole craters, a global database of tadpoles may show clustering of tadpole craters in regions with evidence for widespread volcanic activity.

If intrusions are the primary mechanism for generating liquid water in tadpole craters, this would require widespread intrusions at depths 0.5-5~km beneath both the northern (Fig.~\ref{fig:map}) and southern mid-latitudes \citep{wilson_cold-wet_2016,warren2021overspilling}, at the same time as ice sheets were present. However, it is conceivable that some tadpole craters were fed by basal intrusion-triggered melting while the majority depended on climate-driving supraglacial melting (Section~\ref{sec:supr}). Mapping volcanic features in these regions and comparing their spatial distributions to tadpole craters may provide further insight into the likelihood of intrusion-driven basal melting.

\textbf{\section{Terrestrial analogs}}

\label{sec:analogs}
Here, we examine three possible settings for multiple breach tadpole crater terrestrial analogs:
\begin{itemize}
    \item Bedrock impact structures buried by ice
    \item Subaerial lake overspill from multiple breaches
    \item Subglacial lake overspill
\end{itemize}
However, it remains possible that the actual tadpole crater breach-forming mechanism has not yet been fingerprinted in a terrestrial environment. 

\subsection{Impact Craters Beneath Ice}

Formerly subglacial bedrock impact craters, such as the 24~km diameter Haughton impact structure \citep{osinski2005intra}, have been identified in the Canadian shield. Additionally, radar studies have found craters that are still beneath ice sheets (e.g. \citealt{kjaer2018large,macgregor2019possible}). Most of these impact structures have been extensively eroded and feature inlet breaches, making them poor analogs for tadpole craters. For example, although Hiawatha Crater, Greenland, has an outlet to the northwest, the rim is cut by two subglacial inlet channels in the south \citep{kjaer2018large}.

\begin{figure}[t!]
\centering
\includegraphics[width=\linewidth]{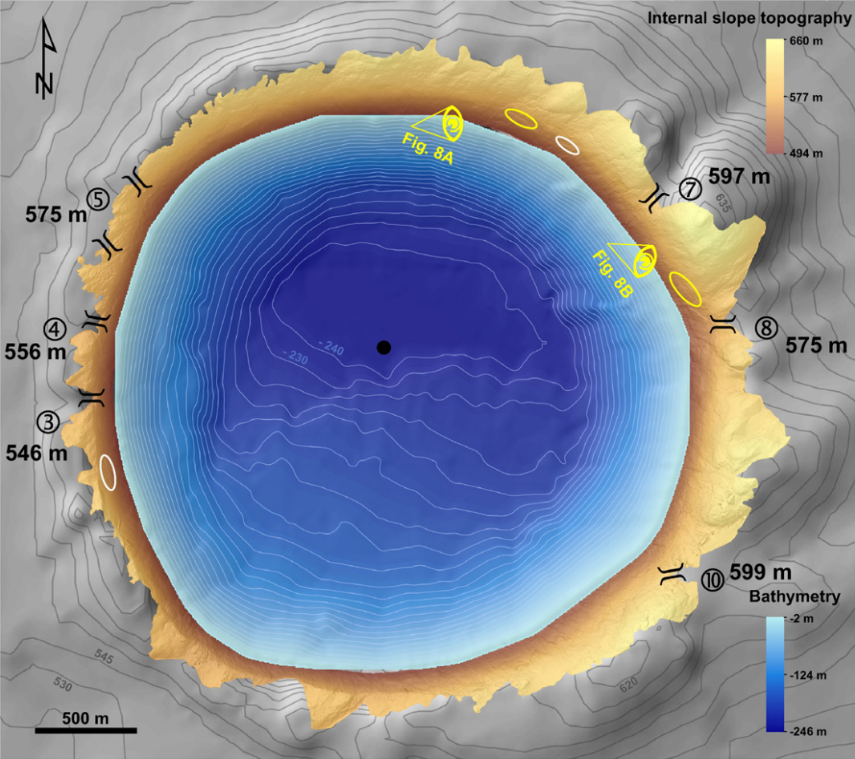}
\caption{Modified from Fig.~7 in \cite{desiage2015deglacial}. Digital elevation model (DEM) of Pingualuit Crater. Black brackets indicate locations of breaches with elevations in meters above sea level. White and yellow ellipses highlight paleoshorelines corresponding to 575~m (the elevation of breach 3). All elevations have an error of $\pm$1~m. }
\label{fig:ping}
\end{figure}

One exception and a strong candidate for a multiple breach tadpole crater terrestrial analog is Pingualuit Crater in Quebec, Canada (Fig.~\ref{fig:ping}). Pingualuit (also referred to as Ungava Crater, Chubb Crater, and New-Quebec Crater in the literature) is a 3.4~km diameter crater dating to 1.4~Ma \citep{bouchard1989histoire_age}. Despite extensive glacial erosion \citep{bouchard1989histoire_glaciation}, the topographic rim remains intact around the full circumference of the crater, extending more than 100~m above the surrounding terrain. The rim is incised by 10 outlet breaches at elevations between 553$\pm$5~m and 608$\pm$5~m above sea level \citep{desiage2015deglacial}. 

As the ice sheet retreated from the NE to SW \citep{gray1985dynamics}, \cite{desiage2015deglacial} propose that the highest elevation breaches on the eastern shore of the crater overspilled first. Then, as the ice retreated, lower points on the crater rim were exposed approximately anticlockwise where later breaches occurred. Evidence for subaerial breach drainage includes one breach coinciding with a paleoshoreline within the lake \citep{desiage2015deglacial}. However, several esker segments close to the crater itself and parallel to downstream portions of the channel networks downstream of rim breaches suggest that subglacial drainage occurred in the vicinity of Pingualuit Crater during the last glaciation \citep{bouchard1989histoire_age,desiage2015deglacial}. \cite{desiage2015deglacial} suggest that the drainage of outlet 10 (Fig.~\ref{fig:ping}) may have been subglacial, therefore not all breach events at Pingualuit or in tadpole craters need to have occurred as a result of the same climate and ice configuration. 

This proposed mechanism for the overspilling of Pingualuit Crater, and the similarities between Pingualuit Crater's rim morphology and that of tadpole craters, provides support for the feasibility of our proposed ``rapidly retreating glacier'' hypothesis (Section~\ref{sec:retreat}) for multiple breach tadpole crater formation. 

\subsection{Subaerial Lakes}
Small terrestrial subaerial lakes with multiple outlet breaches are rare. One example is Lake Colonia in Patagonia, Chile \citep{Dussaillant2010}. Lake Colonia is fed by meltwater generated by the rapid ablation of Colonia Glacier, and has a primary spillway active during a series of 5 j{\"o}kulhlaup events in 2008 and 2009, and an earlier, now inactive spillway with a floor 10~m above the floor of the current spillway. If both spillways were active simultaneously, \cite{Dussaillant2010} calculate that the total flux of water out of the lake would have been 16,000~m$^{3}$~s$^{-1}$, which is of a similar order of magnitude to some of our calculated water fluxes (Table~\ref{tab:qin}). This suggests that the crater inflow rate generated by meltwater routing from a terrestrial glacier is sufficient to activate multiple breaches in a single overspill event, which is likely due to the larger melting rates possible on Earth where the annual average temperature is closer to 273~K, even in high altitude environments. 

On a scale far larger than tadpole craters, multiple spillways formed during the advance and retreat of the Laurentide Ice Sheet, which fed Glacial Lake Agassiz. As the Laurentide Ice Sheet retreated, it exposed lower topography, causing Lake Agassiz to abandon higher elevation breaches to the south, and routing megafloods through lower spillways to the north and east \citep{FISHER2020megaflooding}. However, unlike tadpole craters, Lake Agassiz formed part of a network of connected proglacial lakes flooding into one another, as recorded by fan deposition into Lake Agassiz from lakes to the west \citep{Kehew2009104}. At its maximum extent, Lake Agassiz had a volume of 38,700~km$^{3}$ \citep{Mann1999}, or 90 times the volume of the largest tadpole crater in the 24-52$^{\circ}$~S latitude band (Fig.~\ref{fig:diam}a). Due to the vast difference in scale and the interconnected nature of large terrestrial proglacial lakes and the larger role of lithospheric flexure beneath ice sheets for proglacial flow routing on Earth, these are likely a poor analog for tadpole craters.

\subsection{Subglacial Lakes}

Since the discovery of the first subglacial lake in Antarctica \citep{robin1969interpretation}, almost 800 subglacial lakes have been identified on Earth \citep{livingstone2022subglacial}. Observations suggest that subglacial lakes in Antarctica and Greenland behave differently. In Antarctica, large, stable lakes such as Lake Vostok tend to occur in large depressions in basal topography and can persist for thousands of years \citep{fricker2016decade}. However, there are also smaller, active lakes that drain and refill on timescales of months to decades. These are associated with fast-flowing, wet-based ice streams and are associated with depressions in the ice surface rather than the basal topography \citep{fricker2016decade}. The MacAyeal ice stream, for example, has ten known subglacial lakes \citep{smith2009inventory}. In Greenland, around 20\% of subglacial lakes occur in regions where cold-based ice is expected and are also typically associated with slow-moving ice \citep{bowling2019distribution}. Lakes in cold-based regions of Greenland ice may be better analogies for tadpole craters given the preservation of topographically elevated crater rims in observed tadpole crater areas, with limited evidence for wet-based glacial features.

On Earth, satellite observations of the ice surface above known subglacial lakes can also reveal subglacial drainage and recharge events as the ice surface sinks and inflates respectively \citep{fricker2016decade}, including recharge of a subglacial lake in northeast Greenland by routing of surface meltwater to the base of the ice by crevasses around the lake margins \citep{willis2015recharge}. This may be a good terrestrial analog for providing supraglacial meltwater to subglacial tadpole craters for subsequent overspill. 

Additionally, evidence for catastrophic drainage of subglacial lakes leading to bedrock incision is visible in the Labyrinth in the Antarctic Dry Valleys. The Labyrinth is a network of multiple generations of cross-cutting channels extending 10s of km from the margin of the modern Wright Upper Glacier. Superposed ash fall deposits date date the channels to 12-14~Ma, with each incising 10-50~m into underlying bedrock \citep{lewis2006age}. This is similar to the depths of tadpole breaches (Table~\ref{tab:breach}). The Labyrinth is hypothesized to have drained one or more subglacial lakes at a time when wet-based glaciation was more widespread beneath the Antarctic ice sheet \citep{marchant1993miocene}. Unfortunately, without knowing the configuration of the subglacial lake system(s), it is difficult to assess the similarity of the Labyrinth to multiple breach tadpole craters, as an interconnected network of subglacial lakes is not consistent with the absence of inlets in tadpole craters.

\section{Conclusions}

\begin{figure}[t!]
\centering
\includegraphics[width=0.6\textwidth]{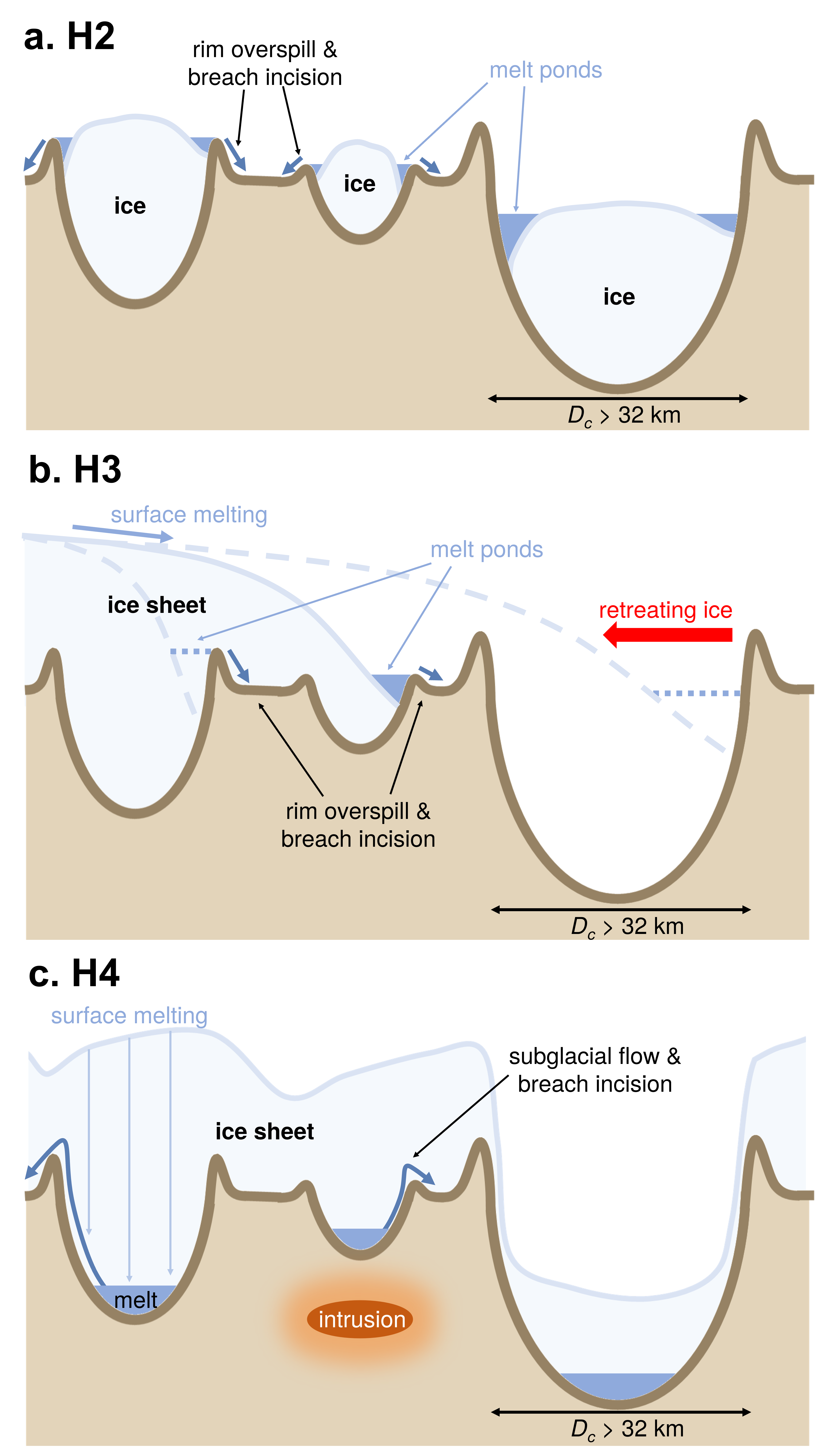}
\caption{Schematic illustration of the three remaining multiple breach tadpole crater forming hypotheses (Fig.~\ref{fig:mech}) that are consistent with our morphological and modeling constraints on tadpole breach forming climate. }
\label{fig:summ_mech}
\end{figure}

Hundreds of tadpole craters exist on Mars \citep{wilson_cold-wet_2016,howard2022distribution} (Fig.~\ref{fig:map}). In the 24-52$^{\circ}$~S latitude band, tadpole crater breaches preferentially form in smaller diameter craters (Fig.~\ref{fig:diam}a). This demonstrates there was not enough water or ice available during tadpole breach formation to fill larger craters to the rim, suggesting a maximum ice thickness of approximately 1 to 2~km. When smoothed topography is subtracted from MOLA topography to generate residual topography, tadpole craters show a preference for occurring on relative highs at all lengthscales, with a steeper slope at shorter length scales (Fig.~\ref{fig:elev}b-f). This favors a high-elevation dependent formation process for tadpole craters, for example accumulation of snow and ice at high altitudes, or orographic precipitation. 

Forming multiple breaches exiting a single crater at different elevations without an inlet is an additional challenge to tadpole formation models. We consider four hypotheses for tadpole breach formation that can explain these features (Fig.~\ref{fig:mech}). 

Among subaerial hypotheses (H1-H3, Fig.\ref{fig:mech}), we find that crater-interior precipitation or melting is too slow to drive simultaneous overspill of multiple breaches, and that appealing to increased melting area on a regional ice sheet routed into the crater requires unrealistically large melting areas. Additionally, we disfavor groundwater as a tadpole crater overspilling water source as this would require unrealistically high aquifer permeabilities, initial aquifer pressures, and aquifer thicknesses for three of the four multiple breach tadpole craters considered in this study. Overall, simultaneous overspill of multiple breaches (Fig.~\ref{fig:mech}a) is unlikely. As a result, the only remaining hypotheses consistent with forming all four multiple breach tadpole craters are: 
\begin{itemize}

    \item \textbf{H2:}~Draining of meltwater ponds smaller than the crater diameter formed on the surface of a crater-filling ice body and overspilling the crater rim (Section~\ref{fig:nevents}a, Figs.~\ref{fig:mech}b~\&~\ref{fig:summ_mech}a) 
    \item \textbf{H3:}~Sequential breach overspill enabled by retreat of an ice sheet progressively exposing lower rim topography and filling proglacial crater lakes with meltwater (Section~\ref{sec:retreat}, Figs.~\ref{fig:mech}c~\&~\ref{fig:summ_mech})
    \item \textbf{H4:}~Subglacial flow with melt supplied by either supraglacial melting or basal melting triggered by a cooling intrusion (Section~\ref{sec:subg}, Figs.~\ref{fig:mech}c~\&~\ref{fig:summ_mech}c)
\end{itemize}

In the subglacial hypothesis (H4), we slightly disfavor intrusion-driven basal melting because crater floor volcanic features are typically associated with craters tens of kilometers in diameter \citep{edwards2014formation,sato2010formation}, whereas tadpole craters are typically $<$10~km in diameter (Fig.~\ref{fig:diam}). 

Finally, we suggest that Pingualuit Crater in Quebec, Canada \citep{bouchard1989histoire_glaciation} is the best currently known terrestrial analog for multiple breach tadpole craters. Pingualuit has a topographically elevated rim, ten exit breaches spanning tens of meters in elevation (Fig.~\ref{fig:ping}), and lacks inlet breaches. The proposed formation mechanism for Pingualuit's series of exit breaches is the retreat of an ice sheet exposing progressively lower rim topography \citep{desiage2015deglacial}. This demonstrates that our proposed ice sheet retreat hypothesis for multiple breach tadpole crater formation (Fig.~\ref{fig:mech}c \& \ref{fig:retreat}) does occur. 

We conclude that the formation of multiple breach tadpole craters in the Martian midlatitudes requires the accumulation of at least hundreds of meters of ice, either as crater-filling ice bodies, or regional ice sheets (Fig.~\ref{fig:summ_mech}). Modeling studies have demonstrated that midlatitude ice accumulation can be achieved under Amazonian conditions if Mars' obliquity is higher \citep{forget2006formation,madeleine2009amazonian,fastook_glaciation_2015}. Our results also suggest that tens to hundreds of days of temperatures exceeding 273~K in the mid-latitudes are likely required to generate the liquid water for multiple breach tadpole crater overspill. This predicts that incised channels should also be found in the midlatitudes outside of the tadpole craters, and some have been reported (e.g., \citet{wilson_cold-wet_2016,adeli2016}). If only small water volumes are required to enable meltwater pond accumulation or subglacial erosion, these melting days are consistent with annual peak temperatures at each multiple breach tadpole crater location on modern Mars \citep{forget1999improved,millour2018exploring} but would require an increase in greenhouse warming to overcome the cooling effects of ice albedo, high thermal inertia, and evaporative cooling \citep{ingersoll1970mars}. If melt needs to accumulate over continuous days, 25-50~K of warming is required.


\begin{acknowledgments}

\textit{Arabia Terra 1} DEM was made by Daniel Robinson of University of Arizona. \textit{Arabia Terra 2} DEM was made by Riley McClurkin of University of Arizona. We would like to thank Frances Butcher and Anna Grau-Galofre for helpful conversations about terrestrial analogs and both terrestrial and Martian subglacial systems, Marc Hesse for productive discussions that guided our groundwater modeling approach, and the HiRISE HiWish program. Grants: NASA NNH19ZDA001N-FINESST (F.I. A.O. Warren), NASA MDAP 80NSSC18K1476 (P.I. S.A. Wilson)

\end{acknowledgments}

%

\vspace{5mm}
\facilities{University of Chicago Research Computing Center}




\bibliography{main}{}

\begin{thebibliography}{}
\expandafter\ifx\csname natexlab\endcsname\relax\def\natexlab#1{#1}\fi
\providecommand{\url}[1]{\href{#1}{#1}}
\providecommand{\dodoi}[1]{doi:~\href{http://doi.org/#1}{\nolinkurl{#1}}}
\providecommand{\doeprint}[1]{\href{http://ascl.net/#1}{\nolinkurl{http://ascl.net/#1}}}
\providecommand{\doarXiv}[1]{\href{https://arxiv.org/abs/#1}{\nolinkurl{https://arxiv.org/abs/#1}}}

\bibitem[{{Adeli} {et~al.}(2016){Adeli}, {Hauber}, {Kleinhans}, {Le Deit},
  {Platz}, {Fawdon}, \& {Jaumann}}]{adeli2016}
{Adeli}, S., {Hauber}, E., {Kleinhans}, M., {et~al.} 2016, \icarus, 277, 286,
  \dodoi{10.1016/j.icarus.2016.05.020}

\bibitem[{Andrews-Hanna \& Phillips(2007)}]{andrews2007hydrological}
Andrews-Hanna, J.~C., \& Phillips, R.~J. 2007, Journal of Geophysical Research:
  Planets, 112

\bibitem[{Arnold {et~al.}(2019)Arnold, Conway, Butcher, \&
  Balme}]{arnold2019modeled}
Arnold, N.~S., Conway, S.~J., Butcher, F. E.~G., \& Balme, M.~R. 2019, Journal
  of Geophysical Research: Planets, 124, 2101,
  \dodoi{https://doi.org/10.1029/2019JE006061}

\bibitem[{Arthur {et~al.}(2022)Arthur, Stokes, Jamieson, Rachel~Carr, Leeson,
  \& Verjans}]{Arthur2022large}
Arthur, J.~F., Stokes, C.~R., Jamieson, S. S.~R., {et~al.} 2022, Nature
  Communications, 13, 1711, \dodoi{10.1038/s41467-022-29385-3}

\bibitem[{Balme {et~al.}(2013)Balme, Gallagher, \&
  Hauber}]{balme2013morphological}
Balme, M., Gallagher, C., \& Hauber, E. 2013, Progress in Physical Geography:
  {{E}arth} and Environment, 37, 289, \dodoi{10.1177/0309133313477123}

\bibitem[{Bamber {et~al.}(2022{\natexlab{a}})Bamber, Goudge, Fassett, \&
  Osinski}]{bamber2022constraining}
Bamber, E.~R., Goudge, T.~A., Fassett, C.~I., \& Osinski, G.~R.
  2022{\natexlab{a}}, Icarus, 378, 114945

\bibitem[{Bamber {et~al.}(2022{\natexlab{b}})Bamber, Goudge, Fassett, Osinski,
  \& Stucky~de Quay}]{bamber2022paleolake}
Bamber, E.~R., Goudge, T.~A., Fassett, C.~I., Osinski, G.~R., \& Stucky~de
  Quay, G. 2022{\natexlab{b}}, Geophysical Research Letters, 49, e2022GL101097,
  \dodoi{https://doi.org/10.1029/2022GL101097}

\bibitem[{Batalha {et~al.}(2015)Batalha, Domagal-Goldman, Ramirez, \&
  Kasting}]{batalha_testing_2015}
Batalha, N., Domagal-Goldman, S.~D., Ramirez, R., \& Kasting, J.~F. 2015,
  Icarus, 258, 337, \dodoi{10.1016/j.icarus.2015.06.016}

\bibitem[{Bintanja \& Reijmer(2001)}]{bintanja_reijmer_2001}
Bintanja, R., \& Reijmer, C.~H. 2001, Journal of Glaciology, 47, 37–50,
  \dodoi{10.3189/172756501781832557}

\bibitem[{Bj{\"o}rnsson(2003)}]{bjornsson2003subglacial}
Bj{\"o}rnsson, H. 2003, Global and Planetary Change, 35, 255

\bibitem[{Bouchard {et~al.}(1989)Bouchard, {M}arsan, P{\'e}loquin, Fortin,
  Saarnisto, Shilts, David, \& Flisz{\'a}r}]{bouchard1989histoire_glaciation}
Bouchard, M., {M}arsan, B., P{\'e}loquin, S., {et~al.} 1989, in L’histoire
  Naturelle du Crat{\`e}re du Nouveau-Qu{\'e}bec, Collection environnement et
  g{\'e}ologie 7, ed. S.~P{\'e}loquin \& M.~A. Bouchard (Universit{\'e} de
  Montr{\'e}al), 101--138

\bibitem[{Bouchard(1989)}]{bouchard1989histoire_age}
Bouchard, M.~A. 1989, in L’histoire Naturelle du Crat{\`e}re du
  Nouveau-Qu{\'e}bec, Collection environnement et g{\'e}ologie 7, ed.
  S.~P{\'e}loquin \& M.~A. Bouchard (Universit{\'e} de Montr{\'e}al), 139–163

\bibitem[{Bowling {et~al.}(2019)Bowling, Livingstone, Sole, \&
  Chu}]{bowling2019distribution}
Bowling, J., Livingstone, S., Sole, A., \& Chu, W. 2019, Nature Communications,
  10, 2810

\bibitem[{Brownlie(1983)}]{brownlie1983flow}
Brownlie, W.~R. 1983, Journal of Hydraulic Engineering, 109, 959

\bibitem[{Butcher {et~al.}(2023)Butcher, Arnold, Balme, Conway, Clark,
  Gallagher, Hagermann, Lewis, Rutledge, Storrar, {et~al.}}]{butcher2023eskers}
Butcher, F.~E., Arnold, N.~S., Balme, M.~R., {et~al.} 2023, Annals of
  Glaciology, 1

\bibitem[{Butcher {et~al.}(2017)Butcher, Balme, Gallagher, Arnold, Conway,
  Hagermann, \& Lewis}]{butcher2017recent}
Butcher, F. E.~G., Balme, M.~R., Gallagher, C., {et~al.} 2017, Journal of
  Geophysical Research: Planets, 122, 2445,
  \dodoi{https://doi.org/10.1002/2017JE005434}

\bibitem[{Carr(1979)}]{carr1979formation}
Carr, M.~H. 1979, Journal of Geophysical Research: Solid {E}arth, 84, 2995

\bibitem[{Chien \& Kuo(2011)}]{chien2011extreme}
Chien, F.-C., \& Kuo, H.-C. 2011, Journal of Geophysical Research: Atmospheres,
  116, \dodoi{https://doi.org/10.1029/2010JD015092}

\bibitem[{Clifford {et~al.}(2010)Clifford, Lasue, Heggy, Boisson, McGovern, \&
  Max}]{clifford2010depth}
Clifford, S.~M., Lasue, J., Heggy, E., {et~al.} 2010, Journal of Geophysical
  Research: Planets, 115

\bibitem[{Clifford \& Parker(2001)}]{clifford_evolution_2001}
Clifford, S.~M., \& Parker, T.~J. 2001, Icarus, 154, 40,
  \dodoi{10.1006/icar.2001.6671}

\bibitem[{Clow(1987)}]{CLOW1987generation}
Clow, G.~D. 1987, Icarus, 72, 95,
  \dodoi{https://doi.org/10.1016/0019-1035(87)90123-0}

\bibitem[{Conway {et~al.}(2018)Conway, Butcher, de~Haas, Deijns, Grindrod, \&
  Davis}]{conway2018glacial}
Conway, S.~J., Butcher, F.~E., de~Haas, T., {et~al.} 2018, Geomorphology, 318,
  26

\bibitem[{Conway {et~al.}(2012)Conway, Hovius, Barnie, Besserer,
  Le~Mou{\'e}lic, Orosei, \& Read}]{conway2012climate}
Conway, S.~J., Hovius, N., Barnie, T., {et~al.} 2012, Icarus, 220, 174

\bibitem[{Cook {et~al.}(2018)Cook, Andermann, Gimbert, Adhikari, \&
  Hovius}]{cook2018glacial}
Cook, K.~L., Andermann, C., Gimbert, F., Adhikari, B.~R., \& Hovius, N. 2018,
  Science, 362, 53

\bibitem[{Costa \& Schuster(1988)}]{costa1988formation}
Costa, J.~E., \& Schuster, R.~L. 1988, GSA Bulletin, 100, 1054,
  \dodoi{10.1130/0016-7606(1988)100<1054:TFAFON>2.3.CO;2}

\bibitem[{Craddock \& Howard(2002)}]{craddock2002case}
Craddock, R.~A., \& Howard, A.~D. 2002, Journal of Geophysical Research:
  Planets, 107, 21

\bibitem[{Craddock {et~al.}(1997)Craddock, Maxwell, \&
  Howard}]{craddock1997crater}
Craddock, R.~A., Maxwell, T.~A., \& Howard, A.~D. 1997, Journal of Geophysical
  Research: Planets, 102, 13321

\bibitem[{{Das} {et~al.}(2008){Das}, {Joughin}, {Behn}, {Howat}, {King},
  {Lizarralde}, \& {Bhatia}}]{das2008}
{Das}, S.~B., {Joughin}, I., {Behn}, M.~D., {et~al.} 2008, Science, 320, 778,
  \dodoi{10.1126/science.1153360}

\bibitem[{Desiage {et~al.}(2015)Desiage, Lajeunesse, St-Onge, Normandeau,
  Ledoux, Guyard, \& Pienitz}]{desiage2015deglacial}
Desiage, P.-A., Lajeunesse, P., St-Onge, G., {et~al.} 2015, Geomorphology, 248,
  327

\bibitem[{Dickson {et~al.}(2018)Dickson, Kerber, Fassett, \&
  Ehlmann}]{dickson2018global}
Dickson, J., Kerber, L., Fassett, C., \& Ehlmann, B. 2018, in 49th Lunar and
  Planetary Science Conference, Vol.~49, 2083

\bibitem[{Domenico \& Schwartz(1998)}]{domenico1998physical}
Domenico, P.~A., \& Schwartz, F.~W. 1998, Physical and chemical hydrogeology,
  Vol. 506 (Wiley New York)

\bibitem[{Dussaillant {et~al.}(2010)Dussaillant, Benito, Buytaert, Carling,
  Meier, \& Espinoza}]{Dussaillant2010}
Dussaillant, A., Benito, G., Buytaert, W., {et~al.} 2010, Natural Hazards, 54,
  469, \dodoi{10.1007/s11069-009-9479-8}

\bibitem[{Edwards {et~al.}(2014)Edwards, Bandfield, Christensen, \&
  Rogers}]{edwards2014formation}
Edwards, C., Bandfield, J., Christensen, P., \& Rogers, A. 2014, Icarus, 228,
  149, \dodoi{https://doi.org/10.1016/j.icarus.2013.10.005}

\bibitem[{Elkins-Tanton \& Hager(2005)}]{elkinstanton2005giant}
Elkins-Tanton, L.~T., \& Hager, B.~H. 2005, {E}arth and Planetary Science
  Letters, 239, 219, \dodoi{https://doi.org/10.1016/j.epsl.2005.07.029}

\bibitem[{Fassett {et~al.}(2010)Fassett, Dickson, Head, Levy, \&
  Marchant}]{fassett2010supraglacial}
Fassett, C.~I., Dickson, J.~L., Head, J.~W., Levy, J.~S., \& Marchant, D.~R.
  2010, Icarus, 208, 86

\bibitem[{Fassett \& Head(2008)}]{fassett2008valley}
Fassett, C.~I., \& Head, J.~W. 2008, Icarus, 198, 37

\bibitem[{Fassett {et~al.}(2014)Fassett, Levy, Dickson, \&
  Head}]{fassett2014extended}
Fassett, C.~I., Levy, J.~S., Dickson, J.~L., \& Head, J.~W. 2014, Geology, 42,
  763, \dodoi{10.1130/G35798.1}

\bibitem[{Fastook \& Head(2013)}]{fastook2013amazonian}
Fastook, J., \& Head, J. 2013, in 44th Annual Lunar and Planetary Science
  Conference, 1256

\bibitem[{Fastook \& Head(2014)}]{FASTOOK201460}
Fastook, J.~L., \& Head, J.~W. 2014, Planetary and Space Science, 91, 60,
  \dodoi{https://doi.org/10.1016/j.pss.2013.12.002}

\bibitem[{Fastook \& Head(2015)}]{fastook_glaciation_2015}
---. 2015, Planetary and Space Science, 106, 82,
  \dodoi{10.1016/j.pss.2014.11.028}

\bibitem[{Fastook {et~al.}(2014)Fastook, Head, \&
  Marchant}]{FASTOOK2014formation}
Fastook, J.~L., Head, J.~W., \& Marchant, D.~R. 2014, Icarus, 228, 54,
  \dodoi{https://doi.org/10.1016/j.icarus.2013.09.025}

\bibitem[{Fisher(2020)}]{FISHER2020megaflooding}
Fisher, T.~G. 2020, {E}arth-Science Reviews, 201, 102974,
  \dodoi{https://doi.org/10.1016/j.earscirev.2019.102974}

\bibitem[{Forget {et~al.}(2006)Forget, Haberle, Montmessin, Levrard, \&
  Head}]{forget2006formation}
Forget, F., Haberle, R., Montmessin, F., Levrard, B., \& Head, J. 2006,
  Science, 311, 368

\bibitem[{{Forget} {et~al.}(2006){Forget}, {Haberle}, {Montmessin}, {Levrard},
  \& {Head}}]{forget2006}
{Forget}, F., {Haberle}, R.~M., {Montmessin}, F., {Levrard}, B., \& {Head},
  J.~W. 2006, Science, 311, 368, \dodoi{10.1126/science.1120335}

\bibitem[{{Forget} {et~al.}(1999){Forget}, {Hourdin}, {Fournier}, {Hourdin},
  {Talagrand}, {Collins}, {Lewis}, {Read}, \& {Huot}}]{forget1999improved}
{Forget}, F., {Hourdin}, F., {Fournier}, R., {et~al.} 1999, Journal of
  Geophysical Research: Planets, 104, 24155, \dodoi{10.1029/1999JE001025}

\bibitem[{{Fricker} {et~al.}(2016){Fricker}, {Siegfried}, {Carter}, \&
  {Scambos}}]{fricker2016decade}
{Fricker}, H.~A., {Siegfried}, M.~R., {Carter}, S.~P., \& {Scambos}, T.~A.
  2016, Philosophical Transactions of the Royal Society of London Series A,
  374, 20140294, \dodoi{10.1098/rsta.2014.0294}

\bibitem[{Gallagher \& Balme(2015)}]{GALLAGHER2015eskers}
Gallagher, C., \& Balme, M. 2015, {E}arth and Planetary Science Letters, 431,
  96, \dodoi{https://doi.org/10.1016/j.epsl.2015.09.023}

\bibitem[{Garvin {et~al.}(2000)Garvin, Sakimoto, Frawley, \&
  Schnetzler}]{garvin2000north}
Garvin, J.~B., Sakimoto, S.~E., Frawley, J.~J., \& Schnetzler, C. 2000, Icarus,
  144, 329

\bibitem[{Goldin \& Melosh(2009)}]{goldin2009self}
Goldin, T.~J., \& Melosh, H.~J. 2009, Geology, 37, 1135,
  \dodoi{10.1130/G30433A.1}

\bibitem[{Golombek \& Bridges(2000)}]{golombek2000erosion}
Golombek, M., \& Bridges, N. 2000, Journal of Geophysical Research: Planets,
  105, 1841

\bibitem[{Golombek {et~al.}(2014)Golombek, Warner, Ganti, Lamb, Parker,
  Fergason, \& Sullivan}]{golombek2014small}
Golombek, M., Warner, N., Ganti, V., {et~al.} 2014, Journal of Geophysical
  Research: Planets, 119, 2522

\bibitem[{Goudge {et~al.}(2023)Goudge, Fassett, Coholich, \&
  Bamber}]{goudge2023assessing}
Goudge, T.~A., Fassett, C.~I., Coholich, M., \& Bamber, E.~R. 2023, Journal of
  Geophysical Research: Planets, 128, e2022JE007443,
  \dodoi{https://doi.org/10.1029/2022JE007443}

\bibitem[{Goudge {et~al.}(2016)Goudge, Fassett, Head, Mustard, \&
  Aureli}]{goudge2016insights}
Goudge, T.~A., Fassett, C.~I., Head, J.~W., Mustard, J.~F., \& Aureli, K.~L.
  2016, Geology, 44, 419

\bibitem[{Grant \& Wilson(2011)}]{grant2011late}
Grant, J.~A., \& Wilson, S.~A. 2011, Geophysical Research Letters, 38

\bibitem[{Grau~Galofre {et~al.}(2020)Grau~Galofre, Jellinek, \&
  Osinski}]{GrauGalofre2020valley}
Grau~Galofre, A., Jellinek, A.~M., \& Osinski, G.~R. 2020, Nature Geoscience,
  13, 663, \dodoi{10.1038/s41561-020-0618-x}

\bibitem[{Grau~Galofre {et~al.}(2018)Grau~Galofre, Jellinek, Osinski, Zanetti,
  \& Kukko}]{galofre2018subglacial}
Grau~Galofre, A., Jellinek, A.~M., Osinski, G.~R., Zanetti, M., \& Kukko, A.
  2018, The Cryosphere, 12, 1461, \dodoi{10.5194/tc-12-1461-2018}

\bibitem[{Gray \& Lauriol(1985)}]{gray1985dynamics}
Gray, J.~T., \& Lauriol, B. 1985, Arctic and Alpine Research, 17, 289

\bibitem[{Haberle {et~al.}(2017)Haberle, Clancy, Forget, Smith, \&
  Zurek}]{haberle2017atmosphere}
Haberle, R.~M., Clancy, R.~T., Forget, F., Smith, M.~D., \& Zurek, R.~W. 2017,
  The Atmosphere and Climate of {M}ars (Cambridge University Press), 497–525

\bibitem[{Hanna \& Phillips(2005)}]{hanna2005hydrological}
Hanna, J.~C., \& Phillips, R.~J. 2005, Journal of Geophysical Research:
  Planets, 110, \dodoi{https://doi.org/10.1029/2004JE002330}

\bibitem[{Harrison \& Grimm(2004)}]{harrison2004tharsis}
Harrison, K.~P., \& Grimm, R.~E. 2004, Geophysical Research Letters, 31,
  \dodoi{https://doi.org/10.1029/2004GL020502}

\bibitem[{Harrison \& Grimm(2009)}]{harrison2009regionally}
---. 2009, Journal of Geophysical Research: Planets, 114

\bibitem[{Hartmann(2005)}]{hartmann2005martian}
Hartmann, W.~K. 2005, Icarus, 174, 294

\bibitem[{Head {et~al.}(2005)Head, Neukum, Jaumann, Hiesinger, Hauber, Carr,
  Masson, Foing, Hoffmann, Kreslavsky, {et~al.}}]{head2005tropical}
Head, J., Neukum, G., Jaumann, R., {et~al.} 2005, Nature, 434, 346

\bibitem[{Head {et~al.}(2010)Head, Fassett, Kadish, Smith, Zuber, Neumann, \&
  Mazarico}]{head2010global}
Head, J.~W., Fassett, C.~I., Kadish, S.~J., {et~al.} 2010, science, 329, 1504

\bibitem[{Head {et~al.}(2003)Head, Mustard, Kreslavsky, Milliken, \&
  Marchant}]{Head2003}
Head, J.~W., Mustard, J.~F., Kreslavsky, M.~A., Milliken, R.~E., \& Marchant,
  D.~R. 2003, Nature, 426, 797, \dodoi{10.1038/Nature02114}

\bibitem[{Hibbard(2021)}]{hibbard2021surface}
Hibbard, S.~M. 2021, PhD thesis, The University of Western Ontario.
\newblock
  \url{https://www.proquest.com/dissertations-theses/surface-morphology-subsurface-ice-content/docview/2787201367/se-2}

\bibitem[{Hobley {et~al.}(2014)Hobley, Howard, \& Moore}]{hobley2014fresh}
Hobley, D.~E., Howard, A.~D., \& Moore, J.~M. 2014, Journal of Geophysical
  Research: Planets, 119, 128

\bibitem[{Holo {et~al.}(2021)Holo, Kite, Wilson, \& Morgan}]{holo2021timing}
Holo, S.~J., Kite, E.~S., Wilson, S.~A., \& Morgan, A.~M. 2021, The Planetary
  Science Journal, 2, 210

\bibitem[{Holt {et~al.}(2008)Holt, Safaeinili, Plaut, Head, Phillips, Seu,
  Kempf, Choudhary, Young, Putzig, {et~al.}}]{holt2008radar}
Holt, J.~W., Safaeinili, A., Plaut, J.~J., {et~al.} 2008, Science, 322, 1235

\bibitem[{{Howard} {et~al.}(2022){Howard}, {Wilson}, \&
  {Moore}}]{howard2022distribution}
{Howard}, A., {Wilson}, S., \& {Moore}, J. 2022, in European Planetary Science
  Congress, EPSC2022--180, \dodoi{10.5194/epsc2022-180}

\bibitem[{Hu {et~al.}(2015)Hu, Kass, Ehlmann, \& Yung}]{hu_tracing_2015-1}
Hu, R., Kass, D.~M., Ehlmann, B.~L., \& Yung, Y.~L. 2015, Nature
  Communications, 6, 10003, \dodoi{10.1038/ncomms10003}

\bibitem[{{Huenges} {et~al.}(1997){Huenges}, {Erzinger}, {K{\"u}ck}, {Engeser},
  \& {Kessels}}]{huenges1997physical}
{Huenges}, E., {Erzinger}, J., {K{\"u}ck}, J., {Engeser}, B., \& {Kessels}, W.
  1997, Journal of Geophysical Research: Solid Earth, 102, 18,255,
  \dodoi{10.1029/96JB03442}

\bibitem[{Ingersoll(1970)}]{ingersoll1970mars}
Ingersoll, A.~P. 1970, Science, 168, 972, \dodoi{10.1126/science.168.3934.972}

\bibitem[{Irwin~III {et~al.}(2013)Irwin~III, Tanaka, \&
  Robbins}]{irwin2013distribution}
Irwin~III, R.~P., Tanaka, K.~L., \& Robbins, S.~J. 2013, Journal of Geophysical
  Research: Planets, 118, 278

\bibitem[{Jakosky \& Phillips(2001)}]{jakosky_mars_2001}
Jakosky, B.~M., \& Phillips, R.~J. 2001, Nature, 412, 237,
  \dodoi{10.1038/35084184}

\bibitem[{Kadish {et~al.}(2010)Kadish, Head, \& Barlow}]{kadish2010pedestal}
Kadish, S.~J., Head, J.~W., \& Barlow, N.~G. 2010, Icarus, 210, 92

\bibitem[{Kehew {et~al.}(2009)Kehew, Lord, Kozlowski, \& Fisher}]{Kehew2009104}
Kehew, A.~E., Lord, M.~L., Kozlowski, A.~L., \& Fisher, T.~G. 2009, Proglacial
  megaflooding along the margins of the Laurentide ice sheet (Cambridge
  University Press), 104 – 127, \dodoi{10.1017/CBO9780511635632.007}

\bibitem[{Kingslake {et~al.}(2017)Kingslake, Ely, Das, \& Bell}]{Kingslake2017}
Kingslake, J., Ely, J.~C., Das, I., \& Bell, R.~E. 2017, Nature, 544, 349,
  \dodoi{10.1038/nature22049}

\bibitem[{Kite(2019)}]{kite_geologic_2019}
Kite, E.~S. 2019, Space Science Reviews, 215, 10,
  \dodoi{10.1007/s11214-018-0575-5}

\bibitem[{Kite {et~al.}(2013)Kite, Halevy, Kahre, Wolff, \&
  Manga}]{kite2013seasonal}
Kite, E.~S., Halevy, I., Kahre, M.~A., Wolff, M.~J., \& Manga, M. 2013, Icarus,
  223, 181

\bibitem[{Kite {et~al.}(2019)Kite, Mayer, Wilson, Davis, Lucas, \& Stucky~de
  Quay}]{kite2019persistence}
Kite, E.~S., Mayer, D.~P., Wilson, S.~A., {et~al.} 2019, {S}cience {A}dvances,
  5, eaav7710

\bibitem[{Kite {et~al.}(2022)Kite, Mischna, Fan, Morgan, Wilson, \&
  Richardson}]{kite2022changing}
Kite, E.~S., Mischna, M.~A., Fan, B., {et~al.} 2022, {S}cience {A}dvances, 8,
  eabo5894, \dodoi{10.1126/sciadv.abo5894}

\bibitem[{Kite \& Noblet(2022)}]{kite2022high}
Kite, E.~S., \& Noblet, A. 2022, Geophysical Research Letters, e2022GL101150

\bibitem[{Kite {et~al.}(2021)Kite, Steele, Mischna, \&
  Richardson}]{kite2021warm}
Kite, E.~S., Steele, L.~J., Mischna, M.~A., \& Richardson, M.~I. 2021,
  Proceedings of the National Academy of Sciences, 118, e2101959118,
  \dodoi{10.1073/pnas.2101959118}

\bibitem[{Kite {et~al.}(2014)Kite, Williams, Lucas, \& Aharonson}]{kite2014}
Kite, E.~S., Williams, J.-P., Lucas, A., \& Aharonson, O. 2014, Nature
  Geoscience, 7, 335, \dodoi{10.1038/ngeo2137}

\bibitem[{Kj{\ae}r {et~al.}(2018)Kj{\ae}r, Larsen, Binder, Bj{\o}rk, Eisen,
  Fahnestock, Funder, Garde, Haack, Helm, {et~al.}}]{kjaer2018large}
Kj{\ae}r, K.~H., Larsen, N.~K., Binder, T., {et~al.} 2018, {S}cience
  {A}dvances, 4, eaar8173

\bibitem[{Kress \& Head(2008)}]{kress2008ring}
Kress, A.~M., \& Head, J.~W. 2008, Geophysical Research Letters, 35

\bibitem[{Kumar(2005)}]{senthil2005structural}
Kumar, P.~S. 2005, Journal of Geophysical Research: Solid {E}arth, 110,
  \dodoi{https://doi.org/10.1029/2005JB003662}

\bibitem[{Kumar \& Kring(2008)}]{senthil2008impact}
Kumar, P.~S., \& Kring, D.~A. 2008, Journal of Geophysical Research: Planets,
  113, \dodoi{https://doi.org/10.1029/2008JE003115}

\bibitem[{Lacelle {et~al.}(2013)Lacelle, Davila, Fisher, Pollard, DeWitt,
  Heldmann, Marinova, \& McKay}]{lacelle2013excess}
Lacelle, D., Davila, A.~F., Fisher, D., {et~al.} 2013, Geochimica et
  Cosmochimica Acta, 120, 280

\bibitem[{Levy {et~al.}(2010)Levy, Head, \& Marchant}]{levy2010concentric}
Levy, J., Head, J.~W., \& Marchant, D.~R. 2010, Icarus, 209, 390

\bibitem[{Levy {et~al.}(2016)Levy, Fassett, \& Head}]{LEVY2016enhanced}
Levy, J.~S., Fassett, C.~I., \& Head, J.~W. 2016, Icarus, 264, 213,
  \dodoi{https://doi.org/10.1016/j.icarus.2015.09.037}

\bibitem[{Levy {et~al.}(2014)Levy, Fassett, Head, Schwartz, \&
  Watters}]{levy2014sequestered}
Levy, J.~S., Fassett, C.~I., Head, J.~W., Schwartz, C., \& Watters, J.~L. 2014,
  Journal of Geophysical Research: Planets, 119, 2188

\bibitem[{Levy {et~al.}(2009)Levy, Head, \& Marchant}]{levy2009concentric}
Levy, J.~S., Head, J.~W., \& Marchant, D.~R. 2009, Icarus, 202, 462

\bibitem[{Lewis {et~al.}(2006)Lewis, Marchant, Kowalewski, Baldwin, \&
  Webb}]{lewis2006age}
Lewis, A.~R., Marchant, D.~R., Kowalewski, D.~E., Baldwin, S.~L., \& Webb,
  L.~E. 2006, Geology, 34, 513, \dodoi{10.1130/G22145.1}

\bibitem[{Livingstone {et~al.}(2022)Livingstone, Li, Rutishauser, Sanderson,
  Winter, Mikucki, Bj{\"o}rnsson, Bowling, Chu, Dow,
  {et~al.}}]{livingstone2022subglacial}
Livingstone, S.~J., Li, Y., Rutishauser, A., {et~al.} 2022, Nature Reviews
  {{E}arth} \& Environment, 3, 106

\bibitem[{MacGregor {et~al.}(2019)MacGregor, Bottke~Jr., Fahnestock, Harbeck,
  Kjaer, Paden, Stillman, \& Studinger}]{macgregor2019possible}
MacGregor, J.~A., Bottke~Jr., W.~F., Fahnestock, M.~A., {et~al.} 2019,
  Geophysical Research Letters, 46, 1496,
  \dodoi{https://doi.org/10.1029/2018GL078126}

\bibitem[{{MacKinnon} \& {Tanaka}(1989)}]{mackinnon1989impacted}
{MacKinnon}, D.~J., \& {Tanaka}, K.~L. 1989, Journal of Geophysical Research:
  Solid Earth, 94, 17359, \dodoi{10.1029/JB094iB12p17359}

\bibitem[{Madeleine {et~al.}(2009)Madeleine, Forget, Head, Levrard, Montmessin,
  \& Millour}]{madeleine2009amazonian}
Madeleine, J.-B., Forget, F., Head, J.~W., {et~al.} 2009, Icarus, 203, 390

\bibitem[{Mahaffy {et~al.}(2015)Mahaffy, Webster, Stern, Brunner, Atreya,
  Conrad, Domagal-Goldman, Eigenbrode, Flesch, Christensen, Franz, Freissinet,
  Glavin, Grotzinger, Jones, Leshin, Malespin, McAdam, Ming, Navarro-Gonzalez,
  Niles, Owen, Pavlov, Steele, Trainer, Williford, Wray, \& the MSL
  Science~Team}]{mahaffy2015imprint}
Mahaffy, P.~R., Webster, C.~R., Stern, J.~C., {et~al.} 2015, Science, 347, 412,
  \dodoi{10.1126/science.1260291}

\bibitem[{Manga(2004)}]{manga2004martian}
Manga, M. 2004, Geophysical Research Letters, 31,
  \dodoi{https://doi.org/10.1029/2003GL018958}

\bibitem[{Mangold(2003)}]{mangold2003geomorphic}
Mangold, N. 2003, Journal of Geophysical Research: Planets, 108,
  \dodoi{https://doi.org/10.1029/2002JE001885}

\bibitem[{Mangold(2005)}]{MANGOLD2005high}
---. 2005, Icarus, 174, 336,
  \dodoi{https://doi.org/10.1016/j.icarus.2004.07.030}

\bibitem[{Mangold {et~al.}(2012{\natexlab{a}})Mangold, Adeli, Conway, Ansan, \&
  Langlais}]{mangold2012chronology}
Mangold, N., Adeli, S., Conway, S., Ansan, V., \& Langlais, B.
  2012{\natexlab{a}}, Journal of Geophysical Research: Planets, 117

\bibitem[{Mangold {et~al.}(2012{\natexlab{b}})Mangold, Kite, Kleinhans, Newsom,
  Ansan, Hauber, Kraal, Quantin, \& Tanaka}]{MANGOLD2012origin}
Mangold, N., Kite, E., Kleinhans, M., {et~al.} 2012{\natexlab{b}}, Icarus, 220,
  530, \dodoi{https://doi.org/10.1016/j.icarus.2012.05.026}

\bibitem[{Mann {et~al.}(1999)Mann, Leverington, Rayburn, \& Teller}]{Mann1999}
Mann, J.~D., Leverington, D.~W., Rayburn, J., \& Teller, J.~T. 1999, Journal of
  Paleolimnology, 22, 71, \dodoi{10.1023/A:1008090015161}

\bibitem[{Marchant {et~al.}(1993)Marchant, Denton, Sugden, \&
  III}]{marchant1993miocene}
Marchant, D.~R., Denton, G.~H., Sugden, D.~E., \& III, C. C.~S. 1993,
  Geografiska Annaler: Series A, Physical Geography, 75, 303,
  \dodoi{10.1080/04353676.1993.11880398}

\bibitem[{Matsuoka \& Murton(2008)}]{matsuoka2008frost}
Matsuoka, N., \& Murton, J. 2008, Permafrost and Periglacial Processes, 19, 195

\bibitem[{Mellon {et~al.}(2008)Mellon, Arvidson, Marlow, Phillips, \&
  Asphaug}]{mellon2008periglacial}
Mellon, M.~T., Arvidson, R.~E., Marlow, J.~J., Phillips, R.~J., \& Asphaug, E.
  2008, Journal of Geophysical Research: Planets, 113

\bibitem[{Michael \& Neukum(2010)}]{michael2010planetary}
Michael, G., \& Neukum, G. 2010, {E}arth and Planetary Science Letters, 294,
  223

\bibitem[{{Millour} {et~al.}(2018){Millour}, {Forget}, {Lopez-Valverde},
  {Lefevre}, {Gonzalez-Galindo}, {Lewis}, {Chaufray}, {Montabone}, {Zakharov},
  {Spiga}, \& {Vals}}]{millour2018exploring}
{Millour}, E., {Forget}, F., {Lopez-Valverde}, M., {et~al.} 2018, in 42nd
  COSPAR Scientific Assembly, Vol.~42, C4.3--6--18

\bibitem[{Mischna {et~al.}(2003)Mischna, Richardson, Wilson, \&
  McCleese}]{mischna2003orbital}
Mischna, M.~A., Richardson, M.~I., Wilson, R.~J., \& McCleese, D.~J. 2003,
  Journal of Geophysical Research: Planets, 108

\bibitem[{Montigny {et~al.}(2022)Montigny, Walwer, \&
  Michaut}]{montigny2022origin}
Montigny, A., Walwer, D., \& Michaut, C. 2022, {E}arth and Planetary Science
  Letters, 600, 117887

\bibitem[{Murphy \& Koop(2005)}]{murphy_review_2005}
Murphy, D.~M., \& Koop, T. 2005, Quarterly Journal of the Royal Meteorological
  Society, 131, 1539, \dodoi{10.1256/qj.04.94}

\bibitem[{Neupane {et~al.}(2019)Neupane, Chen, \& Cao}]{neupane2019review}
Neupane, R., Chen, H., \& Cao, C. 2019, Geomatics, Natural Hazards and Risk,
  10, 1948, \dodoi{10.1080/19475705.2019.1652210}

\bibitem[{Nye(1976)}]{nye1976water}
Nye, J.~F. 1976, Journal of Glaciology, 17, 181

\bibitem[{Osinski \& Lee(2005)}]{osinski2005intra}
Osinski, G.~R., \& Lee, P. 2005, Meteoritics \& Planetary Science, 40, 1887

\bibitem[{Palumbo \& Head(2018)}]{palumbo2018impact}
Palumbo, A.~M., \& Head, J.~W. 2018, Meteoritics \& Planetary Science, 53, 687,
  \dodoi{https://doi.org/10.1111/maps.13001}

\bibitem[{Pierce \& Crown(2003)}]{PIERCE200346}
Pierce, T.~L., \& Crown, D.~A. 2003, Icarus, 163, 46,
  \dodoi{https://doi.org/10.1016/S0019-1035(03)00046-0}

\bibitem[{Polanskey \& Ahrens(1990)}]{POLANSKEY1990impact}
Polanskey, C.~A., \& Ahrens, T.~J. 1990, Icarus, 87, 140,
  \dodoi{https://doi.org/10.1016/0019-1035(90)90025-5}

\bibitem[{Prevec {et~al.}(2000)Prevec, Lightfoot, \&
  Keays}]{prevec2000evolution}
Prevec, S., Lightfoot, P., \& Keays, R. 2000, Lithos, 51, 271,
  \dodoi{https://doi.org/10.1016/S0024-4937(00)00005-0}

\bibitem[{Ramirez(2017)}]{ramirez_volcanic_2017}
Ramirez, R.~M. 2017, Icarus, 297, 71, \dodoi{10.1016/j.icarus.2017.06.025}

\bibitem[{Ramirez {et~al.}(2014)Ramirez, Kopparapu, Zugger, Robinson, Freedman,
  \& Kasting}]{ramirez2014}
Ramirez, R.~M., Kopparapu, R., Zugger, M.~E., {et~al.} 2014, Nature Geoscience,
  7, 59, \dodoi{10.1038/ngeo2000}

\bibitem[{{Rice} {et~al.}(2015){Rice}, {Tsai}, {Fernandes}, \&
  {Platt}}]{rice2015}
{Rice}, J.~R., {Tsai}, V.~C., {Fernandes}, M.~C., \& {Platt}, J.~D. 2015,
  Journal of Applied Mechanics, 82, 071001, \dodoi{10.1115/1.4030325}

\bibitem[{Robbins \& Hynek(2012{\natexlab{a}})}]{robbins2012global}
Robbins, S.~J., \& Hynek, B.~M. 2012{\natexlab{a}}, Journal of Geophysical
  Research: Planets, 117, \dodoi{https://doi.org/10.1029/2011JE003966}

\bibitem[{Robbins \& Hynek(2012{\natexlab{b}})}]{robbins2012new}
---. 2012{\natexlab{b}}, Journal of Geophysical Research: Planets, 117

\bibitem[{Robin {et~al.}(1969)Robin, Evans, \&
  Bailey}]{robin1969interpretation}
Robin, G. d.~Q., Evans, S., \& Bailey, J.~T. 1969, Philosophical Transactions
  of the Royal Society of London. Series A, Mathematical and Physical Sciences,
  265, 437

\bibitem[{Salese {et~al.}(2016)Salese, Ansan, Mangold, Carter, Ody, Poulet, \&
  Ori}]{salese2016sedimentary}
Salese, F., Ansan, V., Mangold, N., {et~al.} 2016, Journal of Geophysical
  Research: Planets, 121, 2239, \dodoi{https://doi.org/10.1002/2016JE005039}

\bibitem[{Sato {et~al.}(2010)Sato, Kurita, \& Baratoux}]{sato2010formation}
Sato, H., Kurita, K., \& Baratoux, D. 2010, Icarus, 207, 248,
  \dodoi{https://doi.org/10.1016/j.icarus.2009.10.023}

\bibitem[{Scanlon {et~al.}(2013)Scanlon, Head, Madeleine, Wordsworth, \&
  Forget}]{scanlon2013orographic}
Scanlon, K.~E., Head, J.~W., Madeleine, J.-B., Wordsworth, R.~D., \& Forget, F.
  2013, Geophysical Research Letters, 40, 4182

\bibitem[{Scheller {et~al.}(2021)Scheller, Ehlmann, Hu, Adams, \&
  Yung}]{scheller2021longterm}
Scheller, E.~L., Ehlmann, B.~L., Hu, R., Adams, D.~J., \& Yung, Y.~L. 2021,
  Science, 372, 56, \dodoi{10.1126/science.abc7717}

\bibitem[{Schultz(1976)}]{Schultz1976floor}
Schultz, P.~H. 1976, The {M}oon, 15, 241, \dodoi{10.1007/BF00562240}

\bibitem[{Segura {et~al.}(2012)Segura, McKay, \& Toon}]{segura2012impact}
Segura, T.~L., McKay, C.~P., \& Toon, O.~B. 2012, Icarus, 220, 144,
  \dodoi{https://doi.org/10.1016/j.icarus.2012.04.013}

\bibitem[{Sizemore {et~al.}(2015)Sizemore, Zent, \&
  Rempel}]{sizemore2015initiation}
Sizemore, H.~G., Zent, A.~P., \& Rempel, A.~W. 2015, Icarus, 251, 191

\bibitem[{Smith {et~al.}(2009)Smith, Fricker, Joughin, \&
  Tulaczyk}]{smith2009inventory}
Smith, B.~E., Fricker, H.~A., Joughin, I.~R., \& Tulaczyk, S. 2009, Journal of
  Glaciology, 55, 573–595, \dodoi{10.3189/002214309789470879}

\bibitem[{Smith {et~al.}(2001)Smith, Zuber, Frey, Garvin, Head, Muhleman,
  Pettengill, Phillips, Solomon, Zwally, {et~al.}}]{smith2001mars}
Smith, D.~E., Zuber, M.~T., Frey, H.~V., {et~al.} 2001, Journal of Geophysical
  Research: Planets, 106, 23689

\bibitem[{Steele {et~al.}(2017)Steele, Balme, \& Lewis}]{steele2017regolith}
Steele, L.~J., Balme, M.~R., \& Lewis, S.~R. 2017, Icarus, 284, 233,
  \dodoi{https://doi.org/10.1016/j.icarus.2016.11.023}

\bibitem[{Stokes {et~al.}(2019)Stokes, Sanderson, Miles, Jamieson, \&
  Leeson}]{Stokes2019widespread}
Stokes, C.~R., Sanderson, J.~E., Miles, B. W.~J., Jamieson, S. S.~R., \&
  Leeson, A.~A. 2019, Scientific Reports, 9, 13823,
  \dodoi{10.1038/s41598-019-50343-5}

\bibitem[{Tanaka {et~al.}(2014)Tanaka, Robbins, Fortezzo, Skinner~Jr, \&
  Hare}]{tanaka2014digital}
Tanaka, K.~L., Robbins, S., Fortezzo, C., Skinner~Jr, J., \& Hare, T.~M. 2014,
  Planetary and Space Science, 95, 11

\bibitem[{Terrell \& Scott(1985)}]{terrell1985oversmoothed}
Terrell, G.~R., \& Scott, D.~W. 1985, Journal of the American Statistical
  Association, 80, 209

\bibitem[{Toon {et~al.}(1997)Toon, Zahnle, Morrison, Turco, \&
  Covey}]{toon1997environmental}
Toon, O.~B., Zahnle, K., Morrison, D., Turco, R.~P., \& Covey, C. 1997, Reviews
  of Geophysics, 35, 41, \dodoi{https://doi.org/10.1029/96RG03038}

\bibitem[{Turbet {et~al.}(2020)Turbet, Gillmann, Forget, Baudin, Palumbo, Head,
  \& Karatekin}]{turbet2020environmental}
Turbet, M., Gillmann, C., Forget, F., {et~al.} 2020, Icarus, 335, 113419

\bibitem[{Walder \& Costa(1996)}]{walder1996outburst}
Walder, J.~S., \& Costa, J.~E. 1996, {E}arth Surface Processes and Landforms,
  21, 701

\bibitem[{Walwer {et~al.}(2021)Walwer, Michaut, Pinel, \&
  Adda-Bedia}]{walwer2021magma}
Walwer, D., Michaut, C., Pinel, V., \& Adda-Bedia, M. 2021, Physics of the
  {{E}arth} and Planetary Interiors, 312, 106658

\bibitem[{Warner {et~al.}(2015)Warner, Gupta, Calef, Grindrod, Boll, \&
  Goddard}]{warner2015minimum}
Warner, N.~H., Gupta, S., Calef, F., {et~al.} 2015, Icarus, 245, 198,
  \dodoi{https://doi.org/10.1016/j.icarus.2014.09.024}

\bibitem[{Warren {et~al.}(2021)Warren, Holo, Kite, \&
  Wilson}]{warren2021overspilling}
Warren, A., Holo, S., Kite, E., \& Wilson, S. 2021, {E}arth and Planetary
  Science Letters, 554, 116671

\bibitem[{Weiss \& Head(2015)}]{weiss2015crater}
Weiss, D.~K., \& Head, J.~W. 2015, Planetary and Space Science, 117, 401

\bibitem[{{Willis} {et~al.}(2015){Willis}, {Herried}, {Bevis}, \&
  {Bell}}]{willis2015recharge}
{Willis}, M.~J., {Herried}, B.~G., {Bevis}, M.~G., \& {Bell}, R.~E. 2015,
  Nature, 518, 223, \dodoi{10.1038/Nature14116}

\bibitem[{{Wilson} \& {Howard}(2021)}]{wilson2021distribution}
{Wilson}, S.~A., \& {Howard}, A.~D. 2021, in 52nd Lunar and Planetary Science
  Conference, Lunar and Planetary Science Conference, 1105

\bibitem[{Wilson {et~al.}(2016)Wilson, Howard, Moore, \&
  Grant}]{wilson_cold-wet_2016}
Wilson, S.~A., Howard, A.~D., Moore, J.~M., \& Grant, J.~A. 2016, Journal of
  Geophysical Research: Planets, 121, 1667, \dodoi{10.1002/2016JE005052}

\bibitem[{{Wilson} {et~al.}(2021){Wilson}, {Morgan}, {Howard}, \&
  {Grant}}]{wilson2021global}
{Wilson}, S.~A., {Morgan}, A.~M., {Howard}, A.~D., \& {Grant}, J.~A. 2021,
  Geophysical Research Letters, 48, e91653, \dodoi{10.1029/2020GL091653}

\bibitem[{Woodley {et~al.}(2022)Woodley, Butcher, Fawdon, Clark, Ng, Davis, \&
  Gallagher}]{WOODLEY2022multiple}
Woodley, S.~Z., Butcher, F.~E., Fawdon, P., {et~al.} 2022, Icarus, 386, 115147,
  \dodoi{https://doi.org/10.1016/j.icarus.2022.115147}

\bibitem[{Wordsworth {et~al.}(2013)Wordsworth, Forget, Millour, Head,
  Madeleine, \& Charnay}]{wordsworth_global_2013}
Wordsworth, R., Forget, F., Millour, E., {et~al.} 2013, Icarus, 222, 1,
  \dodoi{10.1016/j.icarus.2012.09.036}

\bibitem[{Wordsworth {et~al.}(2017)Wordsworth, Kalugina, Lokshtanov, Vigasin,
  Ehlmann, Head, Sanders, \& Wang}]{wordsworth2017}
Wordsworth, R., Kalugina, Y., Lokshtanov, S., {et~al.} 2017, Geophysical
  Research Letters, 44(2), 665, \dodoi{10.1002/2016GL071766}

\bibitem[{Wordsworth {et~al.}(2021)Wordsworth, Knoll, Hurowitz, Baum, Ehlmann,
  Head, \& Steakley}]{Wordsworth2021coupled}
Wordsworth, R., Knoll, A.~H., Hurowitz, J., {et~al.} 2021, Nature Geoscience,
  14, 127, \dodoi{10.1038/s41561-021-00701-8}

\end{thebibliography}
\bibliographystyle{aasjournal}

\appendix

\section{Groundwater Model Output}
\label{sec:gwat_full}

\begin{figure}[t!]
\centering
\includegraphics[width=0.9\textwidth]{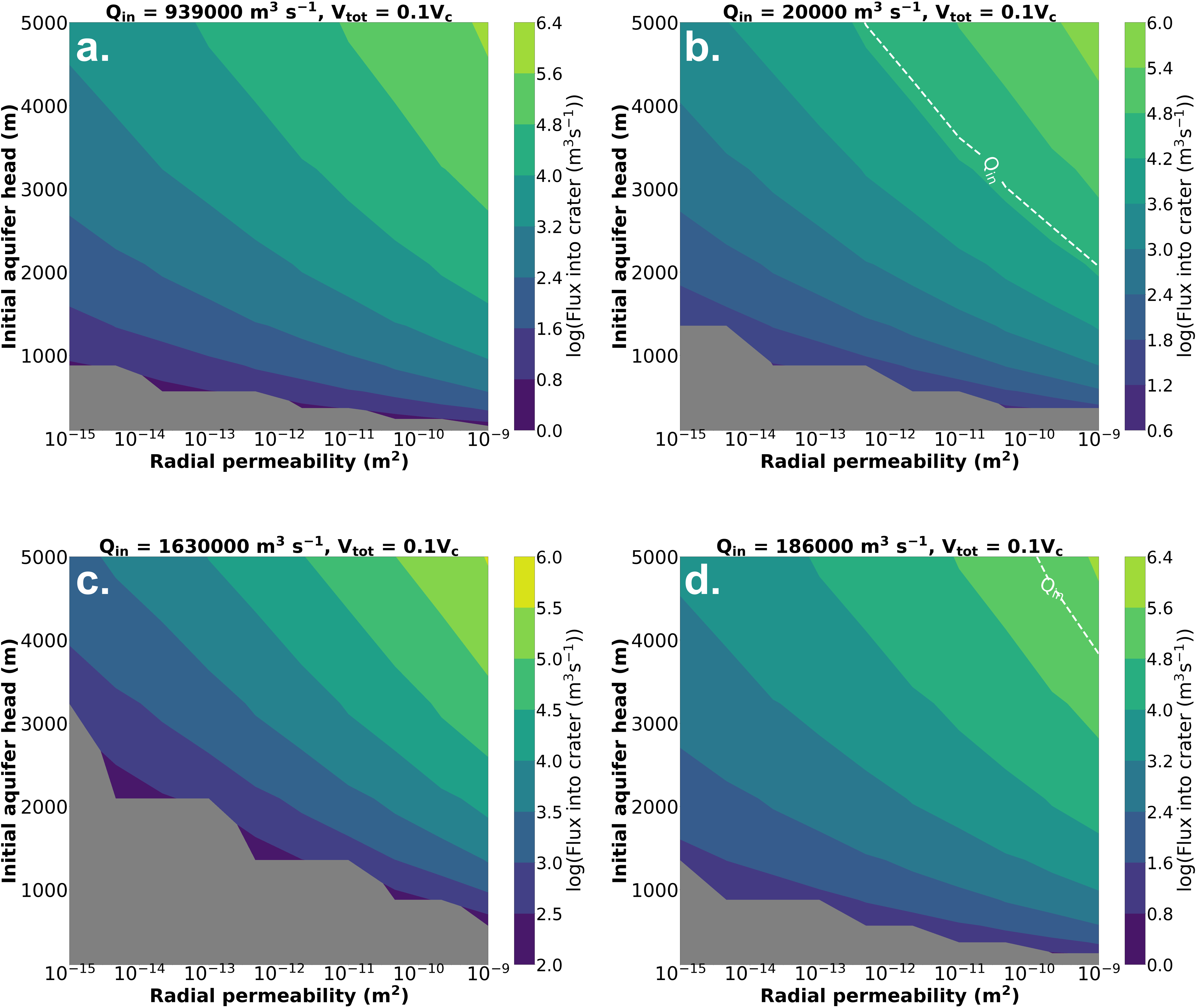}
\caption{Groundwater flux into each multiple exit breach tadpole crater as a function of radial permeability and initial aquifer head when 0.1 crater volumes of water have been discharged from the aquifer. \textbf{a.} \textit{Arabia Terra 1}, \textbf{b.} \textit{Arabia Terra 2}, \textbf{c.} \textit{Hellas}, \textbf{d.} \textit{Noachis}. Dashed white lines indicate the water supply rate necessary to overspill all breaches simultaneously (Table~\ref{tab:qin}). There is no dashed line on c. because the required input flux is not reached in our model. Grey shaded areas indicate negligible groundwater flow into the crater on the timescale of one model timestep.
\label{fig:all_gwat}}
\end{figure}

\newpage
\section{Pond Overspill Model Output}
\label{sec:ovsp_full}

\begin{figure}[t!]
\centering
\includegraphics[width=0.9\textwidth]{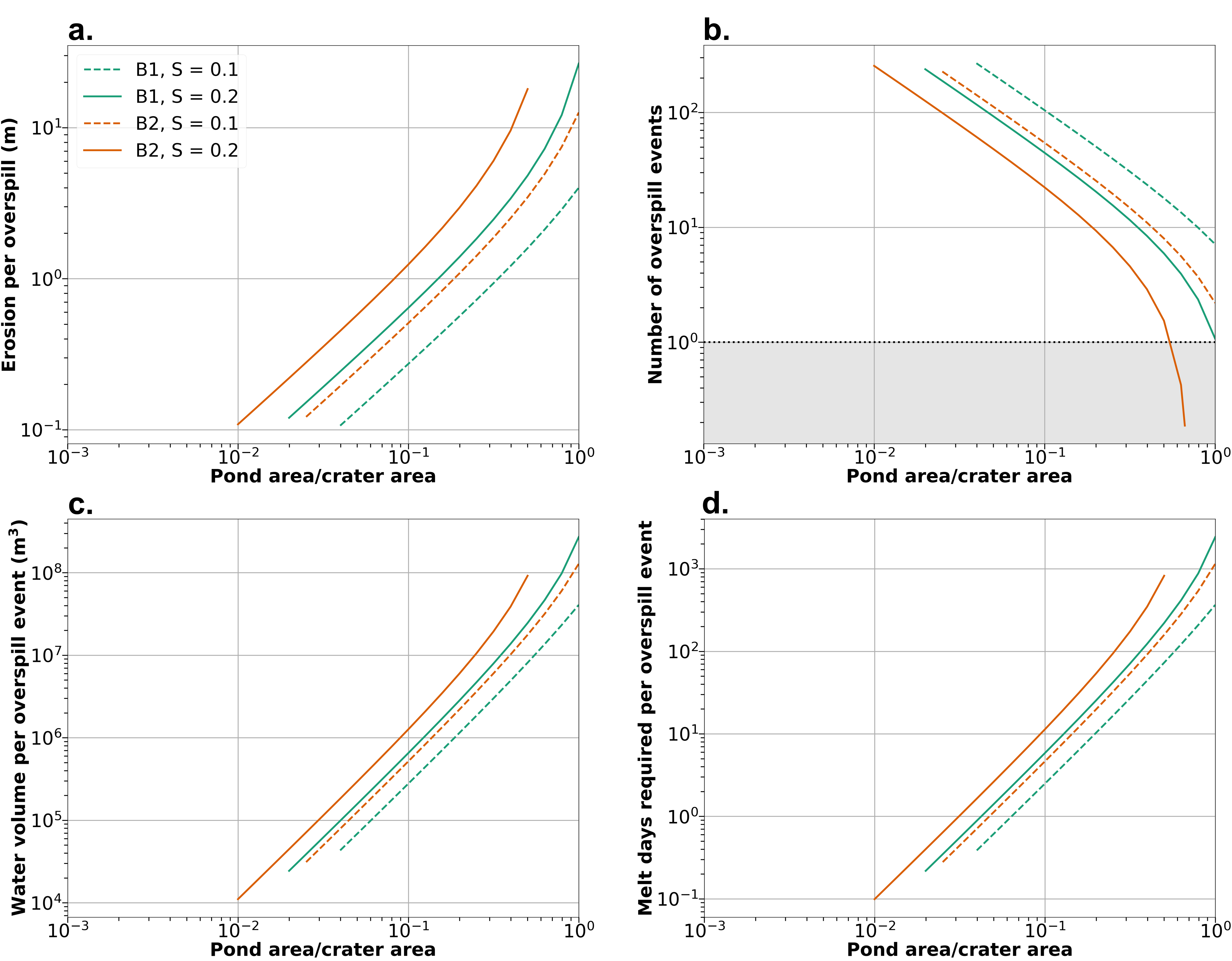}
\caption{Meltwater pond overspill modelling results for \textit{Arabia Terra 1} B1 and B2 (Fig.~\ref{fig:profs}a) for measured breach parameters in Table~\ref{tab:breach} as a function of fractional melt pond area. \textbf{a.} Downward erosion of breach base per individual overspilling event. Smaller melt pond areas generate less erosion when they overspill the rim. \textbf{b.} The number of overspill events required to carve each observed breach. The number of breaches required decreases as melt pond area increases. \textbf{d.} Total water volume required per overspill event in \textbf{b.}. \textbf{d.} Melt days required to generate the water volumes required in \textbf{c.} at a melt rate of 4~m~yr$^{-1}$~m$^{2}$ \citep{kite2013seasonal} across the entire crater area. 
\label{fig:oat1}}
\end{figure}

\begin{figure}[t!]
\centering
\includegraphics[width=0.9\textwidth]{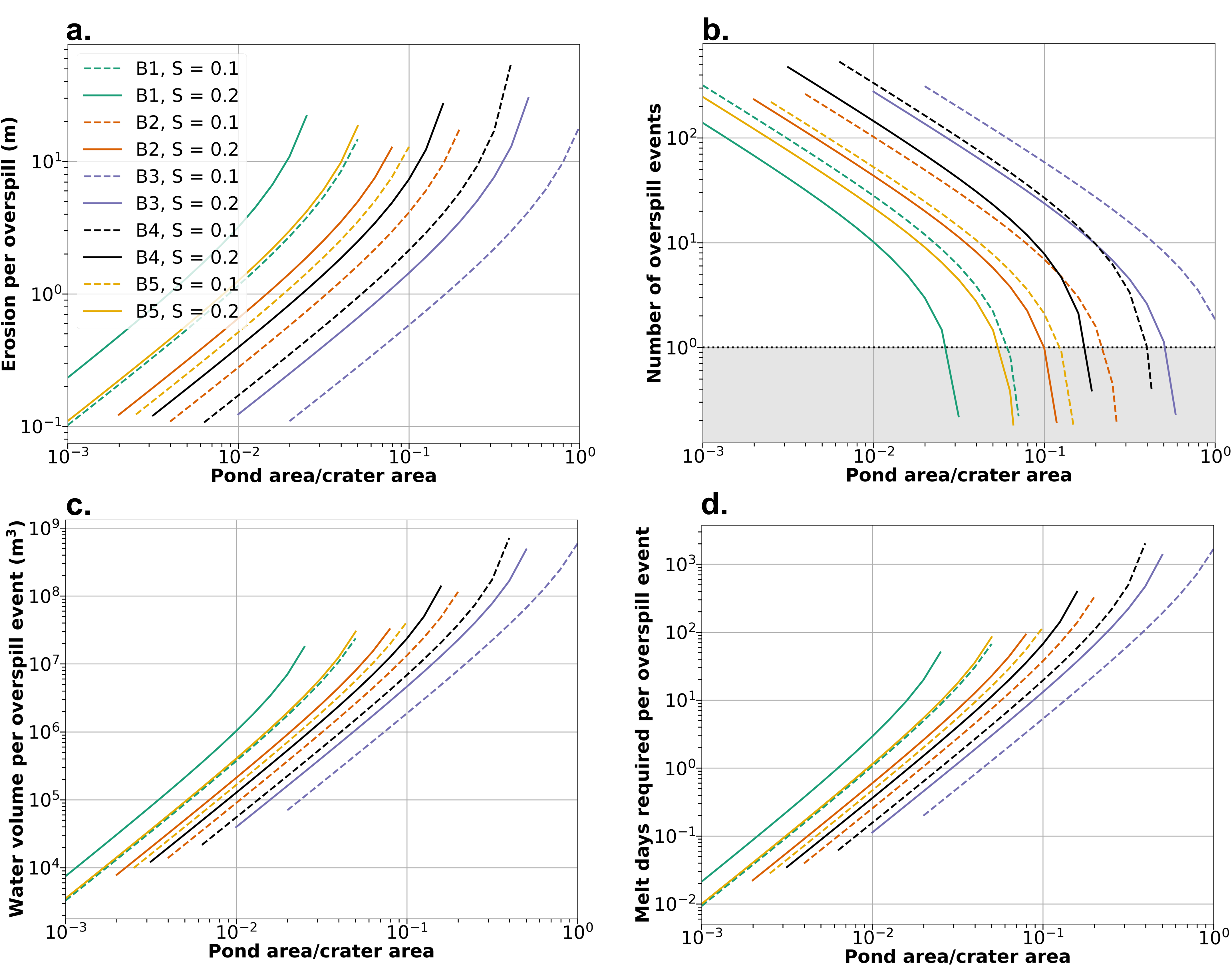}
\caption{Meltwater pond overspill modelling results for \textit{Arabia Terra 2} B1, B2, B3, B4, and B5 (Fig.~\ref{fig:profs}b) for measured breach parameters in Table~\ref{tab:breach} as a function of fractional melt pond area. \textbf{a.} Downward erosion of breach base per individual overspilling event. Smaller melt pond areas generate less erosion when they overspill the rim. \textbf{b.} The number of overspill events required to carve each observed breach. The number of breaches required decreases as melt pond area increases. \textbf{d.} Total water volume required per overspill event in \textbf{b.}. \textbf{d.} Melt days required to generate the water volumes required in \textbf{c.} at a melt rate of 4~m~yr$^{-1}$~m$^{2}$ \citep{kite2013seasonal} across the entire crater area. 
\label{fig:oat2}}
\end{figure}

\begin{figure}[t!]
\centering
\includegraphics[width=0.9\textwidth]{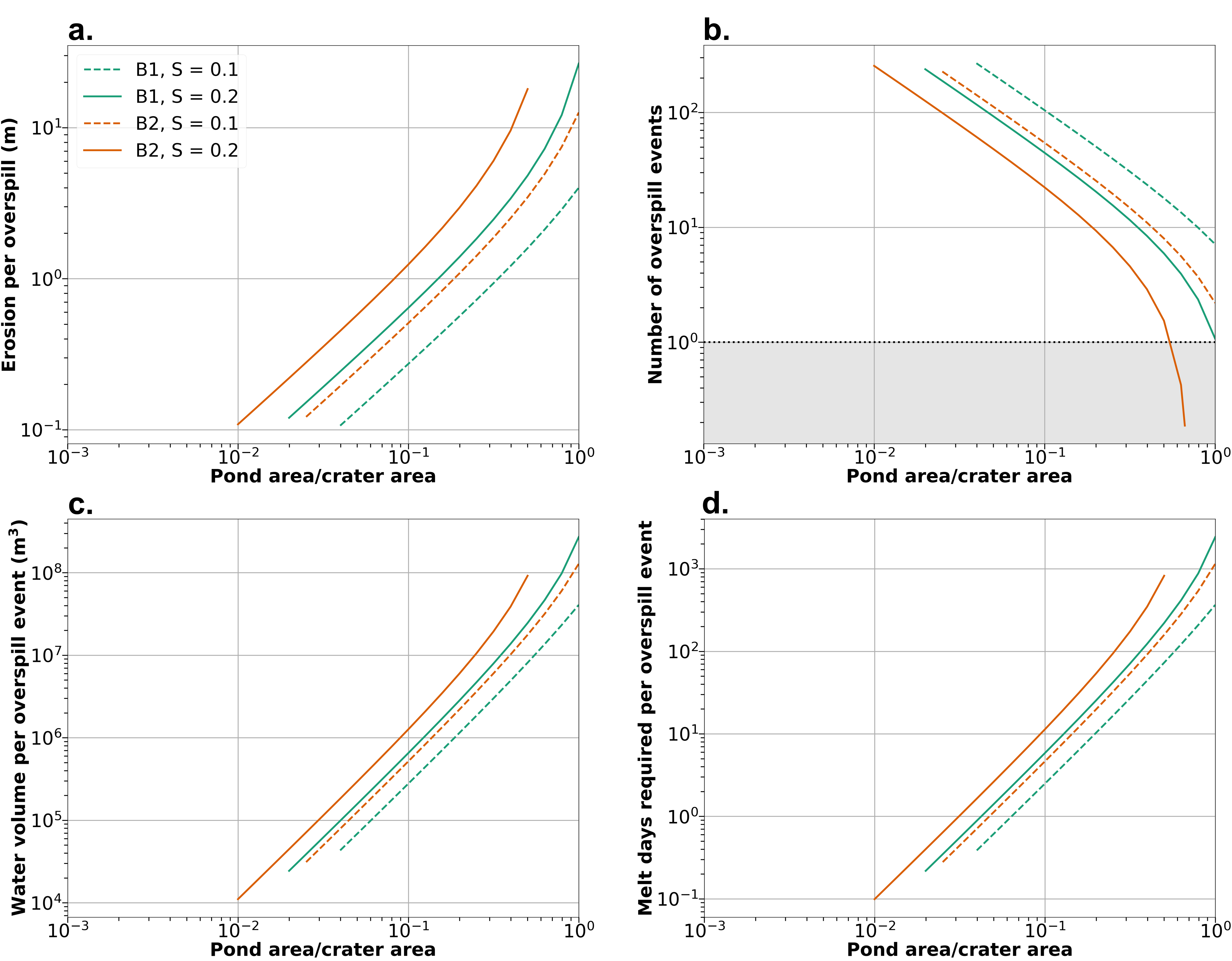}
\caption{Meltwater pond overspill modelling results for \textit{Hellas} B1 and B2 (Fig.~\ref{fig:profs}c) for measured breach parameters in Table~\ref{tab:breach} as a function of fractional melt pond area. \textbf{a.} Downward erosion of breach base per individual overspilling event. Smaller melt pond areas generate less erosion when they overspill the rim. \textbf{b.} The number of overspill events required to carve each observed breach. The number of breaches required decreases as melt pond area increases. \textbf{d.} Total water volume required per overspill event in \textbf{b.}. \textbf{d.} Melt days required to generate the water volumes required in \textbf{c.} at a melt rate of 4~m~yr$^{-1}$~m$^{2}$ \citep{kite2013seasonal} across the entire crater area. 
\label{fig:on}}
\end{figure}




\end{document}